\documentclass[a4paper,12pt]{article}
\pdfoutput=1 

\usepackage{amsmath,amssymb}
\usepackage{mathtools}
\usepackage[pdftex]{hyperref}
\hypersetup{colorlinks=true,linkcolor=blue,urlcolor=blue,filecolor=blue,citecolor=blue}
\usepackage{epsf}
\usepackage{cite}
\usepackage{fancyhdr}
\usepackage{enumerate} 
\usepackage[usenames]{xcolor}
\usepackage{physics}
\usepackage{ytableau}
\usepackage{graphicx}
 
\usepackage{tcolorbox}
\usepackage{makecell}
\usepackage{comment}
\usepackage{tikz}
\usepackage{pgfplots}
\usepackage{pgfplotstable}
\usepgfplotslibrary{fillbetween}
\usepackage{url}
\usetikzlibrary{patterns,patterns.meta}

\makeatletter

\usepackage[dvipsnames]{xcolor}
\definecolor{amethyst}{rgb}{0.54, 0.17, 0.89}
\definecolor{coral}{rgb}{1.0, 0.3, 0.4}

\newcommand{\RAdS}{R_{\rm AdS}}

\newcommand{\NplusNC}{M}

\g@addto@macro\bfseries{\boldmath}

\def\refchecklabelfontsize{\fontsize{5pt}{5pt}\selectfont}
\let\mark@size=\refchecklabelfontsize

\def\half{{\frac{1}{2}}}

\def\p{\partial}
\def\pb{\bar{\partial}}
\def\unit{{1\kern-.65ex {\rm l}}}
\def\1{{1\kern-.65ex {\rm l}}}

\def\Im{\mathop{\mathrm{Im}}\nolimits}

\def\Re{\mathop{\mathrm{Re}}\nolimits}

\def\Tr{\mathop{\mathrm{Tr}}\nolimits}

\def\bra#1{{\langle{#1}|}}

\def\ket#1{{|{#1}\rangle}}
\def\Ket#1{{\left|{#1}\right\rangle}}

\let\ev=\braket
\let\Ev=\Braket

\def\fb{{\bar{f}}}
\def\gb{{\bar{g}}}

\def\wb{{\overline{w}}}

\def\zb{{\overline{z}}}

\def\cA{{\cal A}}
\def\cB{{\cal B}}
\def\cC{{\cal C}}

\def\cF{{\cal F}}

\def\cH{{\cal H}}

\def\cN{{\cal N}}
\def\cO{{\cal O}}
\def\cP{{\cal P}}

\def\bbC{{\mathbb{C}}}

\def\bbR{{\mathbb{R}}}

\def\bbZ{{\mathbb{Z}}}

\newcount\hour \newcount\minute
\hour=\time \divide \hour by 60
\minute=\time
\def\now{%
\ifnum \hour<13
  \ifnum \hour=0 \advance \hour by 12 \number\hour:\else \number\hour:\fi%
     \ifnum \minute<10 0\fi%
     \number\minute%
\ A.M.%
\else \advance \hour by -12 \number\hour:%
  \ifnum \minute<10 0\fi%
  \number\minute%
  \ P.M.%
\fi%
}

\makeatother

\usepackage{xcolor}
\definecolor{amethyst}{rgb}{0.54, 0.17, 0.89}
\definecolor{coral}{rgb}{1.0, 0.3, 0.4}

\begin{document}

\baselineskip=18pt  
\numberwithin{equation}{section}  

%



\thispagestyle{empty}

\vspace*{-2cm} 
\begin{flushright}
YITP-26-163\\
\end{flushright}

\vspace*{1.0cm} 
\begin{center}
%
%
   {\LARGE Superstars, quantum wormholes 
  \\[1ex]
  and von Neumann algebras
  }\\

 \vspace*{1.5cm}
 Vijay Balasubramanian$^{1,2,3}$, Ryo Nemoto$^4$, and Masaki Shigemori$^{4,5}$\\
 \vspace*{0.8cm} 
$^1$ David Rittenhouse Laboratory, University of Pennsylvania, 209 S. 33rd Street, Philadelphia, Pennsylvania 19104, USA\\[1ex]
$^2$
Santa Fe Institute, 1399 Hyde Park Road, Santa Fe, NM 87501, USA \\
[1ex]
$^3$
Theoretische Natuurkunde, Vrije Universiteit Brussel, \\ Pleinlaan 2, B-1050 Brussels, Belgium\\[1ex]
$^4$ 
Department of Physics, Nagoya University\\
Furo-cho, Chikusa-ku, Nagoya 464-8602, Japan\\[1ex]
$^5$
Center for Gravitational Physics,\\
Yukawa Institute for Theoretical Physics, Kyoto University\\
Kitashirakawa Oiwakecho, Sakyo-ku, Kyoto 606-8502, Japan \\
 \vspace*{0.8cm} 
 
\end{center}
\vspace*{0cm}

\noindent
We entangle heavy 1/2-BPS states in two copies of  $\cN=4$ SU($N$) Super-Yang-Mills (SYM) theory.  Each copy is separately dual to an extremal, supersymmetric, singular black hole geometry, in which the horizon coincides with a  singularity.  General ideas in the AdS/CFT correspondence suggest that the entanglement between the two copies should lead to some form of connectivity between the two geometries---a kind of wormhole.  
We use von Neumann algebra methods in the SYM theory to diagnose the possible existence of such wormholes.

\newpage
\setcounter{page}{1} 

\tableofcontents





\section{Introduction}
\label{sec:intro}
Multiple lines of evidence suggest that connectivity of spacetime regions is associated with entanglement between the underlying degrees of freedom, at least in asymptotically AdS spacetimes with a dual conformal field theory (CFT) \cite{VanRaamsdonk2009,VanRaamsdonk2010,Maldacena:2013xja}. 
This relationship is often summarized by the slogan ER=EPR, where ``ER'' refers to an Einstein-Rosen bridge and ``EPR'', derived from Einstein-Podolsky-Rosen pairs, refers to entanglement in the underlying quantum state. However, we do not understand precisely how this connectivity arises, how much entanglement is necessary, and how it should be structured.  Some constraints are known---the entanglement should implement a particular form of quantum error correction \cite{Almheiri:2014lwa}, and should obey an infinite set of inequalities known collectively as the holographic entropy cone \cite{Bao:2015bfa}.  We also know that  multiparty entanglement is necessary for knitting spacetime together \cite{Balasubramanian:2014hda,Akers:2019gcv}.

One way of testing the conjectured relation between connectivity and entanglement is to construct  connected geometries associated to novel entangled states.  For example, if we Higgs the $SU(N)$ $\mathcal{N} =4$ Super-Yang-Mills (SYM) theory to $SU(N/2) \times SU(N/2)$, we can construct a state with substantial entanglement between the two factors, predicting the existence of a wormhole connecting two distant ``throats'' in the dual AdS$_5$ spacetime.  The authors of Ref.~\cite{Balasubramanian:2020ffd}\footnote{See \cite{Begines:2026mlv} for recent related work.} explicitly demonstrated that this wormhole exists.  In this case the SYM configuration has a slow instability towards symmetry restoration which is manifested in the dual geometry as a slow merging of the distant throats accompanied by vanishing of the wormhole.  

Here, we consider a more controlled setting---the 1/2-BPS sector of the $SU(N)$ $\mathcal{N} =4$ SYM theory.  We have a complete account of how to construct these states in SYM theory, and an algorithm for constructing the dual geometries for states that are under semiclassical control \cite{Lin:2004nb}.  If taken literally the map in \cite{Lin:2004nb} would associate Planck-sized topological features in spacetime to many 1/2-BPS states. However, if we want to have reliable geometric descriptions, we should really coarse-grain both the states and the geometry to arrive at effective descriptions appropriate to semiclassical physics.  In this way, \cite{Balasubramanian:2005mg} showed that  typical heavy states in  charge-fixed ensembles of the 1/2-BPS sector of $\mathcal{N} =4$ SYM have coarse-grained dual descriptions as charged, extremal black holes in AdS$_5$, i.e., the ``superstars'' of \cite{Myers:2001aq}, and depending on the choice of ensemble, the related ``hyperstars'' \cite{Balasubramanian:2005mg} which are supersymmetric solutions with structure in the full 10 dimensions of the $\rm{AdS}_5 \times S^5$ geometry that the SYM theory is dual to.\footnote{An analogous coarse-graining procedure was carried out for extremal, asymptotically AdS$_3$ black holes in \cite{Balasubramanian:2005qu}.}  These extremal black hole solutions are singular, in the sense that their horizons have vanishing areas and coincide with the singularity.    In this paper, we entangle two such typical heavy states in two copies of the SYM theory, a procedure previously carried out in \cite{Simon:2018laf}.  If entanglement in the fundamental SYM description leads to connectivity in the dual AdS geometric description, then we expect that the dual description of our states will involve a wormhole connecting two asymptotically AdS$_5$ universes through their apparently singular horizons.  

It is difficult to test this idea directly in the gravitating theory because the singularity at the horizon means that the classical description is breaking down---at best we will have some kind of ``quantum wormhole'' \cite{Simon:2018laf}.  To diagnose what this might mean, we turn to a recently proposed idea: connectivity between regions can be diagnosed by the type of the von Neumann algebra realized by operators acting on  either region \cite{Leutheusser:2021qhd, Engelhardt:2023xer}. In our case the two regions are different asymptotically AdS$_5$ universes, so we will consider operators acting within the half-BPS sectors of the separate CFT factors to which they are dual.  The rule proposed in \cite{Leutheusser:2021qhd} is that if two regions are disconnected the algebra of either region by itself will be Type I\@.  If they are connected, the algebra will not be Type I\@.  A classical connection or wormhole is associated with a Type III$_1$ algebra accompanied by an additional classicality condition.  

Thus, in this paper we explicitly determine the type of the algebras generated by operators acting on either CFT factor in the heavy, supersymmetric entangled states that we discussed above. We consider three kinds of probes, with different interpretations in the dual gravity description:
(i)~coarse BPS probes that IR observers have access to, and that modify the boundary condition in the interior of the classically singular locus of the extremal black hole as seen from either asymptotic region; (ii) fine bosonic BPS probes that modify the boundary conditions at the edge of the singular locus; and (iii) fine fermionic BPS probes that also modify these edge boundary conditions.  We find that coarse algebras are Type~I in the large $N$ limit.  Following \cite{Leutheusser:2021qhd,Engelhardt:2023xer} this indicates that, probed in this coarse way, the two entangled universes appear to be disconnected. The bosonic and fermionic fine probes lead to Type~II$_\infty$ and Type~III$_\gamma$ algebras (with $\gamma \neq 0,1$) in the large $N$ limit depending on the specific state that they are acting on. Again following \cite{Leutheusser:2021qhd,Engelhardt:2023xer} this indicates that, probed in this fine way, the two entangled universes appear to be connected by a quantum wormhole.    Note that we are only considering the algebra of operators acting \emph{within} a BPS sector.\footnote{The complete algebra of operators acting on our half-BPS states considered in perturbation theory in $1/N$ will likely be of Type II$_\infty$ because arbitrary excitations of our states will describe eternal, non-extremal black holes with large classical wormholes behind the horizon with the algebraic type determined by the crossed-product construction of \cite{Witten:2021unn}.}  The idea here is to analyze the algebraic structure of just the half-BPS sector of the $SU(N)$ $\mathcal{N} =4$ SYM theory, and the dual geometries that describe it.

Six sections follow.  In Sec.~\ref{sec:preliminaries} we review the different types of von Neumann algebra and their conjectured relation to connectivity in spacetime \cite{Leutheusser:2021qhd,Engelhardt:2023xer}. 
In Sec.~\ref{sec:halfBPS} we discuss the construction of the 1/2 BPS states of $\mathcal{N} = 4$ SU(N) SYM theory, and their dual asymptotically AdS$_5$ geometries \cite{Lin:2004nb}. We also construct  typical  heavy BPS states that are entangled between two copies of SYM theory \cite{Simon:2018laf}, and explain how they have a  coarse-grained description in spacetime as entangled extremal black holes in two asymptotically AdS$_5$ universes. In Sec.~\ref{sec:coarse} we discuss the algebra of coarse probes of such entangled spacetimes that are accessible to infrared observers. Then in Sec.~\ref{sec:fineBoson}  and Sec.~\ref{sec:fineFermion}  we discuss finer probe algebras of two kinds, bosonic versus fermionic.   We conclude in Sec.~\ref{sec:discussion} with a discussion of our results, and an observation of a possible relation between the algebraic types of perturbations and instabilities of the classical solution describing the background state.

\section{Algebraic preliminaries}
\label{sec:preliminaries}
In this section, we will review concepts concerning von Neumann algebras that we will need for our analysis (see \cite{Sorce:2023fdx, Connes1994, Witten:2018zxz, TakesakibookI,TakesakibookII,TakesakibookIII, Kazama2024, Haag:1992hx, Liu:2025krl,Ueda-sanNote,Bratteli:1979tw,Bratteli:1996xq} for details). 

\subsection{Von Neumann algebras}
\label{sec:vnprelim}

Let $\mathcal{H}$ be the Hilbert space of a physical system, 
and let $\cB(\cH)$ be the set of bounded operators acting on $\cH$.
Now let $\mathcal{A}\subset \cB(\cH)$ be closed under operator multiplication, addition, scalar multiplication, and Hermitian conjugation. 
We regard $\mathcal{A}$ as an algebra of observables associated with a quantum subsystem. 
Let $\mathcal{A}'$ be its commutant, namely the set of bounded operators that commute with all operators in $\mathcal{A}$.  If $\mathcal{A}=(\cA')'=\cA''$, we say that $\mathcal{A}$ forms a von Neumann algebra. 
A von Neumann algebra whose center (operators that commute with both $\mathcal{A}$ and $\mathcal{A}'$) consists only of multiples of the identity operator ($\mathcal{A}\cap\mathcal{A}^{'}=\mathbb{C}\hat{1}$) is called a \emph{factor}; 
von Neumann algebras admit a classification into different types, in which factors play a central role.
In this paper, we focus on hyperfinite von Neumann algebras acting on separable Hilbert spaces (i.e., Hilbert spaces with countable orthonormal bases).

The definition above assumes that a Hilbert space and an algebra of operators acting on it have already been given.  We next describe how they can be constructed from an abstract algebra of observables together with a choice of ``state''. Namely, we start from a C$^*$-algebra\footnote{We do not discuss the definition of  C$^{*}$-algebras in this paper;
see the references above for details.
Von Neumann algebras are special cases of C$^*$-algebras; a von Neumann algebra is a C$^{*}$-algebra, but the converse is not always true.
}~$\cC$
and a state $\omega$, which is a linear functional assigning complex numbers to operators as $\omega\!\!:\mathcal{C}\to \mathbb{C}$, with two conditions: $\omega(c^{\dagger}c)\geq0$ for all $c\in\mathcal{C}$ and $\omega(\hat{1})=1$.
A state is called faithful if $\omega(c^{\dagger}c)=0$ implies $c=0$. Given these,
the procedure to 
construct a Hilbert space and a representation of the algebra is
called the Gelfand-Naimark-Segal (GNS) construction. The GNS construction yields, from $\cC$ and $\omega$,
a Hilbert space~$\cH$, a representation \(\pi\) of \(\mathcal C\) by bounded operators on \(\mathcal H\), and a reference vector \(\ket{\Psi}\) such that \(\omega(c)=\bra{\Psi}\pi(c)\ket{\Psi}\). 
The triple $(\mathcal{H}, \pi, \ket{\Psi})$ is called the GNS triple. 
Here, the vector $\ket{\Psi}$ is cyclic by construction, meaning that $\{\pi(c)\ket{\Psi}\,|\,c\in\mathcal{C}\}$ is dense in $\mathcal{H}$. A von Neumann algebra $\mathcal{A}$ is constructed as the double commutant of $\pi(\mathcal{C})$: $\mathcal{A}=(\pi(\mathcal{C}))''.$\footnote{For more details, for example, see~\cite[Ch.~II, Sec.D]{Liu:2025krl}.} If \(\ket{\Psi}\) is also separating with respect to the von Neumann algebra \(\mathcal A\), meaning that \(a\ket{\Psi}=0\) for \(a\in\mathcal A\) implies \(a=0\), the corresponding vector state on \(\mathcal A\) is faithful.\footnote{For more details, for example, see \cite[Prop.~II.2.]{Liu:2025krl}.}  
In this case, \(\ket{\Psi}\) is called a cyclic separating vector.
In this construction $\ket{\Psi}$ is the GNS representation of $\omega$  through $\omega(c)=\langle\Psi|\pi(c)|\Psi\rangle$. Given a state $\omega$, the GNS triple is unique up to unitary equivalence. In physical terms, when a description by a density matrix $\rho$ is available, the state \(\omega\) can be represented by  \(\rho\), and the GNS vector \(\ket{\Psi}\) can be realized as a purification of \(\rho\).
We additionally require normality, i.e., $\mathop{\mathrm{sup}}(\omega(A_{i}))=\omega(\mathop{\mathrm{sup}}A_{i})$ so that limits are well-defined.
For example, suppose a von Neumann algebra $\mathcal{A}$ consists of a matrix algebra of $n\times n$ matrices with $n$ finite. Then we can define a faithful normal state $\omega$ in terms of the density matrix $\rho$, such that $\omega(a)\equiv\text{Tr}(\rho a)$ ($a\in\mathcal{A}$). In this case faithfulness means that $\rho$ has full rank $n$.

Von Neumann algebras play a fundamental role in characterizing the structure of quantum subsystems. As stated above, a subsystem is specified by the algebra $\mathcal{A}$ of its observables acting on the Hilbert space of the total system, and its complement by the commutant $\mathcal{A}'$. Von Neumann factors are classified into three types: I, II and III\@. A short summary of key characteristics of quantum subsystems associated with each type is given in Table~\ref{TableofvNalgebras}.

\begin{table}[htbp]
  \centering
  \begin{tabular}{c|c@{~~~~~}c@{~~~~~}cc}
    \hline
    Type
    & \makecell{tensor\\ factorization}
    & \makecell{trace}
    & \makecell{density\\ operator}
    & \makecell{modular operator\\ $\Delta_\Psi$}
    \\ \hline\hline
    I   & \checkmark & \checkmark & \checkmark & \checkmark
    \\ \hline
    II  & $\times$  & \checkmark & \checkmark & \checkmark
    \\ \hline
    III & $\times$  & $\times$  & $\times$  & \checkmark
    \\ \hline
  \end{tabular}
  \caption{\sl
    Features of von Neumann factors $\mathcal{A}$,
    with $\mathcal{A}'$ describing the complementary subsystem.
    Here we are considering a system partitioned in two disjoint spaces as $L \cup R$ where $L,R$ are quantum subsystems associated to the algebras  $\mathcal{A}$ and $\mathcal{A}'$. A check mark (cross) indicates that the corresponding structure exists (does not exist). Tensor factorization means $\mathcal{H} =\mathcal{H}_L\otimes\mathcal{H}_R$
    with $\mathcal{A}=\hat{1}_L\otimes\mathcal{B}(\mathcal{H}_R)$, where $\mathcal{B}(\mathcal{H}_R)$ denotes the algebra of all bounded operators on $\mathcal{H}_R$. ``Trace'' means a faithful normal semifinite trace \cite[Sec.~6.1]{Sorce:2023fdx}.
    Density operators describe normal states relative to this trace and may be unbounded in Type~II\@.
    The last column assumes that $\ket{\Psi}$ is cyclic
    and separating for $\mathcal{A}$.
  }
  \label{TableofvNalgebras}
\end{table}

Type I and Type II factors admit a finer classification in terms of the possible range of $\Tr(P)$, where $P$ is an operator  in the algebra that projects onto the Hilbert subspace under consideration, and $\Tr$ denotes a faithful normal semifinite trace on the algebra (see Table~\ref{TableofTypeIandII}, \cite[Sec.~6.1]{Sorce:2023fdx} and \cite[Ch.~V]{Connes1994}). In Type~I factors, with the standard normalization, $\Tr(\hat{1})$ is a non-negative integer or $\infty$ and is interpreted as the dimension $n$ of $\mathcal{H}_R$ where the total Hilbert space factorizes $\mathcal{H}=\mathcal{H}_L\otimes\mathcal{H}_R$  with $\mathcal{H}_{R,L}$ associated to $\mathcal{A},\mathcal{A}'$ respectively. Depending on the range, Type I factors are either Type~I$_n$ or Type~I$_\infty$. 
In Type II factors, $\Tr(P)$  can take continuous values. In the Type II$_1$ case, the trace is finite: although the Hilbert space is infinite-dimensional, it can be normalized so that $\Tr(\hat{1})=1$.  $\Tr(P)$ then takes values in $[0,1]$ and can be interpreted as a continuous dimension. A Type II$_{\infty}$ factor is isomorphic to a tensor product of a Type I$_{\infty}$ factor with a Type II$_{1}$ factor, which implies that for Type II$_\infty$ the trace of a projection takes values in $[0,\infty]$.

\begin{table}[htpb]
  \begin{center}
    \begin{tabular}{c|c} \hline
       Types & \makecell{Range of Tr($P$)
       }
       \\ \hline\hline
        I$_{n}$  & $\{0,1,\cdots,n \}$ \\ \hline
                I$_{\infty}$ & $\{0,1,\cdots,\infty \}$ \\ \hline
        II$_{1}$ & $[0,1]$ \\ \hline
II$_{\infty}$& $[0,\infty]$ \\ \hline
     \end{tabular}
     \end{center} 
     \caption{\sl A finer classification of Type I and Type II factors, based on the range of $\Tr(P)$ where $P$ is a projection operator in the algebra onto a Hilbert subspace.
     For details, see the main text.
    }
     \label{TableofTypeIandII}
\end{table}

A finer classification of Type III von Neumann algebras, developed by Connes, is based on Tomita--Takesaki theory.
It divides Type III factors into three subclasses: Type III$_{1}$, Type III$_{\lambda}$, and Type III$_0$. When clarity is necessary, we will refer to the second subclass as \emph{intermediate Type III}, although this is not a generally accepted term. This classification is based on the spectral properties of modular operators \(\Delta_\Psi\), which depend on the choice of cyclic and separating vector \(\ket{\Psi}\) and play a central role in Tomita--Takesaki theory.
We refer the reader to \cite{Witten:2018zxz}  for the precise definition of the modular operator, but physically $\Delta_{\Psi}$ generates a kind of time evolution through its imaginary powers, leaving the reference vector invariant: $\Delta_{\Psi}\Ket{\Psi}=\ket{\Psi}$.  The construction of the modular operator is complicated in general, but we will later explain how to do it in the situation of interest to us. The modular operator $\Delta_{\Psi}$ generates a one-parameter group of automorphisms implemented by unitary operators, parametrized by $s\in\mathbb{R}$, called the ``modular flow'' on $\mathcal{A}$ and $\mathcal{A}^{'}$:\,\footnote{The latter is the modular flow of $\mathcal{A}'$, whose modular operator is $\Delta_{\Psi}^{-1}$.}
\begin{align}
    \sigma_{s}(a)&\equiv
    \Delta_{\Psi}^{-is}\,a\, \Delta_{\Psi}^{is}
    =e^{iK_{\Psi}s}\,a\, e^{-iK_{\Psi}s}\qquad(a\in\mathcal{A}),\\
    \sigma_{s}(a')&
    \equiv\Delta_{\Psi}^{is}\, a'\, \Delta_{\Psi}^{-is}
    =e^{-iK_{\Psi}s}\,a'\, e^{iK_{\Psi}s}\qquad(a'\in\mathcal{A}').
\end{align}
Here, $K_{\Psi}=-\log{\Delta_{\Psi}}$ is called the modular Hamiltonian. 
For a given $s\in\mathbb R$, the modular automorphism $\sigma_s$ is called inner if there exists a unitary $U(s)\in\mathcal A$ such that $\sigma_s(a)=U^\dagger(s)\,a\,U(s)$ for all $a\in\mathcal A$, and similarly for $\mathcal{A}'$. This is the case if $\Delta_{\Psi}^{-is}$ is a product of unitary operators that separately act within $\cA$ and $\cA'$.\footnote{\label{ModularOperator}For example, later we will discuss the finite dimensional case such that we can write $\Delta_{\Psi}$ in terms of the density matrices: $\Delta_{\Psi}=\rho_{L}^{-1}\otimes\rho_{R}$ (where $\rho_{R}\in\mathcal{A}$ and $\rho_{L}\in\mathcal{A}'$). Then, for $a\in\mathcal{A}$, the action of $\sigma_{s}$ can be written as $\sigma_{s}(a)=\rho_{R}^{-is}\, a\, \rho_{R}^{is}$ which is the form of $U^{\dagger}(s)\, a\, U(s)$.} If no such unitary exists, the automorphism is called outer. In the Type I and II cases, modular flows are always inner.  For Type III factors, the modular automorphism is outer for at least some values of $s$ \cite{Liu:2025krl}. 
Such factors appear, for example, in the
semiclassical limit of thermofield double states
in the AdS/CFT correspondence \cite{Witten:2021unn}.

The modular operator $\Delta_{\Psi}$ and the corresponding flow $\sigma_{s}$ depend on the specific choice of the cyclic separating vector $\ket{\Psi}$.   If  we want to classify the types of algebras, independent of such choices, we need an invariant that reflects the intrinsic features of the algebra. Connes \cite{Connes1973,Connes1974} found that the ``spectral invariant'' defined by
\begin{align}
\mathcal{S(A)}=\bigcap_{\Psi}\, \text{Spec}(\Delta_{\Psi})\label{SpectralInv}
\end{align}
can be used for the finer classification of Type III factors. Here, ${\rm Spec}(\Delta_\Psi)$ denotes the spectrum of $\Delta_\Psi$, and the intersection runs over all vectors $\ket{\Psi}\in\cH$ that are cyclic and separating for~$\cA$.
The resulting classification for Type III factors is summarized in Table~\ref{TableofTypeIII}.  See \cite{Basteiro:2024cuh} for an instructive spin chain realization of algebras of various types.
\begin{table}[htpb]
 \begin{center}
    \begin{tabular}{c|c} \hline
       Types & \makecell{ spectral invariant $\mathcal{S(A)}$}  \\ \hline\hline
        III$_{1}$ & $[0,\infty)$ \\ \hline
        \makecell{III$_{\lambda}$ \footnotesize{$(0<\lambda<1)$}\\ ``intermediate Type III'' }& $\{\lambda^i\}^{}_{i\in \mathbb{Z}}\cup\{0\}$ \\ \hline
        III$_{0}$ & $\{0, 1\}$ \\ \hline
     \end{tabular}
     \end{center} 
     \caption{\sl A classification of Type III von Neumann factors by the spectral invariant.}
      \label{TableofTypeIII}
\end{table}

\subsection{Algebras and spacetime connectivity }
\label{ss:alg_ER=EPR_conj}

As we discussed, it has been conjectured that entanglement in CFT corresponds to spacetime connectivity in the dual AdS spacetime  \cite{VanRaamsdonk2009,VanRaamsdonk2010,Maldacena:2013xja}.  In this context, the authors of \cite{Engelhardt:2023xer} suggested that the algebra of operators acting on a state can be used to diagnose if the subsystems are geometrically connected in the gravitational description.  Translated to our setting, their formulation assumes:
\begin{itemize}
  \item The  AdS/CFT correspondence and  the large $N$ limit, in which  the  Newton constant  vanishes  ($G_{N}\rightarrow0$).
  \item Two copies of the boundary CFT denoted by $L$ and $R$.  Each boundary is complete and the total system is in a pure state $\ket{\Psi}$.
  \item The total state $\ket{\Psi}$ has a semiclassical bulk description in the large $N$ limit. 
  \item $\mathcal{A}_{L}$ and $\mathcal{A}_{R}$ are respectively the relevant operator algebras of the boundary subsystems in the large $N$ limit.
  \item Both $\mathcal{A}_{L}$ and $\mathcal{A}_{R}$ are factors and $\mathcal{A}_{L}'=\mathcal{A}_{R}$.
\end{itemize}
Under these assumptions, the proposal in \cite{Engelhardt:2023xer}  asserts that the type of von Neumann algebra is related to the presence or absence of a wormhole between the geometries dual to the $L$ and $R$ copies of the CFT, as follows.

\begin{table}[htpb]
\begin{center}
        \begin{tabular}{c|c|c} \hline 
           & Types of $\mathcal{A}_{L,R}$ & classical condition \\ \hline\hline
           classical wormhole & III$_{1}$ & satisfied \\ 
           quantum wormhole & not I &  not satisfied \\ 
           disconnected spacetime & I & -- \\ \hline
         \end{tabular}
\end{center}
\caption{\sl The relation between algebra types and the presence of wormholes proposed in \cite{Engelhardt:2023xer}.}
\label{tab:AlgebrasConnectedness}
\end{table}
The ``classical condition'' means that in the $N \to \infty$, $G_N \to 0$ limit spacetime diffeomorphism invariants (such as volume, area, and length) do not scale as $(G_N)^a$ with $a<0$, while their fluctuations vanish as $(G_N)^{a}$ with $a>0$. In the context of holography, the classical condition determines whether bulk quantities admit a  classical description in the large $N$ limit. This condition is consistent with other works, for instance \cite{Balasubramanian:2007zt,Leutheusser:2021frk, Leutheusser:2021qhd}. ``Classical wormholes'' are spacetimes that contain classical Einstein-Rosen bridges in the large $N$ limit, while ``quantum wormholes'' (or ``quantum volatile spacetimes'')  cannot be described as containing such classical bridges  in the large $N$ limit.  Note that Table~\ref{tab:AlgebrasConnectedness} says that algebraic subtypes of von Neumann algebras play a role in diagnosing spacetime connectivity.

We could say that quantum wormholes are constructed from an insufficient number of entangled pairs to support classical geometries from the viewpoint of the relation between entanglement and connectivity, although the amount of entanglement is not a sufficient indicator---the entanglement needs to be structured correctly relative to the observables in question \cite{Balasubramanian:2014gla}.  It is possible that the idea of a quantum wormhole can be interpreted \cite{Simon:2018laf} through the ``bit thread picture'' \cite{Freedman:2016zud}, although it is not clear that bit threads adequately capture the multiparty entanglement structures that partly knit spacetime together
\cite{Balasubramanian:2014hda,Bao:2015bfa,Akers:2019gcv,Hayden:2021gno
,Gadde:2022cqi,Harper:2024ker,Iizuka:2025ioc,Balasubramanian:2024ysu}.

Alternatively, it could be that the quantumness arises because the wavefunction is spread out over many configurations that separately have interpretations as semiclassical ER bridges
\cite{Nandy:2024zcd,Balasubramanian:2024lqk,Miyaji:2024ity,Miyaji:2025yvm}, and relatedly because the state can be expressed in terms of many alternative superpositions of semiclassical objects (which provide an overcomplete basis), thus making the interpretation in terms of a specific geometry ambiguous \cite{Balasubramanian:2022gmo,Balasubramanian:2025zey}.

\section{Half-BPS states of $\mathcal{N} = 4$ super-Yang-Mills theory}
\label{sec:halfBPS}

We are going to try to construct quantum wormholes by entangling half-BPS states in two asymptotically AdS universes.  Via the AdS/CFT correspondence, 
Type IIB supergravity in global AdS$_5\times S^5$ is dual to 
four-dimensional $\cN=4$ $SU(N)$ super-Yang-Mills (SYM) theory on spatial $S^3$.
This theory has  32 supercharges, and its 1/2-BPS sector preserves  16 supercharges.  This system can be described in terms of non-interacting fermions, and their description on both sides of the holographic duality is well known \cite{Berenstein:2004kk,Lin:2004nb, Balasubramanian:2005mg}.

In this section, we first review the 1/2-BPS sector of the AdS$_5\times S^5$ correspondence, starting with the CFT side in which 1/2-BPS states are described in terms of $N$ free fermions in a harmonic potential, and then the dual gravity picture of the states in terms of the so-called LLM geometries.  We will discuss various statistical ensembles of half-BPS states in a single asymptotically AdS universe and then describe how to entangle two of these, by way of purification of the ensembles.  We argue that the resulting entangled state should represent a quantum wormhole because of its large entanglement.  In later sections, we will build von Neumann algebras on the state and explore connectivity of the corresponding spacetime by studying the type of the algebras.

\subsection{CFT description: free fermions}

On the field theory side, highest weight 1/2-BPS states in the $(0,1,0)$ representation of the SO(6) R-symmetry group of the $\cN=4$ $SU(N)$ SYM theory can be constructed by exciting   a particular 
combination of scalar fields, $Z=X^5+i X^6$.  (We can similarly construct 1/2-BPS states built on other representations of SO(6).) To preserve supersymmetry, we must not have any spatial momentum, so we smear $Z$ uniformly over the boundary $S^3$.  The dynamics of this sector reduces to a matrix model for~$Z$ \cite{Berenstein:2004kk,Takayama:2005yq}.  Standard techniques further reduce the problem to one of understanding the eigenvalue dynamics.  The latter  can be described in terms of $N$ fermions in a harmonic potential.  The eigenvalues act as fermions because the Vandermonde determinant appearing in the path integral measure when we diagonalize $Z$ creates an effective statistical repulsion between the eigenvalues which can be absorbed by writing them as fermions.  The quadratic potential arises directly from the spherically symmetric reduction of the potential terms in the $\mathcal{N} = 4$ SYM action.
Let us write the energy of the $i$th fermion ($i=1,2,\dots,N$) as
\begin{align}
 E_i=e_i\hbar \omega,\qquad\text{with}\qquad
 e_i=0,1,2,\dots,\quad
 0\le e_1< e_2< \dots < e_N,
\end{align}
where we have dropped the zero-point energy $\hbar\omega/2$.  We will henceforth set $\hbar\omega=1$. 
For example, the ground state is given by $e^g_i=i-1$ and the ground state energy is
\begin{math}
 E_{g}=\sum_{i=1}^N e_i^g
 ={1\over 2}(N-1)N.
\end{math}
We will refer to $i$ as the (energy) level.

Another useful set of $N$ variables to parametrize states is
\begin{align}
 r_i\equiv e_i-e_i^g=e_i-i+1,\qquad\qquad
 0\le r_1\le r_2\le \dots \le r_N,
\end{align}
which indicates how much the $i$th fermion is excited above the ground state.  We can represent~$\{r_i\}$ by a Young diagram $Y$ with row lengths
$r_N,r_{N-1},\dots,r_1$ (from top to bottom).  The number of rows is $N$
and therefore $Y$ has height $N$, although we can have $r_i=0$, i.e., some
rows can be empty.   An example with $N=5$ is
\begin{align}
 \ytableausetup{boxsize=1.5ex}
 (r_5,r_4,r_3,r_2,r_1)=(4,3,1,1,0)
 ~~\to~~
 \vcenter{\hbox{
\begin{tikzpicture}[scale=0.25]
\draw (0,0) rectangle +(1,1);
\draw (1,0) rectangle +(1,1);
\draw (2,0) rectangle +(1,1);
\draw (3,0) rectangle +(1,1);
\draw (0,-1) rectangle +(1,1);
\draw (1,-1) rectangle +(1,1);
\draw (2,-1) rectangle +(1,1);
\draw (0,-2) rectangle +(1,1);
\draw (0,-3) rectangle +(1,1);
\draw (0,0) -- (0,-4);
\draw (-0.1,-4) -- +(0.2,0);
\end{tikzpicture}}}
\end{align}

If we define
\begin{align}
 c_N=r_1,\quad
 c_{N-1}=r_2-r_1,\quad
 c_{N-2}=r_3-r_2,\quad,\dots,\quad
 c_1=r_{N}-r_{N-1}\label{isko28Sep26}
\end{align}
then $c_j\ge 0$ counts the number of columns of height $j$.  
Unlike $e_i$ or $r_i$, the value of $c_j$ is unconstrained (not ordered). The number of boxes is equal to the total excitation energy above the ground state:
\begin{align}
 E&-E_{g}
 =\sum_{i=1}^N (e_i-e_i^g)
 =\sum_{i=1}^N r_i
 =\sum_{j=1}^N j c_j.
\end{align}
Because there are no constraints in this presentation, we can regard $c_j$ as counting  bosons with energy $j$.  A useful relation is
\begin{align}
 r_{i+1}=e_{i+1}-i=c_{N-i}+\dots+c_N.\label{mgra14Nov23}
\end{align}
We denote the quantum state specified by the bosonic variables $\{c_j\}$ as
\begin{align}
    \ket{c_1,c_2,\dots,c_N}
    =\prod_{j=1}^N \ket{c_j}_j,
\end{align}
where $\ket{c}_j$ denotes the state with $c$ columns of length $j$. We will refer to $j$ as a \emph{column-height index}.
For later use, let us introduce annihilation and creation operators $D_j,D_j^\dagger$ of the $j$th boson, satisfying
\begin{align}
[D_j,D_{j'}^\dagger]=\delta_{jj'},\qquad
D_j^\dagger D_j=\hat{c}_j,
\label{eq:D_algebra_def}
\end{align}
where $\hat{c}_j$ is the number operator for columns of height $j$, namely, the operator version of $c_j$.

Yet another set of variables that we can use to specify states is the occupation number $f_n=0,1$ at oscillator level $n$ ($n=0,1,2\dots$),  which satisfies $\sum_{n=0}^\infty f_n=N$ and $\sum_{n=0}^\infty n f_n =E$.
In terms of the fermionic variables $\{f_n\}$, the state of the system can be written as
\begin{align}
    \bigotimes_{n=0}^\infty \ket{f_n}_n,
\end{align}
where $\ket{f}_n$ denotes the state of the fermion at level $n$ (with $\ket{0}_n$ and $\ket{1}_n$ denoting unoccupied and occupied states, respectively).  Let $C_n$ and $C_n^\dagger$ be the annihilation and creation operators, respectively, of the fermion at level $n$, which satisfy
\begin{align}
\{C_n,C_{m}^\dagger\}=\delta_{nm},\qquad
C_n^\dagger\ket{0}_n=\ket{1}_n,\quad
C_n\ket{1}_n=\ket{0}_n,
\label{eq:C_algebra_def}
\end{align}
where $n,m=0,1,2,\dots$.

A crucial difference between the bosonic algebra of $D_j,D_j^\dagger$ and the fermionic algebra of $C_n,C_n^\dagger$ is that the former preserves the total fermion number $N$ while the latter does not.  This difference will be important in the algebraic classifications in later sections.

The construction described above can also be understood as defining Schur polynomials in $Z$ associated to the Young diagrams that act on the vacuum to create states \cite{Balasubramanian:2001nh,Corley:2001zk,Bhattacharyya:2008xy,Brown:2007xh}.   In this language a single column of height $N$ corresponds to the operator $\det Z$ in the CFT which creates a state dual to maximal-size spherical D3-brane (giant gravitons) on the $S^5$ in ${\rm AdS}_5 \times S^5$;  shorter columns correspond to sub-determinants and smaller giant gravitons \cite{Myers:1999ps,McGreevy:2000cw,Balasubramanian:2001nh}.  This identification has been verified by inserting defects on the operator corresponding to string excitations on the branes \cite{Balasubramanian:2004nb,deMelloKoch:2007rqf}.  This is all to say that the Young diagram states that we are discussing are very well understood, and can be interpreted as bound states of D-branes in AdS$_5$.

\subsection{Effective thermodynamic ensembles and thermo-field doubles}
\label{sss:eff_thermodyn_ensembles}

\medskip
\subsubsection*{The hyperstar ensembles}  

\vspace*{-1ex}
We can consider various ensembles of the fermionic system we discussed above.  Because the states are BPS, the excitation energy $E$ is equal to the R-charge $\Delta$.
We can introduce an inverse \emph{effective temperature} $\beta$ conjugate to $E=\Delta$ and consider a ``canonical ensemble'' with respect to $\beta$.  The canonical partition function is easiest to compute in terms of the unconstrained variables $\{c_j\}$ and is given by
\begin{align}
 Z_N(q)=
 q^{E_g}
\prod_{j=1}^N \sum_{c_j=0}^\infty  q^{j c_j}
 =
 {q^{E_g}\over \prod_{j=1}^N(1-q^j)},\qquad\quad q=e^{-\beta}.
 \label{mftu14Nov23}
\end{align}
The bosonic excitation number $c_j$ obeys Bose-Einstein statistics,
\begin{align}
    \ev{\hat{c}_j}={q^j\over 1-q^j}={1\over e^{\beta j}-1}.
    \label{c_j_BE_dist}
\end{align}
The density matrix for this canonical ensemble is
\begin{align}
    \rho_N(q)=\bigotimes_{j=1}^N\, (1-q^j)\sum_{c_j=0}^\infty q^{jc_j} \ket{c_j}_j \bra{c_j}_j,
    \label{eq:rho(q)}
\end{align}
which is a tensor product of density matrices for each $j$. 

Just as one does for systems with a physical temperature, we can purify this density matrix. Namely, we consider two copies of the  LLM system and introduce a pure, entangled state of the total system which reduces to \eqref{eq:rho(q)} upon tracing out one copy.  We call the original copy the right ($R$) system and the other copy the left ($L$) system. 
By abuse of language we call such a pure state a thermo-field double (TFD) state.  
In the present case, the TFD hyperstar state is
\begin{align}
    \ket{\Psi_N(q)}=\bigotimes_{j=1}^N
    \, \sqrt{1-q^j}\sum_{c_j=0}^\infty q^{jc_j/2} \ket{c_j,c_j}_j,
    \label{eq:TFD_hyperstar_bos}
\end{align}
where $\ket{c,c'}_j=\ket{c}_{L,j}\otimes \ket{c'}_{R,j}$.  This is a tensor product of states for each $j$. 
We can interpret the TFD state \eqref{eq:TFD_hyperstar_bos} as a superposition of states in which the left and right bosons are excited to exactly the same occupation number.  In terms of Young diagrams, we consider exactly the same Young diagram on the left and right, and then take a quantum superposition over all possible Young diagrams (with weights).

\bigskip
We can also consider a grand canonical ensemble in which we sum over $N$ with a fugacity~$p=e^{\beta\mu}$, where $\mu$ is the chemical potential conjugate to $N$.  In the thermodynamic limit ($N,\Delta\to\infty$), the canonical and grand canonical ensembles are expected to become indistinguishable.  In the present context  we will call both of these   ``hyperstar'' ensembles \cite{Balasubramanian:2005mg}, for reasons that we will explain later. As usual, the values of $q,p$ are related to the mean values of $\Delta,N$ in the ensemble.  The grand canonical ensemble is particularly suited to the fermionic description. For example, the grand partition function can be computed easily using the fermionic variable $f_n$ and takes a factorized form:
\begin{align}
 Z(q,p)
 &=\prod_{n=0}^\infty \sum_{f_n=0,1} q^{n f_n}p^{f_n}
 =\prod_{n=0}^\infty (1+q^n p).
 \label{Z(q,p)_hyp}
\end{align}
This factorized expression allows us to write down the grand-canonical density matrix in terms of fermions as
\begin{align}
 \rho(q,p)=\bigotimes_{n=0}^\infty {1\over  1+pq^{n}}\Bigl(\ket{0}_n\bra{0}_n+ pq^{n}\ket{1}_n\bra{1}_n\Bigr),
 \label{eq:rho(p,q)}
\end{align}
which is a tensor product of $2\times 2$ fermion density matrices at each level.
The TFD hyperstar state in this fermion language is
\begin{align}
 \Ket{\Psi(q,p)}
 \equiv
 \bigotimes_{n=0}^\infty
 {1\over\sqrt{1+pq^{n}}}
 \left(\ket{00}_n+ p^{1\over 2}q^{n\over 2}\ket{11}_n\right),
 \label{TFD_hyperstar_rho}
\end{align}
where $\ket{ff'}_n=\ket{f}_{L,n}\otimes \ket{f'}_{R,n}$ with $f,f'=0,1$.
Again, this factorizes into states at each level.
We can say that we consider states in which the occupation numbers of all the fermions are exactly the same on the left and right, and take a quantum superposition over all such possibilities.

The statistical entropy for this ensemble \eqref{eq:rho(p,q)}, which is equal to the entanglement entropy of the TFD state \eqref{TFD_hyperstar_rho} between the left and right, can be evaluated to be
\begin{align}
S=(1-\beta\partial_\beta)\ln Z(q,p)
=-\sum_{n=0}^\infty 
\left[ P_n \log P_n + (1-P_n)\log (1-P_n)\right],
\label{lkbf8Nov26}
\end{align}
where
\begin{align}
    P_n={q^n p\over 1+q^n p}
\end{align}
is the probability for level $n$ to be occupied (and thus $1-P_n$ is the probability for level $n$ to be unoccupied).  Therefore,  \eqref{lkbf8Nov26} is simply a sum of the Shannon entropy over all levels.

\subsubsection*{The superstar ensembles}

Now consider an ensemble in which the number of columns  $N_C$  of the Young diagram  representation is fixed:
\begin{align}
\vcenter{\hbox{\scalebox{0.9}{\begin{tikzpicture}[scale=0.25]
\draw (0,0) -- +(10,0);
\draw (0,-1) -- +(10,0);
\draw (0,-2) -- +(7,0);
\draw (0,-3) -- +(5,0);
\draw (0,-4) -- +(1,0);
\draw (0,-5) -- +(1,0);
\draw (0,0) -- +(0,-7);
\draw (1,0) -- +(0,-5);
\draw (2,0) -- +(0,-3);
\draw (3,0) -- +(0,-3);
\draw (4,0) -- +(0,-3);
\draw (5,0) -- +(0,-3);
\draw (6,0) -- +(0,-2);
\draw (7,0) -- +(0,-2);
\draw (8,0) -- +(0,-1);
\draw (9,0) -- +(0,-1);
\draw (10,0) -- +(0,-1);
\draw (-0.2,-7) -- +(0.4,0);
\draw [decorate,decoration={brace,amplitude=7,raise=4}]
  (0,0) -- (10,0) node[midway,above,yshift=9]{$N_C$};
\draw [decorate,decoration={brace,mirror,amplitude=7,raise=4}]
  (0,0) -- (0,-7) node[midway,left,xshift=-9]{$N$};
\end{tikzpicture}}}}
\end{align}
This means that the length of the top row must be $N_C$, so we have
\begin{align}
r_N=c_1+\cdots+c_N=N_C,
\label{iwnc20Nov23}
\end{align}
where we used \eqref{mgra14Nov23}.
This ensemble was designed in \cite{Balasubramanian:2005mg} to match the properties of the superstar geometries in \cite{Myers:2001aq} in a certain limit.\footnote{The grand canonical superstar ensemble that we consider below has a relatively large column number variance. The authors of \cite{DErrico:2007qch} proposed a more restricted ensemble with a smaller variance which seeks to be compatible with the Bekenstein area law for a stretched horizon. We will stick to the grand canonical ensemble of \cite{Balasubramanian:2005mg} here, but it would be interesting to investigate the ensemble of \cite{DErrico:2007qch} in the future. }

We will refer to this ensemble and the other equivalent ensembles discussed below as superstar ensembles.

As before, we can alternatively consider an ensemble in which $N_C$ is not fixed by introducing a fugacity $\zeta=e^{-\gamma}$ that counts $N_C$.  The  partition function for this ensemble is
\begin{align}
 Z_N(q,\zeta)
 =
 q^{E_g}
 \prod_{j=1}^N \sum_{c_j=0}^{\infty} q^{j c_j}\zeta^{c_j}
 =
 {q^{E_g}\over \prod_{j=1}^N (1-q^j\zeta)}.\label{ggwx15Nov23}
\end{align}
The occupation number $\ev{\hat{c}_j}$ is now given by
\begin{align}
    \ev{\hat{c}_j}={q^j\zeta\over 1-q^j\zeta}={1\over e^{\beta j+\gamma}-1}.
    \label{c_j_superstar}
\end{align}
The corresponding density matrix and the TFD state are
\begin{subequations}
\begin{align}
 \rho_N(q,\zeta)
 &=
 \bigotimes_{j=1}^N \, (1-q^j\zeta)\sum_{c_j=0}^{\infty} q^{j c_j}\zeta^{c_j}
 \ket{c_j}_j\bra{c_j}_j,
 \label{fkah10Nov26}
 \\
 \ket{\Psi_N(q,\zeta)}
 &=
 \bigotimes_{j=1}^N \sqrt{1-q^j\zeta}\sum_{c_j=0}^{\infty} q^{j c_j/2}\zeta^{c_j/2}
 \ket{c_j,c_j}_j.\label{fkab10Nov26}
\end{align}
\end{subequations}
The ``superstar'' solution in the dual geometry that we will discuss later corresponds to the $\beta\to 0$ limit of this ensemble.  In this limit, we have $q=1$, and the density matrix and TFD state become
\begin{subequations}\label{TFD_superstar_rho_Psi_boson}
\begin{align}
 \rho_N(q=1,\zeta)
 &=
 \bigotimes_{j=1}^N \, (1-\zeta)\sum_{c_j=0}^{\infty} \zeta^{c_j}
 \ket{c_j}_j\bra{c_j}_j,
 \label{TFD_superstar_rho_boson_1}
 \\
 \ket{\Psi_N(q=1,\zeta)}
 &=
 \bigotimes_{j=1}^N \sqrt{1-\zeta}\sum_{c_j=0}^{\infty} \zeta^{c_j/2}
 \ket{c_j,c_j}_j.
 \label{TFD_superstar_rho_Psi_1}
\end{align}
\end{subequations}

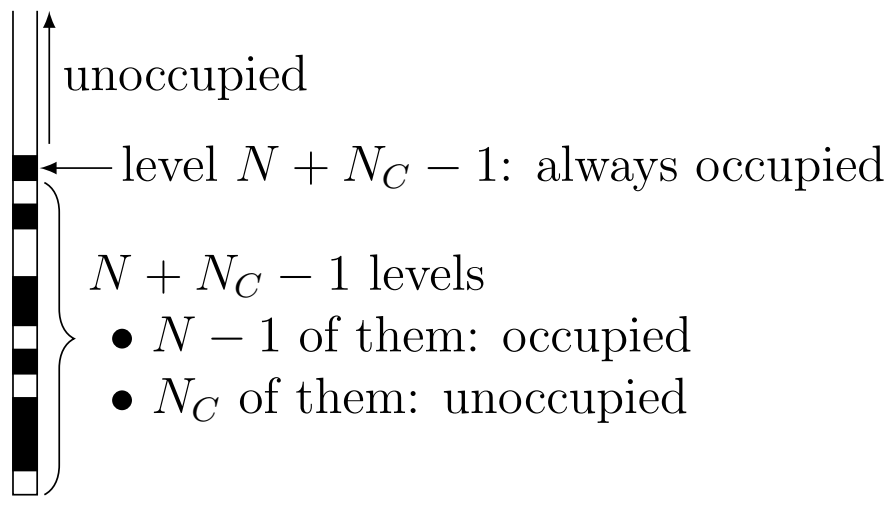
\begin{figure}[tbh]
\begin{quotation}
\begin{center}
\scalebox{0.8}{
 \begin{tikzpicture}[scale=0.2]
 \draw (0,20) -- (0,0) -- (1,0) -- (1,20);
 \draw[fill=black] (0,1) rectangle +(1,3);
 \draw[fill=black] (0,5) rectangle +(1,1);
 \draw[fill=black] (0,7) rectangle +(1,2);
 \draw[fill=black] (0,11) rectangle +(1,1);
 \draw [decorate,decoration={brace,mirror,amplitude=7,raise=-4}]
  (2,0) -- (2,12.9) node[midway,right,xshift=2]{
  \begin{minipage}{5cm}$N+N_C-1$ levels\\
  \hspace*{1ex}$\bullet$ $N-1$ of them: occupied\\
  \hspace*{1ex}$\bullet$ $N_C$ of them: unoccupied
  \end{minipage}
  };
 \draw[fill=black] (0,13) rectangle +(1,1);
  \draw[latex-] (1.1,13.5) -- +(3,0) node [right,xshift=-2] {level $N+N_C-1$: always occupied};
  \draw[-latex] (1.5,14.5) -- +(0,5.5) node [right,midway,xshift=-1] {unoccupied};
\end{tikzpicture}}
\caption{\sl The superstar ensemble in fermion variables.  Occupied levels are shown in black and unoccupied levels are shown in white.
 \label{fig:superstar_ens_levels}}
\end{center}
\end{quotation}
\end{figure}

In terms of the fermionic variables $\{f_n\}$, the superstar ensemble corresponds to fixing  the highest occupied level to be level $M-1$ where $\NplusNC\equiv N+N_C$; namely, $f_{\NplusNC-1}=1$ and $f_{n\ge \NplusNC}=0$.  Among the $\NplusNC-1$ levels ($n=0,1,\dots,M-2$) below it, $N-1$ are occupied and $N_C$ are unoccupied (see Figure~\ref{fig:superstar_ens_levels}).

In order to write down a fermionic density matrix, it is useful to go to another thermodynamically equivalent ensemble in which $\NplusNC=N+N_C$ is fixed while $N$ is unfixed. 
Level $\NplusNC-1$ is the highest occupied level. The $\NplusNC-1$ levels below it may be occupied or unoccupied. Letting $p$ count the number of occupied levels, we find
\begin{align}
 Z_\NplusNC(q,p)
 &=\left[\prod_{n=0}^{\NplusNC-2} \sum_{f_n=0,1}q^{nf_n} p^{f_n}\right] q^{\NplusNC-1}p
 =\left[\prod_{n=0}^{\NplusNC-2} (1+q^{n} p)\right] q^{\NplusNC-1}p.\label{mfgb20Nov23}
\end{align}
From this, we see that the density matrix and the associated TFD state are
\begin{subequations}
\begin{align}
 \rho_\NplusNC(q,p)
 &=
 \left[\bigotimes_{n=0}^{\NplusNC-2}
{ \ket{0}_n\bra{0}_n+q^{n}p\ket{1}_n\bra{1}_n\over 1+q^{n}p}\right]
 \otimes
 \ket{1}_{\NplusNC-1}\bra{1}_{\NplusNC-1}
    \otimes 
    \left[\bigotimes_{n=\NplusNC}^\infty\ket{0}_n\bra{0}_n\right]
    ,\label{kiyu29Mar24}
    \\
 \Ket{\Psi_\NplusNC(q,p)}
 &=
 \left[\bigotimes_{n=0}^{\NplusNC-2}
 {\ket{00}_n+q^{n\over 2}p^{1\over 2}\ket{11}_n\over\sqrt{1+q^{n}p}}\right]
 \otimes 
 \ket{11}_{\NplusNC-1}
 \otimes
 \left[\bigotimes_{n=\NplusNC}^\infty \ket{00}_n\right]
 .\label{TFD_gen_superstar_rho}
\end{align}
\end{subequations}
In the $\beta\to 0$ ($q=1$) limit, the density matrix and  TFD state become
\begin{subequations}\label{TFD_superstar_rho_Psi_fermion}
\begin{align}
 \rho_\NplusNC(q=1,p) &= 
    \left[\bigotimes_{n=0}^{\NplusNC-2}{\ket{0}_n\bra{0}_n+p\ket{1}_n\bra{1}_n\over 1+p}
    \right]
        \otimes \ket{1}_{\NplusNC-1}\bra{1}_{\NplusNC-1}
        \otimes
    \left[\bigotimes_{n=\NplusNC}^{\infty}\ket{0}_n\bra{0}_n
    \right],\label{TFD_superstar_rho}
\\
 \Ket{\Psi_\NplusNC(q=1,p)}
 &=
 \left[\bigotimes_{n=0}^{\NplusNC-2}
 {\ket{00}_n+p^{1\over 2}\ket{11}_n\over\sqrt{1+p}}\right]
 \otimes 
 \ket{11}_{\NplusNC-1}
 \otimes
 \left[\bigotimes_{n=\NplusNC}^\infty \ket{00}_n\right].
 \label{TFD_superstar_state}
\end{align}
\end{subequations}
All these density matrices and TFD states are simple tensor products of those at each level.

\subsubsection*{Limit curves}

In the large $N$ limit, the boundaries of almost all Young diagrams in the ensembles above approach a smooth curve.  Define  $x$ and $y$ axes as below:
\begin{align}
\ytableausetup{boxsize=0.75ex}
\vcenter{\hbox{\begin{tikzpicture}[scale=0.25]
\draw[-latex] (0,-5) -- +(15,0) node [right] {$y$};
\draw[-latex] (0,-5) -- +(0,10) node [right] {$x$};
\node () at (0,0) [xshift=1.525cm,yshift=-0.335cm] {$\ydiagram{20,17,15,10,8,5,4,3,2,2,1,1}$};
\node () at (0,2.2) [left] {$N$};
\node () at (0,-5) [below left,yshift=5,xshift=2] {O};
\end{tikzpicture}}}
\label{eq:Ydiag_xy}
\end{align}
In the fermion language, $x$ corresponds to the particle label $i$, while $y$ corresponds to its excitation above the ground state, $r_i$.
The relation \eqref{mgra14Nov23} gives
\begin{align}
    y(x)=\sum_{j=N-x}^N c_j
    \approx \int_{N-x}^N dj\, \ev{\hat{c}_j},
\end{align}
where in the second equality we approximated the sum by an integral and replaced $c_i$ by its expectation value,
as appropriate to the large $N$ limit.\footnote{See \cite{Balasubramanian:2005mg} for an estimate of the error in this approximation.}

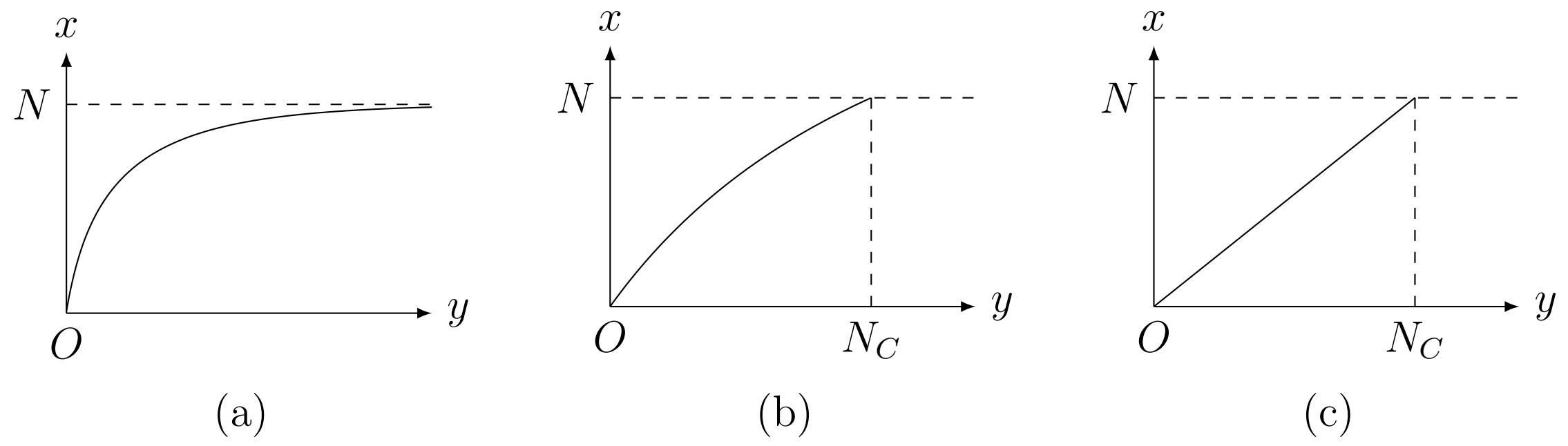
\begin{figure}[tbh]
\begin{center}
\begin{tabular}{c@{~~~~}c@{~~~~}c}
\begin{tikzpicture}
  \pgfmathsetmacro{\beta}{1}
  \pgfmathsetmacro{\N}{2.0}
  \draw[-latex] (0,0)--(3.5,0) node [right] {$y$};
  \draw[-latex] (0,0)--(0,2.5) node [above] {$x$};
  \draw[dashed] (0,2)--(3.5,2);
  \draw[domain=0:3.5, samples=300, smooth]
    plot (\x, {\N+(1/\beta)*ln(1-(1-exp(-\beta*\N))*exp(-\beta*\x))});
\node at (0,0) [below] {$O$};
\node at (0,\N) [left] {$N$};
\end{tikzpicture}
&
\begin{tikzpicture}
  \pgfmathsetmacro{\beta}{0.25}
  \pgfmathsetmacro{\N}{2.0}
  \pgfmathsetmacro{\NC}{2.5}
  \draw[-latex] (0,0)--(3.5,0) node [right] {$y$};
  \draw[-latex] (0,0)--(0,2.5) node [above] {$x$};
  \draw[dashed] (0,2)--(3.5,2);
  \draw[dashed] (\NC,\N)--(\NC,0);
  \draw[domain=0:\NC, samples=300, smooth]
    plot (\x, {\N+(1/\beta)*ln((1-exp(-\beta*(\N+\NC))-(1-exp(-\beta*\N))*exp(-\beta*\x))/(1-exp(-\beta*\NC)))});
\node at (0,0) [below] {$O$};
\node at (\NC,0) [below] {$N_C$};
\node at (0,\N) [left] {$N$};
\end{tikzpicture}
&
\begin{tikzpicture}
  \pgfmathsetmacro{\N}{2.0}
  \pgfmathsetmacro{\NC}{2.5}
\draw[-latex] (0,0) -- (3.5,0) node [right] {$y$};
\draw[-latex] (0,0) -- (0,2.5) node [above] {$x$};
\draw[dashed] (0,2) -- (3.5,2);
\draw (0,0) -- (\NC,\N);
 \draw[dashed] (\NC,\N)--(\NC,0);
\node at (0,0) [below] {$O$};
\node at (\NC,0) [below] {$N_C$};
\node at (0,\N) [left] {$N$};
\end{tikzpicture}
\\
(a) & (b) & (c)
\end{tabular}
\caption{\sl Examples of the limit curves for (a) the hyperstar ensemble, (b) the superstar ensemble with $\beta>0$, and (c) the superstar ensemble with $\beta=0$.\label{fig:limit_shapes}}
\end{center}
\end{figure}

For the ``hyperstar'' ensemble with \eqref{c_j_BE_dist}, we find 
\begin{align}
q^N q^{-x}+(1-q^N)q^y=1.
\label{eq:hyperstarlimit}
\end{align}
The existence of this limit curve means that in the thermodynamic, large system  limit, almost all half-BPS states are built from Young diagrams whose boundaries are small fluctuations of the curve in \eqref{eq:hyperstarlimit}.  In this limit, simple probes will not be able to tell the different states apart (see the conceptual explanation in \cite{Balasubramanian:2005kk}) and they can all be regarded as microstates  of the effective, coarse-grained object described by the limiting diagram with boundary \eqref{eq:hyperstarlimit}.  Similar comments apply to the limiting diagram for the superstar ensemble that we discuss below. We will also later explain how to construct the dual gravitational description of the limiting diagram, which will turn out to be an extremal black hole.   See Fig.~\ref{fig:limit_shapes}(a) for an example of the limit curve in this ensemble. 

In the large $N$ limit, we need to scale the quantities in \eqref{eq:hyperstarlimit} as
\begin{align}
x,y\sim N,\qquad
\beta=-\log q\sim {1\over N},\qquad
\text{so that}\quad 
q^N,q^x,q^y
=O(1).
\end{align}
As shown in Appendix \ref{app:limit_curves}, 
the fugacity $p$ appearing in \eqref{Z(q,p)_hyp}--\eqref{TFD_hyperstar_rho} can be written at large $N$ as
\begin{align}
    {1\over p+1}=q^{N}=O(1).
    \label{p_ito_q,N}
\end{align}

For the superstar ensemble \eqref{c_j_superstar}, we find
\begin{align}
q^N \zeta q^{-x}+(1-q^N \zeta)q^y=1.
\end{align}
Requiring that this curve passes through $(x,y)=(N,N_C)$ determines the value of $\zeta$ to be
$\zeta=(1-q^{N_C})/(1-q^{N+N_C})$, and the limit curve becomes
\begin{align}
    (1-q^{N_C})q^{N-x}+(1-q^N)q^y=1-q^{N+N_C}.
    \label{limitcurve_superstar_beta>0}
\end{align}
See Figure~\ref{fig:limit_shapes}b for an example of the limit curve.

In the infinite temperature limit $\beta\to 0$ ($q\to 1$), the curve \eqref{limitcurve_superstar_beta>0} reduces to
\begin{align}
    y=\lambda x, \qquad \lambda\equiv {N_C\over N}.
    \label{limitshape_superstar}
\end{align}
This is a straight line connecting $(x,y)=(0,0)$ and $(N,N_C)$; see Fig.~\ref{fig:limit_shapes}(c) for an example of this limit curve.  We will see that this limiting case  corresponds to the superstar geometry.
In this limit, we can show that
\begin{align}
\begin{split}
\ev{r_i}&\approx \lambda i,\qquad 
    \ev{e_i}\approx (\lambda+1)i,\qquad
    \ev{f_i}\approx{1\over \ev{e_{i+1}}-\ev{e_i}}\approx {1\over \lambda+1},\\
    \zeta&={\lambda\over \lambda+1},\qquad p={1\over \lambda},
\end{split}
\label{values_superstar}
\end{align}
where we used approximations valid for large $N\sim N_C$.

Let us comment on the cyclic and separating properties of the TFD states we constructed above, because these properties are important for type classification based on the modular operator.
First, as mentioned before, the vector $\ket{\Psi}$ constructed from the GNS construction is automatically cyclic for the Hilbert space $\cH$.  $\ket{\Psi}$ must be separating in order for the finer classification of Type III algebras using the Tomita-Takesaki theory and the modular operator.   It is easy to show that  bosonic states such as \eqref{eq:TFD_hyperstar_bos},
\eqref{fkab10Nov26}, and \eqref{TFD_superstar_rho_Psi_1} and the fermionic hyperstar state \eqref{TFD_hyperstar_rho} are cyclic and separating.  
However, one can also show that the fermionic superstar states, \eqref{TFD_gen_superstar_rho}  and \eqref{TFD_superstar_state}, are not  separating, because the part of the state with level $n\ge M-1$ is pure.  
However, this will not be a problem for our type classification, as we will see in detail in Sec.~\ref{sec:fineFermion}.  This is because the $n<M-1$ part is separating and we can use the Tomita-Takesaki theory, while the $n\ge M-1$ part is of Type~I and we do not need the Tomita-Takesaki theory for its classification.

\subsubsection*{Summary: $\beta=0$ superstar}
 
For later convenience, we summarize the density matrix and  TFD state for the superstar ensemble at $\beta=0$.  In the bosonic, canonical ensemble description, we have (see~\eqref{TFD_superstar_rho_Psi_boson})%
\begin{subequations} \label{TFD_superstar_rho_Psi_boson_2}%
\begin{align}
 \rho^{(s)}(\zeta)
 &=
 \bigotimes_{j=1}^N (1-\zeta)\sum_{c_j=0}^{\infty} \zeta^{c_j}
 \ket{c_j}_j\bra{c_j}_j,
     \label{TFD_superstar_rho_boson_2}
 \\
 \ket{\Psi^{(s)}(\zeta)}
 &=
 \bigotimes_{j=1}^N \sqrt{1-\zeta}\sum_{c_j=0}^{\infty} \zeta^{c_j/2}
 \ket{c_j,c_j}_j.
     \label{TFD_superstar_Psi_boson_2}
\end{align}   
\end{subequations}

On the other hand, in the fermionic, grand-canonical description, we have (see \eqref{TFD_superstar_rho_Psi_fermion})
\begin{subequations}\begin{align}
 \rho^{\text{(s)}}(\lambda) &= 
    \left[\bigotimes_{n=0}^{\NplusNC-1}
    {1\over 1+\lambda}
    \Bigl(\lambda\ket{0}_n\bra{0}_n+\ket{1}_n\bra{1}_n\Bigr)
    \right]
        \otimes
    \left[\bigotimes_{n=\NplusNC}^{\infty}\ket{0}_n\bra{0}_n
    \right],\label{TFD_superstar_rho_2}
\\
 \Ket{\Psi^{\text{(s)}}(\lambda)}
 &=
 \left[\bigotimes_{n=0}^{\NplusNC-1}{1\over \sqrt{1+\lambda}}
  \left(\sqrt{\lambda}\,\ket{00}_n+\ket{11}_n\right)
  \right]
 \otimes
 \bigotimes_{n=\NplusNC}^\infty \ket{00}_n,
 \label{TFD_superstar_state_2}
\end{align}\label{TFD_superstar_rhoPsi_2}%
\end{subequations}%
where we used $\NplusNC=N+N_C=(1+\lambda)N$ and \eqref{values_superstar}.
Furthermore, we dropped the factor for level $n=N+N_C-1$ in \eqref{TFD_superstar_rho_Psi_fermion}, which is always occupied.  This is justified in the large $N$ limit; in any case, such a change in the density matrix and the TFD state at a finite number of levels will not affect the algebra classification.  The entanglement entropy for the state \eqref{TFD_superstar_state_2} is, just as \eqref{lkbf8Nov26},
\begin{align}
    S_{\rm superstar}&=-M\left(
        {\lambda\over 1+\lambda}\log {\lambda\over 1+\lambda}
       + {1\over 1+\lambda}\log {1\over 1+\lambda}
    \right)\notag\\
    &=N[(1+\lambda)\log(1+\lambda)-\lambda\log\lambda].
    \label{S_superstar}
\end{align}

The density matrices and TFD states that we described above corresponding to different ensembles are of course not precisely the same states,  but in the thermodynamic limit they are expected to give identical answers to macroscopic questions.
Note that,
from \eqref{values_superstar}, the parameters $\zeta$ and $\lambda$ in the two descriptions are related by
\begin{align}
    \zeta={\lambda\over1+\lambda}.\label{eq:lambda_and_zeta}
\end{align}

\subsection{Gravity description}
\label{sss:bulk_side}

Next we turn to the dual gravity description of the states described above.  A large class of half-BPS solutions of 10-dimensional Type IIB supergravity with asymptotically ${\rm AdS}_5 \times S^5$ boundary conditions, dubbed LLM geometries, can be written as  \cite{Lin:2004nb}:
\begin{subequations}
 \label{mcoi21Jul24}
 \begin{align}
 ds_{10}^2&=-h^{-2}(dt+V_i dx^i)^2+h^2(d\eta^2+dx^idx^i)+\eta e^G d\Omega_3^2+\eta e^{-G}d\tilde{\Omega}_3^2\label{lnej18Jul24},\\
 \eta \p_\eta  V_i &=\epsilon_{ij}\p_j z,\qquad \eta (\p_i V_j-\p_j V_i)=\epsilon_{ij}\p_\eta  z,\\
 h^2&={\sqrt{{1/4}-z^2}\over \eta },
\qquad e^G=\sqrt{1/2+z\over 1/2-z},\\
 z(x_1,x_2,\eta )&={\eta ^2\over \pi}\int{z(x_1',x_2',0)dx_1'dx_2'\over [({\bf x}-{\bf x}')^2+\eta ^2]^2}, \label{eq:zinsolution}\\
 V_i(x_1,x_2,\eta )&={\epsilon_{ij}\over \pi}
 \int{z(x_1',x_2',0)\,(x_j-x_j')dx_1'dx_2'\over [({\bf x}-{\bf x}')^2+\eta ^2]^2},\qquad \label{eq:Vinsolution}
\end{align}
\end{subequations}
The solution is entirely specified by data on the $\eta=0$ plane: $z(x_1,x_2,\eta=0)$.  We call the $\eta=0$ plane the LLM plane.  Solutions with $z(x_1,x_2,0) = \pm 1/2$ turn out to be smooth everywhere \cite{Lin:2004nb}.  The two values of $z$ indicate which of the two 3-spheres appearing in the last two terms of \eqref{lnej18Jul24} remains of finite size as we approach $\eta = 0$ at a given $(x_1,x_2)$.   With these boundary conditions the solution is non-singular, but topologically complex.  It is also helpful to use the alternative parameterization 
\begin{align}
    u(x_1,x_2) \equiv \frac12-z(x_1,x_2,0) \, .
\end{align}

The $u$ parametrization facilitates  identification between the geometry and the CFT states described above.  This identification is carried out as follows \cite{Lin:2004nb,Balasubramanian:2005mg}.  First, we work out the Wigner density in the single fermion phase space for one of the half-BPS CFT states described above.  Then we coarse-grain that density by convolving with a Gaussian to derive the Husimi  distribution which is non-negative everywhere.\footnote{The Wigner distribution can be negative.  Indeed, this negativity is  related to ``quantum magic'', a quantity that controls how difficult it is to simulate the system on a classical computer in this representation of the physics.}  We then identify the Husimi phase space distribution of the state with the function $u(x_1,x_2)$ which sets the boundary conditions for the solution described above.  An identification between the phase space distribution in the CFT and the boundary condition at $\eta = 0$ was proposed in \cite{Lin:2004nb}; here we are describing the precise formulation in \cite{Balasubramanian:2005mg}.  Described in this way, locations in phase space (now identified with $(x_1,x_2)$) that are occupied by a fermion correspond to $u=1$, and $u=0$ elsewhere.  So the fermions of different energies associated to different rows of the Young diagrams described above produce concentric rings with $u=1$ in the $x_1$-$x_2$ plane at $\eta = 0$. Concretely, a fermion of level $n$ corresponds to a ring of radius $r_n$ with
\begin{align}
    r_n^2=2n\hbar,\qquad \text{where}\quad \hbar=2\pi l_p^4,\quad 4\pi N l_p^4=R_{\rm AdS}^4.
    \label{eq:r_vs_n}
\end{align}
Here, $l_p$ is the Planck length in the AdS$_5$ gravity and $R_{\rm AdS}$ is the AdS$_5$ radius.
See \cite{Balasubramanian:2018yjq} for more discussion.

Following from this description, one finds that in the thermodynamic, large system limit, a fermion distribution corresponding to a Young diagram with a limit curve $y(x)$ will correspond to a gravity solution with \cite[Eqs.~(110) and (111)]{Balasubramanian:2005mg}
\begin{align}
\begin{split}    u(r)&={1\over 1+y'} \\
   {\rm with} \ \  y(x)+x&={r^2\over 2\hbar} ~~~;~~~ r^2 = x_1^2 + x_2^2
   \end{split}
   \label{y(x)_and_u(x)}
\end{align}
where in the second line we took polar coordinates $(r,\phi)$ in the $x_1$-$x_2$ plane.

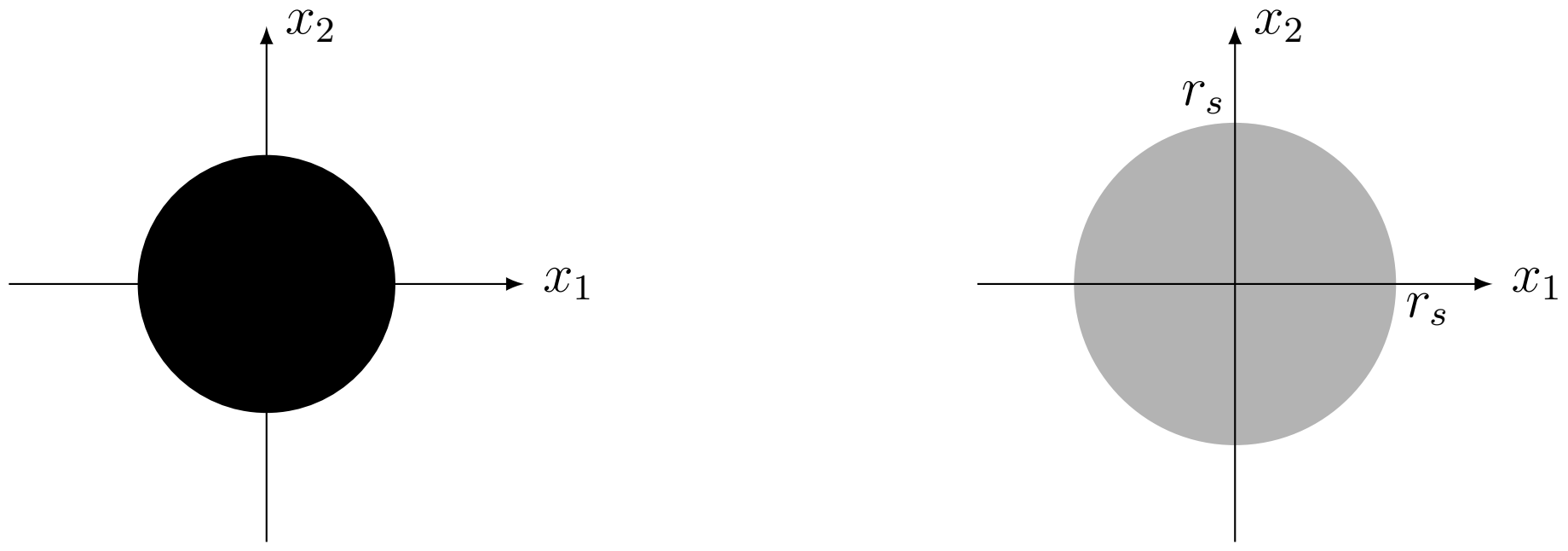
\begin{figure}
\begin{quotation}
    \begin{center}
    \begin{tabular}{c@{\hspace{2cm}}c}
    ~~~~~
    \begin{tikzpicture}
        \draw[-latex] (-2,0) -- (2,0) node [right] {$x_1$};
        \draw[-latex] (0,-2) -- (0,2) node [right] {$x_2$};
        \fill[black] (0,0) circle (1);
    \end{tikzpicture}
    &
    ~~~~~
    \begin{tikzpicture}
        \fill[white!70!black] (0,0) circle (1.25);
        \draw[-latex] (-2,0) -- (2,0) node [right] {$x_1$};
        \draw[-latex] (0,-2) -- (0,2) node [right] {$x_2$};
        \node at (0,1.25) [above left,xshift=2,yshift=-2] {$r_s$};
        \node at (1.25,0) [below right,xshift=-2,yshift=2] {$r_s$};
    \end{tikzpicture}
    \\
    (a) Empty AdS$_5\times S^5$ 
    &
    (b) Superstar
    \end{tabular}
        \caption{\label{fig:LLM_diag} \sl The LLM diagrams describing empty AdS$_5\times S^5$ and the $\beta=0$ superstar geometries, corresponding to \eqref{u_and_lambda} with $\lambda=0$ and generic $\lambda>0$, respectively.  Regions where $u(x_1,x_2)=1$ are shown in black, regions where $u(x_1,x_2)=0$ are shown in white, and regions where $0<u<1$ are shown in gray.  We will sometimes refer to $u(x_1,x_2)$ as the {\it grayness field}. }
    \end{center}
\end{quotation}
\end{figure}

From the above relations, it is easy to show that the infinite temperature ($\beta = 0$) superstar ensemble     \eqref{limitshape_superstar}
corresponds to the following $\eta=0$ data:
\begin{align}
u(r)=\begin{cases}
    \frac{1}{\lambda+1} & (0\le r\le r_{\rm s})\\[1ex]
    0 & (r_{\rm s}\le r)\\
\end{cases}~,
\label{u_and_lambda}
\end{align}
where $\lambda$ was defined in \eqref{limitshape_superstar} and $r_{\rm s}^2\equiv (1+\lambda)\RAdS^4$. Examples of the $\eta=0$ data are depicted in Figure~\ref{fig:LLM_diag}. Such diagrams are often called the LLM diagrams. We will refer to the LLM diagram for the superstar in Figure~\ref{fig:LLM_diag}(b) as the ``gray disk.''
With the boundary conditions \eqref{u_and_lambda} the integrals in \eqref{eq:zinsolution} and \eqref{eq:Vinsolution} can be carried out explicitly:
\begin{align}
\label{z,V_superstara}
\begin{split}
z(\eta,x_1,x_2)&=\frac{1}{ 2(\lambda+1)}
    \biggl(\frac{r^2-r_s^2+\eta^2}{\sqrt{(r^2+r_s^2+\eta^2)^2-4r_s^2r^2}}+\lambda\biggr),
\\
    V_\phi(\eta,x_1,x_2)&=-\frac{1}{2(\lambda+1)}
    \bigg(\frac{r^2+r_s^2+\eta^2}{\sqrt{(r^2+r_s^2+\eta^2)^2-4r_s^2r^2}}-1\bigg),
    \end{split}
\end{align}
where $x_1+ix_2=re^{i\phi}$
and the authors of \cite{Balasubramanian:2005mg} found  a singular geometry with vanishing horizon.  This geometry turns out to precisely match the explicit, closed form superstar solutions of \cite{Myers:2001aq}, which are extremal limits of the black hole solutions of 5d gauged supergravity \cite{Gunaydin:1985cu} found in \cite{Behrndt:1998jd,Cvetic:1999ne,Cvetic:1999xp,Cvetic:2004ny}.  For the hyperstar, $u(r)$ can be found from \eqref{eq:hyperstarlimit} and \eqref{y(x)_and_u(x)}, but no closed-form solution is known because the integrals in 
\eqref{eq:zinsolution} and \eqref{eq:Vinsolution} are more challenging in that case, although perturbative solutions exist \cite{Giombi:2005zq}. The hyperstar solution will also have a singularity on the $x_1$-$x_2$ plane at $\eta = 0$.  We will sometimes refer to the function $u(x_1,x_2)$ as the  {\it grayness field} where the grayness is the value of $0\leq u\leq1$ with $u=1$ referred to as ``black'' and $u=0$ referred to as ``white''.

\subsection{The case for quantum ER bridges in entangled superstars}
\label{sec:QuantumWormholes}

The authors of \cite{Engelhardt:2023xer} proposed algebraic criteria for determining when the entanglement between  degrees of freedom in distinct regions of spacetime is sufficient to induce a physical connection. The algebraic approach also allowed for the definition of a quantum notion of physical connectivity that applied, for example, to the relation between an evaporating black hole and its Hawking radiation before the Page time,  and to TFD states at very late times of order the inverse Newton constant. In these situations, and in the example studied in \cite{Herderschee:2025nsb}, the criteria of  \cite{Engelhardt:2023xer} suggest that a kind of quantum wormhole can be present even in the absence of a conventional classical bridge. Our goal in this paper is to develop explicit constructions of states of gravity that would be considered to contain quantum wormholes according to the algebraic criteria of \cite{Engelhardt:2023xer}.

The LLM framework \cite{Lin:2004nb} that we described above allows us to study the possibility of a quantum wormhole since the microscopic degrees of freedom and their relation to bulk geometry are explicitly known.  We will focus on  TFD-entangled superstar states in two copies of the half-BPS sector of $\mathcal N=4$ SYM that we described above.   The  reduced state in either copy is then described by the density matrix of a superstar ensemble which has a gravity description as an extremal black hole \cite{Balasubramanian:2005mg}.  But what is the geometric dual of the pure entangled state itself?  The pure  state \eqref{TFD_superstar_state}, interpreted literally, should be understood as a superposition of geometries.  Each element of the superposition has a separate description as a non-singular geometry which has two disconnected components of the form described above, each associated to the same Young diagram, which in turn should be separately dual to a non-singular, but topologically complex, LLM geometry.    However, in the large system limit, as we discussed, almost all the diagrams will lie close to certain limit curves.  So we are really summing over small fluctuations of this limit curve, which is the coarse-grained description of the typical state in this system, and is described by a (classically singular) black hole geometry \cite{Balasubramanian:2005mg}.   The entangled state is then a superposition of such states, consisting of two microscopically non-singular and disconnected factors that separately coarse-grain to the same (classically singular) black hole geometry.

The setup we are describing  was first analyzed in  \cite{Simon:2018laf}, and the idea that entanglement implies connectivity then says that the complete system should act as if there is a wormhole between two apparently disconnected extremal black holes.  It is certainly possible  that a superposition of semiclassical geometries can have a description as a different semiclassical geometry in this way \cite{Balasubramanian:2022gmo,Balasubramanian:2025jeu,Balasubramanian:2025hns}.
As we will show in the following sections, the associated von Neumann algebras of fine-grained probes are of Type~$\mathrm{III}_\lambda$, suggesting quantum connectivity~\cite{Engelhardt:2023xer}, while low-energy, coarse-grained probes yield Type~I algebras, suggesting that the two sides are disconnected at this level of description.  In other words, whether the two asymptotic regions act as if they are connected or disconnected depends on the fineness of the probes.

Note that, following \cite{Simon:2018laf}, we should not expect a classical Einstein-Rosen bridge between the two asymptotic regions of our construction, even though they are entangled with entropy of order $N$.   One reason for this is that the entanglement between the two systems is parametrically smaller than that associated with a macroscopic classical bridge:
the entanglement entropy $S_{\text{superstar}}$ \eqref{S_superstar} of the superstar TFD is order $N$ rather than $N^2$ as we would expect for a classical ER bridge and an associated horizon \cite{Simon:2018laf}.
Similarly, the mutual information for the TFD superstar scales as
\begin{align}
    I(L;R)=2S_{\text{superstar}}\sim N,
\end{align}
which scales as $N$ rather than $N^2$. The mutual information also constrains correlations between the two sides. In particular, defining the connected correlator by
\begin{align}
    \mathcal C_\rho
    \equiv
    \operatorname{tr}\!\left(\rho\,\mathcal O_L\otimes\mathcal O_R\right)
    -
    \operatorname{tr}(\rho_L\mathcal O_L)\,
    \operatorname{tr}(\rho_R\mathcal O_R),
\end{align}
one has, for bounded operators $\mathcal O_L$ and $\mathcal O_R$ \cite{PhysRevLett.100.070502},
\begin{align}
    I(L;R)\geq\frac{|\mathcal C_\rho|^2}
    {2\|\mathcal O_L\|^2\|\mathcal O_R\|^2}.
\end{align}
Here, $\rho$ is the total density matrix, $\rho_L$ and $\rho_R$ are its reduced density matrices, and $\|\cdot\|$ denotes the operator norm. As argued in \cite{Simon:2018laf}, the entanglement and correlations of the entangled superstar state are insufficient to support a classical ER bridge.

In addition, the gravitational description of the eternal two-sided black hole exhibits a degeneration of the bridge in the BPS limit. Superstars arise as such BPS limits of non-extremal, single R-charged black holes in five dimensions \cite{Cvetic:1999xp}. Following \cite{Simon:2018laf}, one can approach this limit from the non-extremal geometries and examine the behavior of the ER bridge: its length diverges while the horizon area vanishes. This provides a complementary indication that the limiting state does not admit a conventional classical bridge. The diverging length also suggests a qualitative parallel with recent discussions of long wormholes in eternal black holes, where the growing bridge length has been related to quantum states that become delocalized over the Hilbert space and cease to admit a classical geometric description \cite{Nandy:2024zcd,Miyaji:2024ity,Balasubramanian:2024lqk,Miyaji:2025yvm,Balasubramanian:2025xkj}.

Thus, our goal below is to use algebraic methods to test whether the two entangled universes nevertheless have some kind of quantum wormhole between them.

\section{Coarse probes} 
\label{sec:coarse}

We want to understand the physics of low-energy, coarse probes of the background defined by a typical state that coarse-grains to a superstar geometry.  To this end we first study the algebra of the bosonic $D_j$ operators.    In the large $N$ limit, as explained in Secs.~\ref{sss:eff_thermodyn_ensembles} and \ref{sss:bulk_side}, a typical LLM geometry coarse-grained as appropriate to probes that cannot resolve features at the Planck scale, is described by the boundary data (the grayness field) $u(r)$ with $0\le u\le 1$, which corresponds to a limit curve of Young diagrams describing the associated microstates.  As discussed above, almost all the underlying microstates  become indistinguishable at large~$N$.  This motivates us to look not at the full algebra of $D_j$ but at the subsector corresponding to long-wavelength fluctuations about the limit curve.  Such fluctuations are accessible to an IR observer, who can only see macroscopic, coarse-grained physics (also see \cite{Balasubramanian:2018yjq}).

\subsection{IR deformations}
\label{ss:IR_deformations}

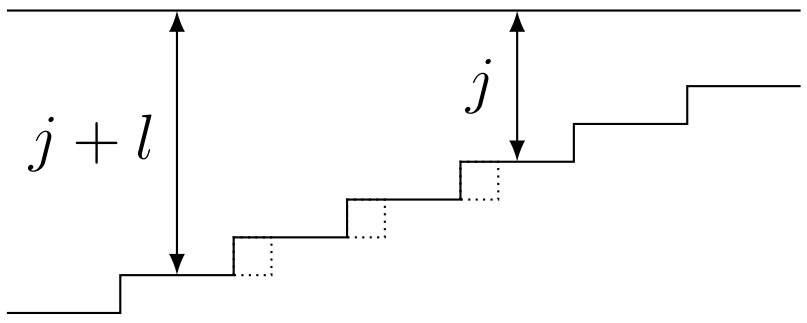
\begin{figure}[tbh]
\begin{quotation}
\begin{center}
\vspace*{3ex}
 \begin{tikzpicture}[scale=0.25]
 \draw (0,0) 
 -- ++(3,0) -- ++(0,1)
 -- ++(3,0) -- ++(0,1)
 -- ++(3,0) -- ++(0,1)
 -- ++(3,0) -- ++(0,1)
 -- ++(3,0) -- ++(0,1)
 -- ++(3,0) -- ++(0,1)
 -- ++(3,0);
 \draw (0,8) -- +(21,0);
 \draw[densely dotted] (6,1) rectangle +(1,1);
 \draw[densely dotted] (9,2) rectangle +(1,1);
 \draw[densely dotted] (12,3) rectangle +(1,1);
 \draw[latex-latex] (13.5,4) -- +(0,4) node [midway,left] {$j$};
 \draw[latex-latex] (4.5,1) -- +(0,7) node [midway,left] {$j+l$};
 \end{tikzpicture}
 \caption{\label{fig:DdaggerD} \sl The operator $D_{j+l}^\dagger D_{j}$ adds one box to each of the
 $(j+1)$th through $(j+l)$th rows. The dotted boxes represent $l$ added boxes.}
\end{center}
\end{quotation}
\end{figure}

What are the operators available to an IR observer in the large $N$ limit? 
Concretely, let us consider the limit curve of the $\beta=0$ superstar in Figure~\ref{fig:limit_shapes}(c), whose microscopic state is given by~\eqref{TFD_superstar_rho_boson_2}, and think about operators that induce long-wavelength fluctuations of the ``interior'' part of the curve, $0<x<N$.
The operators $D_j$ and $D^\dagger_j$ themselves  in the interior part, where $j=O(N)$, are not allowed because they create or annihilate a column of height $j=O(N)$, and change R-charge (and thus energy) by an amount of order $N$, which is taken to infinity.  Thus they cannot be regarded as IR modes.  On the other hand, the combination $D_{j+l}^\dagger D_j$ with $l=O(1)$ has an $O(1)$ R-charge, and is therefore allowed (see Figure~\ref{fig:DdaggerD} for a graphical explanation of the action of this operator).  
Having R-charge $l$, this operator describes an excitation that is not rotationally symmetric; namely, it creates waves along the angular $\phi$ direction on the LLM plane (see Appendix \ref{app:A_are_fluct}).

So, we can consider the following family of  operators that is well-defined in the large $N$ limit:
\begin{align}
 A_{f,l}={1\over\sqrt{N}}\sum_{j=1}^{N-l} D_{j+l}^\dagger D_j\, f(\tfrac{j}{N}),
 \label{mgns28Sep26}
\end{align}
where $f(v)$ ($0\le v\le 1$) is a smooth complex function and $l>0$, $l=O(1)$, is a positive integer. Let us give some justification for this choice.  A macroscopic observer knows how much R-charge is being added, but not exactly which bosons are being manipulated; this is why we are summing over the boson label $j$.   Although we have no microscopic control over the amplitude of the fluctuation, it is possible to modulate the amplitude on a large scale of order~$N$, which is why we have kernel $f(\tfrac{j}{N})$.
In Appendix \ref{app:A_are_fluct}, we show  that the operator $A_{f,l}$ indeed corresponds to a fluctuation of the grayness field defined in Sec.~\ref{sss:bulk_side}, with $f(v)$ related to the radial fluctuation profile and with $l$ giving the angular frequency.
Finally, the overall normalization of
\eqref{mgns28Sep26} is chosen so that its two-point function is finite in the large $N$ limit, as we will see below.

We do not have to consider the case with $l<0$ separately, because for $l>0$,
\begin{align}
 A_{f,l}^\dagger
 ={1\over\sqrt{N}}\sum_{j=1}^{N-l} D_{j}^\dagger D_{j+l}\, \fb(\tfrac{j}{N})
 ={1\over\sqrt{N}}\sum_{j=l+1}^{N} D_{j-l}^\dagger D_{j}\, \fb(\tfrac{j-l}{N})
 \approx {1\over\sqrt{N}}\sum_{j=l+1}^{N} D_{j-l}^\dagger D_{j}\, \fb(\tfrac{j}{N}),
\end{align}
where we approximated $\fb(\tfrac{j-l}{N})=\fb(\tfrac{j}{N})-\tfrac{l}{N}\fb'(\tfrac{j}{N})+\cdots\approx  \fb(\tfrac{j}{N})$ at large $N$.  Therefore, $A^\dagger_{f,l}$ can be identified with $A_{\fb,-l}$ at large $N$.  The $l=0$ case, $A_{f,0}$, requires a separate treatment and is discussed in Appendix \ref{app:l=0_modes}.

If we only require that the operator have $O(1)$ R-charge (and energy), we can also consider operators such as $D_{2j}^\dagger D_j^2$.  However, if $j=O(N)$ such an operator is nonlocal in that it acts simultaneously on macroscopically distant values of~$j$.  So we discard such operators, keeping the large $N$ limit in mind. 

We can also ask about IR modes living near the disk edge, which corresponds to bosons with $j=O(1)$.  In particular, from Fig.~\ref{fig:DdaggerD}, it is natural to include the operator that adds one box to each of the rows $1$ through $l$, which is $D_l^\dagger$.  
However, there are other operators with the same charge that are macroscopically indistinguishable in their effects.
For example, $D_{l-k}^\dagger (D_1^\dagger)^{k}$ 
and $D_{l-k_1-2k_2}^\dagger (D_1^\dagger)^{k_1}(D_2^\dagger)^{k_2}$ both have charge $l$, and it would seem that we should sum over all such operators with a kernel, like in~\eqref{mgns28Sep26}.
However, we dropped similar operators in the interior by requiring locality.
Exactly what operators to include in the sum, and with what relative coefficients, requires detailed knowledge about how IR observables (namely, the grayness field $u$) respond to perturbations they induce---we need to sum over UV operators that lead to identical IR responses, and drop ones that lead to singular IR responses in the large $N$ limit.  We leave it to future work to sort this out, and here do not consider the small number of near-edge modes, and instead focus on the interior modes $A_{f,l}$.

Below, we will discuss the algebra of the operators $A_{f,l}$ on the superstar TFD state, \eqref{TFD_superstar_Psi_boson_2}.  Actually, because we have left and right universes, we have left and right copies of the operator:
\begin{align}
 A_{f,l}^{\alpha}={1\over\sqrt{N}}\sum_{j=1}^{N-l} D_{j+l}^{\alpha\dagger} D_j^\alpha\, f(\tfrac{j}{N}),\qquad
 \alpha=R,L.\label{llfj10Nov26}
\end{align}
These operators represent coarse fluctuations of the left and right LLM systems in our construction.

\subsection{Single-sided algebra of IR deformations}

We are interested in the algebra of $A_{f,l}^\alpha$ on the superstar TFD state $\ket{\Psi^{(s)}}$ of \eqref{TFD_superstar_Psi_boson_2}.  First, the one-sided correlation function in the state $\ket{\Psi^{(s)}}$ is
\begin{align}
&\ev{A_{f,l}^{R\dagger}\, A_{g,m}^R}_{\Psi^{(s)}}    
=\ev{ A_{g,m}^R \,A_{f,l}^{R\dagger}}    _{\Psi^{(s)}}    
    ={\delta_{lm}\over N}\sum_{j=1}^{N-l}\fb(\tfrac{j}{N}) g(\tfrac{j}{N})(\ev{\hat{c}^R_{j}}+1)\ev{\hat{c}^R_{j+l}}
    \notag\\
    &\qquad
    ={\delta_{lm}\over N}{\zeta\over (1-\zeta)^2}\sum_{j=1}^{N-l}
    \fb(\tfrac{j}{N}) g(\tfrac{j}{N})
    \approx \delta_{lm}\,(f,g),
    \label{<ALAL>_on_superstar}
\end{align}
where
\begin{align}
    \ev{\hat{c}_j^R}={\zeta\over 1-\zeta}\equiv \ev{c}
    \label{eq:<c>}
\end{align}
is the expectation value of the occupation number of the $j$th boson, which is independent of~$j$ for the state $\ket{\Psi^{(s)}}$, and
\begin{align}
    (f,g)\equiv {\zeta\over (1-\zeta)^2}\int_0^1 dv\,\fb(v)\,g(v).
\end{align}
Here we are focusing on the right-side  ($\alpha=R$) correlators, but the left-side ones are the same.
Assuming that $f,g=O(1)$, the correlator \eqref{<ALAL>_on_superstar} is $O(1)$, justifying the normalization in \eqref{mgns28Sep26}.
Consider an orthonormal basis of functions $\{f_k\}$, $k=1,2,\dots,$ such that
\begin{align}
    (f_k,f_{k'})=\delta_{kk'}.\label{eq:ff_orthogonality}
\end{align}
We can take any orthonormal basis in principle, but let us say that the basis is indexed such that larger $k$ corresponds to shorter wavelengths or smaller scale fluctuations.  Then in order to stay in the IR, we must not let $k$ scale with $N$; namely, we must take $k=O(1)$. 
Likewise, the two-sided correlator is given by
\begin{align}
\ev{A_{f,l}^R\, A_{g,m}^L}_{\Psi^{(s)}}
    &={\delta_{lm}\over N}{\zeta\over (1-\zeta)^2}\sum_{j=1}^{N-l}
    f(\tfrac{j}{N}) g(\tfrac{j}{N})
    \approx \delta_{lm}\, (\fb,g). \label{<ALAR>_on_superstar}
\end{align}
All other combinations vanish,
$\ev{A_{f,l}^{\alpha}\, A_{g,m}^{\alpha}}_{\Psi^{(s)}}
=\ev{A_{f,l}^{\alpha\dagger}\, A_{g,m\mathstrut}^{\alpha\dagger}}_{\Psi^{(s)}}
=\ev{A_{f,l}^{\alpha}\, A_{g,m}^{\beta\dagger}}_{\Psi^{(s)}}=0$, where $\alpha\neq \beta$.
Higher point correlation functions at large $N$ can be computed by Wick contraction using the 2-point functions \eqref{<ALAL>_on_superstar} and \eqref{<ALAR>_on_superstar}. This can be easily shown because a contraction---relating the summation variables in two factors of $A$---reduces the number of independent sums by one and is down by a factor of $1/N$. The dominant contribution at large $N$ is then obtained by pairwise contractions, as long as the number of operators is $O(1)$.  For example,
\begin{align}
&\ev{A_{f_1,l_1}^{R\dagger} A_{f_2,l_2}^{R\dagger}
A_{g_1,m_1}^R A_{g_2,m_2}^R}_{\Psi^{(s)}}
\notag\\
&\qquad \approx 
 \delta_{l_1 m_1}\delta_{l_2 m_2}(f_1,g_1)(f_2,g_2)
+ \delta_{l_1 m_2}\delta_{l_2 m_1}(f_1,g_2)(f_2,g_1)
+O(1/N).\label{eq:large_N_fact'n}
\end{align}
This large $N$ factorization is true even in the case with non-constant grayness, as discussed in Appendix~\ref{app:nonconst_grayness}.

Now let us turn to commutators.  The one-sided commutators are, for $l,m>0$,
\begin{subequations}\label{AA_comm_op}
\begin{align}
 [A_{f,l}^R,A_{g,m}^R]
 &={1\over N}\sum_{j=1}^{N-l-m}
    [f(\tfrac{j+m}{N})g(\tfrac{j}{N})-f(\tfrac{j}{N})g(\tfrac{j+l}{N})]D_{j+l+m}^{R\dagger} D_{j}^R,
    \label{AA_comm_op_1}
    \\
 [A_{f,l}^R,A_{g,m}^{R\dagger}]
 &=
 {1\over N}\sum_{j=1}^{N-{\rm max}(l,m)}\!\!
 f(\tfrac{j}{N})\gb(\tfrac{j}{N}) D_{j+l}^{R\dagger} D_{j+m}^R
 \notag\\[-1ex]
 &\qquad
 -
 {1\over N}\sum_{j=1+{\rm max}(l,m)}^N\!\!
 f(\tfrac{j-l}{N})\gb(\tfrac{j-m}{N}) D_{j-m}^{R\dagger} D_{j-l}^R.\label{AA_comm_op_2}
\end{align}
\end{subequations}
On the other hand, the two-sided commutators vanish,
\begin{align}
    [A_{f,l}^R,A_{g,m}^L]=
    [A_{f,l}^R,A_{g,m}^{L\dagger}]=0.
\end{align}

Now we argue that the commutators \eqref{AA_comm_op} vanish as far as the IR algebra is concerned.  More precisely, let $\cH^R_{\rm IR}$ be the right-side Hilbert space generated by finite products of the IR operators $A_{f,l}^R,A_{g,m}^{R\dagger}$.  The number of the $A_{f,l}^R,A_{g,m}^{R\dagger}$ operators and the sum of angular harmonics $l$ in the products are held fixed as $N\to\infty$. 
First, the commutators \eqref{AA_comm_op} vanish in the superstar TFD state:
\begin{subequations}
\begin{align}
 \ev{[A_{f,l}^R,A_{g,m}^R]}_{\Psi^{(s)}}
 &={1\over N}\sum_{j=1}^{N-l-m}
    [f(\tfrac{j+m}{N})g(\tfrac{j}{N})-f(\tfrac{j}{N})g(\tfrac{j+l}{N})]\ev{D_{j+l+m}^{R\dagger} D_{j}^R}_{\Psi^{(s)}}=0
 ,
 \\
 \ev{[A_{f,l}^R,A_{g,m}^{R\dagger}]}_{\Psi^{(s)}}
 &=
 {1\over N}\delta_{lm}\sum_{j=1}^{N-l}
 f(\tfrac{j}{N})\gb(\tfrac{j}{N})
 (\ev{\hat{c}_{j+l}^R}-\ev{\hat{c}_{j}^R})=0,
\end{align}%
\end{subequations}%
because $\ev{D_{j+l}^{R\dagger} D_{j}^R}_{\Psi^{(s)}}=0$ for $l\neq 0$ and 
because $\ev{\hat{c}_j^R}$ is independent of $j$ (see \eqref{eq:<c>}).
For more general states, note that, for $l,m=O(1)$, the right-hand side of the commutator \eqref{AA_comm_op_1} can be expanded as
\begin{align}
 [A_{f,l}^R,A_{g,m}^R]
 &={1\over N^2}\sum_{j=1}^{N-l-m}
    \Bigl[    mf'(\tfrac{j}{N})g(\tfrac{j}{N})
    -lf(\tfrac{j}{N})g'(\tfrac{j}{N})+O({1/N})
    \Bigr]
    D_{j+l+m}^{R\dagger} D_{j}^R .
    \label{kmcn12Oct26}
\end{align}
Because for states in $\cH^R_{\rm IR}$ the expectation value 
$\ev{D_{j+l+m}^{R\dagger} D_{j}^R}=O(1)$, the expectation value of \eqref{kmcn12Oct26} is $O(1/N)$ and vanishes in the $N\to\infty$ limit.  Similarly, one can show that the expectation value of the right-hand side of \eqref{AA_comm_op_2} is $O(1/N)$ and vanishes in the $N\to\infty$ limit.
Thus, in the IR Hilbert space $\cH_{\rm IR}^R$, the generators $A^R_{f,l},A_{g,m}^{R\dagger}$ all commute in the $N\to\infty$ limit. Another argument is as follows.
The right-hand side of \eqref{kmcn12Oct26} can be written as $N^{-3/2}A^R_{mf'g-lfg',l+m}$.  Since $A^R\sim O(\sqrt{N})$ from \eqref{AA_comm_op}, the right-hand side is $O(1/N)$.
We have shown this here for the case of the $\beta=0$ superstar with constant grayness field, but we show in Appendix~\ref{app:nonconst_grayness} that the $A^R_{f,l},A_{g,m}^{R\dagger}$ generators all commute at large $N$ even for non-constant grayness.

The operators $A_{f,l},A_{g,m}^{\dagger}$ correspond to the long-wavelength fluctuation of the grayness field $u(x_1,x_2)$ in supergravity, which is an IR theory.  It was shown in~\cite{Balasubramanian:2018yjq} that, in the IR theory, $u(x_1,x_2)$ has a conjugate field $\pi(x_1,x_2)$ with the canonical Poisson bracket, $\{u(x_1,x_2), \pi(x_1',x_2')\}_{\rm PB} = \delta(x_1-x_1')\delta(x_2-x_2')$.  This seems to suggest that, even in the current setting, there are right-side IR operators $B_{f,l}^R,B_{g,m}^{R\dagger}$ that are conjugate to $A_{f,l}^R,A_{g,m}^{R\dagger}$ and satisfy canonical commutation relations $[A_{f,l}^R,B_{g,m}^{R\dagger}]\propto i\delta_{lm}\int dv\,f\gb$.

However, in fact, we can prove a ``no-go theorem'' that says that there is no right-side operator $\cO^R$ that has non-vanishing commutator with $A_{f,l}^R,A_{g,m}^{R\dagger}$ on $\ket{\Psi^{(s)}}$.  Here $\cO^R$ is not just a coarse-grained IR operator but any operator in the UV theory.
To show the no-go theorem, first note that the $\beta=0$ superstar density operator \eqref{TFD_superstar_rho_boson_2}, here regarded as the reduced density operator on the right, can be written as
\begin{align}
 \rho^{(s)}_R
&\propto \zeta^{N_{\rm col}},\qquad
N_{\rm col}\equiv \sum_{i=1}^N D_i^{R\dagger} D_i^R,
\end{align}
where $N_{\rm col}$ is the total number of columns (on the right).  Because $D_{i+l}^{R\dagger} D_i^R$ conserves the number of columns, $[N_{\rm col},A_{f,l}^R]=0$ and therefore  $[\rho^{(s)}_R,A_{f,l}^R]=0$.
Then, for any right-side operator $\cO^R$, using cyclicity of trace,
\begin{align}
 \ev{[A_{f,l}^R,\cO^R]}_{\rho^{(s)}_R}
=
 \Tr_R(\rho^{(s)}_R[A_{f,l}^R,\cO^R])
=
 \Tr_R(\cO^R[\rho^{(s)}_R,A_{f,l}^R])=0.\label{hotu2Oct26}
\end{align}
Namely, there is no operator $\cO^R$ that has non-vanishing commutator with $A_{f,l}^R,A_{f,l}^{R\dagger}$ when evaluated on $\rho^{(s)}_R$.
Furthermore,
we can show the same thing even if we change the density operator $\rho^{(s)}_R$ as
\begin{align}
    \rho^{(s)}_R \to \rho_R' =  e^{iG_R} \rho^{(s)}_R  e^{-iG_R}
    \label{eq:rho_R'}
\end{align}
(or equivalently $\ket{\Psi^{(s)}}\to e^{iG_R}\ket{\Psi^{(s)}}$), where $G_R$ is an operator made of the IR fields  $A^R_{f,l},A^{R\dagger}_{f,l}$.  This follows from the fact that $\rho^{(s)}_R$ commutes with $A^R_{f,l},A^{R\dagger}_{f,l}$ (as operators, with no $O(1/N)$ corrections) as stated above, and hence in fact $\rho_R'=\rho^{(s)}_R$.

So, the above ``no-go theorem'' says that the commutator should vanish at large $N$ evaluated on $\rho_R^{(s)}$ and on a perturbed density matrix of the form 
\eqref{eq:rho_R'}.  This does not say that the commutator should vanish under arbitrary IR perturbations that are not of the form of \eqref{eq:rho_R'}.  However, it does say that, when evaluated on $\rho_R^{(s)}$, the commutator vanishes exactly.  This means that there is no right-side operator, even a UV one, satisfying $[A^R_{f,l},\cO^R]\propto \hat{1}$ in $\cH^R_{\rm IR}$, because that would contradict the identity \eqref{hotu2Oct26}.
This presents a tension with the result of \cite{Balasubramanian:2018yjq}, which showed that we can construct a field conjugate to the IR field $u^R$, which is essentially our $A^R_{f,l}$,\footnote{It is logically possible that $A^R_{f,l}$ is not equivalent to the IR gravitational field $u$, although the calculation in Appendix \ref{app:A_are_fluct} is strong evidence for that.} using UV data.  We will discuss this point further in Sec.~\ref{sec:discussion}.

The absence of right-side IR operators conjugate to the grayness fluctuation operators $A^R_{f,l},A^{R\dagger}_{f,l}$ means that a right-side IR observer cannot dynamically change the grayness field by adding any IR perturbation to the system's right-side Hamiltonian.  Of course, this does {\it not} mean that there are no gravity solutions with $u(x_1,x_2)$, i.e.~the grayness, varying over the LLM plane. Rather it means that agents acting in the IR cannot change the singular boundary conditions at the LLM plane. In effect, these boundary conditions define the asymptotic structure of the vacuum state around which IR perturbations act.

What is the algebra of the IR physics?  In the right-side IR Hilbert space $\cH_{\rm IR}^R$, the operators $A_{f,l}^R,A_{f,l}^{R\dagger}$ are all mutually commuting.  By switching the function label $f$ to the discrete label $k=O(1)$ of an orthogonal basis \eqref{eq:ff_orthogonality}, we can say that we have a countably infinite collection of mutually commuting operators $A_{k,l}^R$ and $A_{k,l}^{R\dagger}$, with $k,l=O(1)$.  
For fixed $k$ and $l$, this algebra of $A^R_{k,l}$ is like the algebra $\cA_x$ of the position operator $\hat{x}$ in one-dimensional quantum mechanics.  Each eigenspace, spanned by $\ket{x}$ with $x\in\bbR$, furnishes an algebra of $1\times 1$ matrices, namely $\bbC$, which is a Type I$_1$ algebra. Because we have a Type~I$_1$ algebra for each eigenvalue $x\in\bbR$, the algebra of $\hat{x}$ is a direct integral\footnote{The continuous version of the direct sum.} of Type~I$_1$ factors, which can be denoted by
\begin{align}
    \int_{\bbR}^{\oplus} \bbC\, dx,
    \label{direct_int_I_1}
\end{align}
where $\int^\oplus$ denotes a direct integral.\footnote{%
Another explanation is as follows.  Consider an observable $O=O^\dagger$ in one-dimensional quantum mechanics, and let its eigenvalues be $\lambda_n\in\bbR$ ($n=1,2,\dots$), assumed to be discrete and non-degenerate.  The algebra generated by $O$ includes $f(O)$ for all appropriately bounded complex functions $f$. In the diagonal basis, this means we should consider all matrices of the form \begin{align}
\mqty(f(\lambda_1)&0&\\
0&f(\lambda_2)&\\
&&\ddots).
\end{align}
Because $\lambda_n$ are all distinct, $f(\lambda_n)$ for different $n$ are independent.  So, the algebra of each diagonal element is~$\bbC$, namely the algebra of $1\times 1$ matrices, which is Type I$_1$. Therefore, the algebra of $O$ is a direct sum of $\bbC$, namely Type I$_1$ algebras, over $n=1,2,\dots$. Eq.~\eqref{direct_int_I_1} is the continuous version of this statement.}
Another way to say this is the following: the center of the algebra $\cA_x$ is composed of $\hat{x}$ itself, and $\cA_x$ can be decomposed into factors labeled by the value $x$ of the central element $\hat{x}$, as in \eqref{direct_int_I_1}.
Likewise, for fixed $k,l$, the algebra of $A_{k,l}$  is the direct integral~\eqref{direct_int_I_1} over all possible eigenvalues of $A^R_{k,l}$.  If we consider all $k,l=O(1)$, including the conjugate $A^{R\dagger}_{k,l}$, we have an infinite tensor product of those algebras.  This algebra is equivalent to the direct integral over all mode configurations $\{A^R_{k,l},A^{R\dagger}_{k,l}\}$, namely, over all their joint eigenvalues.
In this way, the algebra of the IR operators is decomposed into a direct integral of Type~I$_1$ factors, labeled by the values of the central elements $\{A^R_{k,l},A^{R\dagger}_{k,l}\}$.

According to the algebraic ER=EPR conjecture \cite{Engelhardt:2023xer}, if the one-sided algebra is of Type~I, the spacetime is disconnected.  So, the algebra of the IR probes $\{A^R_{k,l},A^{R\dagger}_{k,l}\}$ being of Type~I$_1$ as we saw above suggests that the spacetime is disconnected as far as the IR physics is concerned.  This is consistent with the fact that the superstar geometry is a limit of a classical black hole in which the horizon vanishes, the ER bridge becomes infinitely long, and hence classical communication between the left and right universes is impossible.

Strictly speaking, however, the algebraic ER=EPR conjecture as reviewed in  Sec.~\ref{ss:alg_ER=EPR_conj} requires that the one-sided algebra be a factor. As we discussed above, the algebra of the IR probes is not a factor, but decomposes into sectors each of which is a Type~I$_1$ factor labeled by the values of $\{A^R_{k,l},A^{R\dagger}_{k,l}\}$. Nevertheless, if we restrict to one subsector, it is a Type~I$_1$ factor.
If we assume that the algebraic ER=EPR is applicable in this generalized setting, then it does say that the right and left universes are disconnected.  
Or, perhaps more physically, the factor assumption of the conjecture is reasonable if the left and right systems are quantum fields in bulk causal wedges.  When the two systems are disconnected spacetimes, the associated algebras are each expected to be of Type~I$_\infty$.  Now, if one takes a classical limit by sending $\hbar \to 0$, the quantum fields become commutative and the algebra reduces to a direct integral of Type~I$_1$ factors, while the spacetime stays disconnected.  So, it is physically reasonable that the IR algebra $\{A^R_{k,l},A^{R\dagger}_{k,l}\}$, as a direct integral of Type~I$_1$ factors, implies that the asymptotic regions of the geometry are classically disconnected.

\subsection{The mirror universes}

The no-go theorem \eqref{hotu2Oct26} suggests that we need to combine operators from both the left and right universes to construct operators that have non-vanishing commutators with $A_{f,l}^R,A_{f,l}^{R\dagger}$.  Here we construct such operators and discuss their physical significance.

The origin of this situation  can be found in Eqs.~\eqref{<ALAL>_on_superstar} and \eqref{<ALAR>_on_superstar}, which imply that the operators $A_{f,l}^R$ and $A_{\fb,l}^{L\dagger}$ behave the same way in the TFD state $\ket{\Psi^{(s)}}$.  In fact, it is not difficult to show that
\begin{align}
    A_{f,l}^{R}\ket{\Psi^{(s)}}=A_{\fb,l}^{L\dagger}\ket{\Psi^{(s)}},\qquad
    A_{f,l}^{R\dagger}\ket{\Psi^{(s)}}
    =A_{\fb,l}^{L}\ket{\Psi^{(s)}}.
    \label{muxi14Oct26}
\end{align}
This implies that, although $A^R$ and $A^{L\dagger}$ are different operators in the UV theory, they behave as identical operators when acting on~$\ket{\Psi^{(s)}}$.  More generally, using \eqref{muxi14Oct26} and following the same reasoning as in the proof of large-$N$ factorization \eqref{eq:large_N_fact'n}, we can show that the replacements $A_{f,l}^{R}\to A_{\fb,l}^{L\dagger}$ and $A_{f,l}^{R\dagger}\to A_{\fb,l}^{L}$ leave any correlation function in this state unchanged in the large-$N$ limit if the number of operators is $O(1)$ (this holds also in the case of non-constant grayness).
Because of this, we cannot think of the left and right universes independently if we work with the kind of IR algebra that we are considering here.

We can also see this as follows.  Define a ``coherent state'' for the $A,A^\dagger$ operators by
\begin{align}
 \ket{\chi}= 
 {1\over\sqrt{Z}}
 \ket{\chi_0},\qquad
 \ket{\chi_0}=
 \exp[
 z_+^R A_R+z_-^R A_R^\dagger 
 + z_+^L A_L^\dagger + z_-^L A_L
 ]\ket{\Psi^{(s)}},\label{efmx7Oct26}
\end{align}
where we are writing 
$A_R=A^R_{f,l}, A_L=A^L_{\fb,l}$ for simplicity. Also,
$Z\equiv \ev{\chi_0|\chi_0}$ and $z_+^R,z_-^R,z_+^L,z_-^L\in \bbC$
are all independent complex parameters.   Much as the ordinary coherent state for a harmonic oscillator, this state represents a classical, IR state in which operators can be effectively replaced by their expectation values \cite{Yaffe:1981vf}.
From \eqref{<ALAL>_on_superstar} and \eqref{<ALAR>_on_superstar}, we can compute, at leading order in the small $|z|$ expansion,
\begin{align}
 Z=\ev{\chi_0|\chi_0}
 &=1+\kappa\Bigl(|z_+^R+z_+^L|^2+|z_-^R+z_-^L|^2
 \notag\\
 &\qquad\qquad
 +(z_+^R+z_+^L)(z_-^R+z_-^L)+(\zb_+^R+\zb_+^L)(\zb_-^R+\zb_-^L)
 \Bigr)+\cdots,
\end{align}
where
\begin{align}
    \kappa\equiv (f,f).
\end{align}
The symplectic form of the IR theory \cite{Belin:2018fxe} is then, for small $|z|$,
\begin{align}
i\p_i \pb_i \log Z\, dz^i\wedge d\bar{z}^i
&\sim 
 i\kappa [d(z_+^R+z_+^L)\wedge d(\zb_+^R+\zb_+^L)
 +d(z_-^R+z_-^L)\wedge d(\zb_-^R+\zb_-^L)]
 \notag\\
&=i \kappa (dw_+\wedge d\wb_+ + dw_-\wedge d\wb_-),\label{iex7Oct26}
\\
    w_\pm &\equiv z_\pm ^R + z_\pm ^L.
\end{align}
This symplectic form means that there are only two real canonical pairs, $(\Re w_+,\Im w_+)$ and $(\Re w_-,\Im w_-)$, not four as expected from the four complex variables
$z_\pm^L,z_\pm^R$.  This is because $A_R$ and $A_L^{\dagger}$ are identical
on $\ket{\Psi^{(s)}}$, as we have seen above.  In the coherent state
\eqref{efmx7Oct26}, the expectation values of the $A,A^\dagger$ operators can be
computed from \eqref{<ALAL>_on_superstar} and \eqref{<ALAR>_on_superstar}, at leading order in small $w$ expansion, as
\begin{align}
 \ev{A_R}_\chi = \ev{A_L^\dagger}_\chi =\kappa(\wb_++w_-),\qquad
 \ev{A_R^\dagger}_\chi  =\ev{A_L}_\chi  =\kappa(w_+ + \wb_-).\label{whb7Oct26}
\end{align}
The left and right operators take the same expectation value, and the
value depends only on the combination $w_\pm$ and not separately on
$z_\pm^R,z_\pm^L
$, again reflecting the fact that left and right fields have become identical on the state $\ket{\Psi^{(s)}}$.

The symplectic form \eqref{iex7Oct26} and the expectation values \eqref{whb7Oct26} imply that the operators conjugate to $A_R$ and $A_L^\dagger$ take expectation value $\sim i(\wb_+-w_-)$.  Let us explicitly construct such operators.

\subsubsection*{Constructing the conjugate}

The construction starts by defining the following operator that involves both left and right:
\begin{align}
 \Delta=\half\sum_{j=1}^N (\hat{c}_j^R-\hat{c}_j^L
 )^2.
\label{eq:Delta_def}
\end{align}
This operator satisfies:
\begin{align}
 \Delta\ket{\Psi^{(s)}}=0,\qquad
 \Delta A_\alpha \ket{\Psi^{(s)}}=A_\alpha\ket{\Psi^{(s)}},\qquad
 \Delta A_\alpha^\dagger\ket{\Psi^{(s)}}=A_\alpha^\dagger\ket{\Psi^{(s)}},
 \label{itu7Oct26}
\end{align}
where $\alpha=R,L$. The first relation is because left and right occupation numbers are always the same in the state $\ket{\Psi^{(s)}}$.  The other two relations hold because the operator $D^{\alpha\dagger}_{j+l}D^\alpha_{j}$ makes the left and right occupation numbers different by one for two (the $j$th and $(j+l)$th) bosons.
Now define, for $\alpha=R,L$,
\begin{align}
 B_\alpha\equiv i[\Delta,A_\alpha],\qquad
 B_\alpha^\dagger\equiv i[\Delta,A_\alpha^\dagger].
 \label{eq:B_alpha_def}
\end{align}

We can show, at leading order in the small $z$ expansion,
\begin{align}
 \ev{B_R}_\chi 
 &=i\ev{\chi|[\Delta,A_R]|\chi}
 \notag\\
 &\approx i\ev{
 (\zb^R_+ A_R^\dagger+\zb^L_+ A_L)(\Delta A_R- A_R \Delta )
 +(\Delta A_R-A_R \Delta  )(z^R_- A_R^\dagger+z^L_- A_L)
 }_{\Psi^{(s)}}
 \notag\\
 &=i\ev{
 (\zb^R_+ A_R^\dagger+\zb^L_+ A_L) A_R
 -A_R (z^R_- A_R^\dagger+z^L_- A_L)
 }_{\Psi^{(s)}}
 \notag\\
 &=i\kappa(\wb_+-w_-).
\end{align}
In the second line we expanded $\ket{\chi}$ in \eqref{efmx7Oct26} and kept leading non-vanishing terms, while in the third line we used \eqref{itu7Oct26}.
Similarly, we find
\begin{align}
 \ev{B_R}_\chi =\ev{B_L^\dagger}_\chi=i\kappa(\wb_+-w_-),\qquad
 \ev{B_R^\dagger}_\chi =\ev{B_L}_\chi=-i\kappa(w_+-\wb_-),\label{wiw7Oct26}
\end{align}
which are correct values for operators conjugate to $A_{L,R}$.
Indeed, we can show that
\begin{align}
\ev{[A_R,B_R^\dagger]}_{\Psi^{(s)}}
=\ev{[A_L,B_L^\dagger]}_{\Psi^{(s)}}
=\ev{[A_R,B_L]}_{\Psi^{(s)}}
=\ev{[A_L,B_R]}_{\Psi^{(s)}}
= 2i\kappa.\label{wji7Oct26}
\end{align}
For example,
again using \eqref{itu7Oct26},
\begin{align}
\ev{[A_R,B_R^\dagger]}_{\Psi^{(s)}}
&=i\ev{A_R(\Delta A_R^\dagger- A_R^\dagger \Delta )-(\Delta A_R^\dagger-  A_R^\dagger \Delta )A_R}_{\Psi^{(s)}}\notag\\
&=i\ev{A_R A_R^\dagger+  A_R^\dagger A_R}_{\Psi^{(s)}}
=2i\kappa.
\end{align}
Because in the IR $A_R$ and $A_L^\dagger$ are identified, they do not have separate canonical conjugate operators. Namely, $A_R\cong A_L^\dagger$ and $B_R^\dagger\cong B_L$ form canonical pairs, where ``$\cong$'' means identification in the IR large-$N$ theory.

One may still wonder if one can find operators canonically conjugate to
$A_R$ and $A_L^\dagger$ separately, because they are really different
operators in the UV\@.  However, we can show that this is not possible on the state $\ket{\Psi^{(s)}}$ as follows.  First, recall that $A_R$ and $A_L^\dagger$
are the same on $\ket{\Psi^{(s)}}$ as in \eqref{muxi14Oct26}. So, if we define
\begin{align}
 \delta\equiv A_R-A_L^\dagger,
\end{align}
it satisfies
\begin{align}
 \delta\ket{\Psi^{(s)}}= \delta^\dagger\ket{\Psi^{(s)}}=0.
\end{align}
This means that, for \emph{any} operator $\cO$ (which can involve both left and right),
\begin{align}
 \ev{\Psi^{(s)}|[\delta,\cO]|\Psi^{(s)}}=0.
\end{align}
In other words,
\begin{align}
 \ev{\Psi^{(s)}|[A_R,\cO]|\Psi^{(s)}}=
 \ev{\Psi^{(s)}|[A_L^\dagger,\cO]|\Psi^{(s)}}.
\end{align}
Therefore, it is not possible to find an operator $\cO$ which is
conjugate only to $A_R$ (and not $A_L^\dagger$), satisfying
\begin{align}
 \ev{\Psi^{(s)}|[A_R,\cO]|\Psi^{(s)}}\neq 0,\qquad
 \ev{\Psi^{(s)}|[A_L^\dagger,\cO]|\Psi^{(s)}}=0,
\end{align}
proving the claim.

This situation is in contrast to that of the Rindler modes $a_R,a_L^\dagger$ on the Minkowski vacuum $\ket{M}$, which satisfies
\begin{align}
 a_R\ket{M}= \gamma  a_L^\dagger\ket{M},\qquad
 a_R^\dagger\ket{M}= \gamma^{-1}a_L\ket{M},\qquad 0<\gamma<1.
\end{align}
In this case, the operator
\begin{align}
\delta\equiv a_R-\gamma a_L^\dagger 
\end{align}
satisfies
\begin{align}
 \delta \ket{M}=0,\qquad 
 \delta^\dagger \ket{M}
 =(\gamma^{-1}-\gamma)a_L\ket{M}\neq 0.
\end{align}
Therefore, we generally have
\begin{align}
 \ev{M|[\delta,\cO]|M}
 =(\gamma^{-1}-\gamma)\ev{M|a_L^\dagger\cO|M}
 \neq 0,
\end{align}
and therefore
\begin{align}
 \ev{M|[a_R,\cO]|M}\neq 
 \gamma \ev{M|[a_L^\dagger,\cO]|M}.
\end{align}
So, in this case, it is possible to find operators conjugate to $a_R$ and
$a_L^\dagger$ separately (which are $a_R^\dagger$ and
$a_L$, respectively).

Note that, because of \eqref{itu7Oct26}, we have
\begin{align}
 (A_R+iB_R)\ket{\Psi^{(s)}}=(A_L+iB_L
 )\ket{\Psi^{(s)}}=0.
 \label{eq:fluct_vac}
\end{align}
In this sense, we can say that $\ket{\Psi^{(s)}}$ is the ``fluctuation vacuum'' with respect to the  IR operators $A,B$, just as the harmonic-oscillator vacuum $\ket{0}$ is annihilated by $a=x+ip$.

\subsubsection*{Mirror universes}

The above means that, although we have left and right universes, the
effective IR theory on $\ket{\Psi^{(s)}}$ has no such thing.  Namely, we only
have one complex operator
\begin{align}
 A\equiv A_R\cong A_L^\dagger,
 \qquad
 A^\dagger \equiv A_R^\dagger \cong A_L,
\label{eoqq8Oct26}
\end{align}
and its conjugate
\begin{align}
 B\equiv B_R\cong B_L^\dagger,
 \qquad
 B^\dagger\equiv B_R^\dagger \cong B_L.
\label{eoqs8Oct26}
\end{align}
The correlation functions \eqref{<ALAL>_on_superstar}, \eqref{<ALAR>_on_superstar}, the expectation values
\eqref{whb7Oct26}, \eqref{wiw7Oct26}, and the canonical commutation relations \eqref{wji7Oct26} are all consistent with that. When the grayness field of the disk is dynamically changed by adding the $B$ operator to the system's Hamiltonian, the grayness fields of the disks in the left and right universes change the same way at the same time. The relations \eqref{eoqq8Oct26} and \eqref{eoqs8Oct26} are not UV identities but are valid only in the IR theory of fluctuations around the state $\ket{\Psi^{(s)}}$.

What is the algebra type of the IR physics?
We have only one field theory, common to both left and right.  The
correlators \eqref{<ALAL>_on_superstar}, \eqref{<ALAR>_on_superstar} and the canonical commutators \eqref{wji7Oct26} are both $O(1)$. So, we can safely take the large $N$ limit.  By taking the orthonormal basis \eqref{eq:ff_orthogonality}, the $A$ and $B$ operators can be thought of as a countably infinite collection of fluctuation modes, in the vacuum state
\eqref{eq:fluct_vac}; note that this is a pure state and not a mixed state.
Therefore, the algebra of the $A,B$ operators is a Type I$_\infty$ algebra.\footnote{In the vacuum Fock representation, the von Neumann algebra generated by the creation and annihilation operators of infinitely many oscillators is a Type~$\mathrm{I}_\infty$ algebra (see Secs.~2.2 and 2.3 of~\cite{Witten:2021jzq}).} 

What is the relation to the algebraic ER=EPR conjecture?
Because the conjecture concerns the one-sided algebra of two potentially connected universes, it has nothing to say about this left-right algebra being Type I$_\infty$.  However, if we regard the combined left-right field theory as one side, which can potentially be connected to yet another universe, we can say that there is no classical wormhole connecting to it.

\section{Fine probes: bosonic realization} 
\label{sec:fineBoson}

In the previous section, we found that the one-sided algebra of IR observables probing the gray disk is of Type I, suggesting that there is no classical wormhole connecting left and right universes, as far as those coarse probes are concerned.  However, microscopically the system is in a TFD state with a large amount of entanglement between the two universes.  Therefore, a natural next step is to consider an observer who does have access to microscopic information, such as the individual column-height index (or  boson index) $j$, and investigate the corresponding algebras---fine-probe algebras---and look into the relation between their algebra type and connectivity of spacetime.

In the bosonic description, the microscopic theory contains $N$ bosonic modes and the associated operator algebra of $D_j,D_j^\dagger$ introduced in \eqref{eq:D_algebra_def}.  Studying the $N\to\infty$ limit of this UV algebra in full generality is technically challenging.  Here let us instead focus on a subsector of the full UV algebra which is under good algebraic control---the ``local'' algebra of excitations near the disk edge, with $O(1)$ energies.  In terms of the bosonic operators $D_j,D_j^\dagger$, this corresponds to the subsector in which excitations are restricted to bosons with $j=O(1)$, and the total number of excitations is $O(1)$.
Our goal here is to classify the type of this algebra.

\subsection{Construction of a bosonic fine-probe algebra}
\label{ConstructionOfBosons}

Recall that the bosonic representation of the superstar TFD state associated with the canonical ensemble is
given in \eqref{TFD_superstar_Psi_boson_2}:
\begin{align}
 \ket{\Psi^{(s)}(\zeta)}
 &=
 \bigotimes_{j=1}^N \left(
 \sqrt{1-\zeta}\sum_{c_j=0}^{\infty} \zeta^{c_j/2}
 \ket{c_j,c_j}_j
 \right)
 \equiv\bigotimes_{j=1}^N\ket{\Psi^{(s)}_{j}(\zeta)},
 \label{TFD_superstar_Psi_boson_3}
\end{align}
where we initially assume $0<\zeta<1$.
We are interested in the operator algebras acting on either the left or right subsystem in this TFD state.  Here we will consider the algebra of $D_j,D_j^\dagger$ on the right subsystem in the $N\to\infty$ limit (we will explain later how to consider the corresponding algebra acting on the left subsystem).  We restrict to finite products of operators with the column-height index $j=O(1)$, so that the total energy is $O(1)$.

Because the radius of the gray disk in Fig.~\ref{fig:LLM_diag}, i.e., the region where $0<u(x_1,x_2)<1$ in the LLM plane, is given by $r_s^2=(1+\lambda)R_{\rm AdS}^4=2\hbar (N+N_C)$, an operation involving the column-height index $j$ affects the outermost $j$ fermions, whose radial region is characterized by the radius $r^2_{N+N_C-j}$, satisfying (see \eqref{eq:r_vs_n})
\begin{align}
    r_{N+N_C-j}^2\sim 2\hbar(N+N_C-j),\qquad
    {r_s^2-r_{N+N_C-j}^2\over r_s^2}\sim {j\over N+N_C}.\label{eq:radius}
\end{align}
Thus, in the large $N$ limit with fixed $j$ and $\lambda=N_C/N$, the relative radial distance \eqref{eq:radius} from the edge vanishes, and the operator algebra of $D_j,D_j^\dagger$ with $j=O(1)$ is associated with the region very near the edge of the gray disk.  In this sense, this is a ``local'' algebra.

\bigskip
Let us proceed with the construction of this edge algebra.\footnote{The construction here follows \cite[Sec.~V.4]{Connes1994}. Although the construction there is formulated for finite-dimensional matrix algebras, it extends to our bosonic case~\cite[Cor.~2.9]{Araki1968}} Assuming first that $N$ is large but finite, let $\mathcal{H}_{j,R}$ denote the Hilbert space of the $j$th boson on the right. We introduce the corresponding algebra $\Xi_{j,R}=\mathcal{B}(\mathcal{H}_{j,R})$, where
$\mathcal{B}(\mathcal{H}_{j,R})$ includes all bounded operators on $\mathcal{H}_{j,R}$. Roughly, this is generated by the $D_j^\dagger$ and $D_j$ operators.\footnote{Strictly speaking, $D_j^\dagger$ and $D_j$ are unbounded operators and therefore cannot be elements of von Neumann algebras. One must instead use bounded operators constructed out of these operators.
Physically, this can be understood as follows.  In $\cH_{j,R}$, the operators $D_j,D_j^\dagger$ can be regarded as $\infty\times\infty$ matrices, but considering their $(m,n)$th entries with arbitrarily large $m,n$ is not physically meaningful, because we cannot excite one boson arbitrarily many times.  Therefore, physically, one must truncate them to $L\times L$ matrices with some large cutoff $L$ (or,  impose an appropriate falloff condition on the matrix elements as $m,n\to\infty$). Considering only bounded operators can be regarded as a mathematical implementation of this process.
}
For a finite number $K$ ($<N$), we define
\begin{align}
    \cC_{K,R}
    \equiv\Xi_{1,R}\otimes\Xi_{2,R}\otimes\cdots
    \otimes\Xi_{K,R}.
    \label{generalop}
\end{align}
Consider an element of $\cC_{K,R}$ of the form
\begin{align}
    a= a_1\otimes a_2\otimes\cdots\otimes a_K,
    \qquad a_j\in\Xi_{j,R},\qquad j=1,\dots,K
    \label{a=a1x...xam}
\end{align}
and define the state on the $j$th algebra by
\begin{align}
    \omega_{j,R}(a_j)
    \equiv\operatorname{Tr}(\rho_{j,R}\, a_j),
\end{align}
where $\rho_{j,R}$ is the reduced density matrix for the $j$th boson of the right subsystem. Explicitly, in our bosonic TFD
case, it is
\begin{align}
    \rho_{j,R}
    =(1-\zeta)\sum_{c_j=0}^{\infty}
    \zeta^{c_j}\ket{c_j}\bra{c_j}
    .
\end{align}
The state on $\cC_{K,R}$ is then given by the product state
\begin{align}
    \Omega_{K,R}(a)
    \equiv\prod_{j=1}^{K}\omega_{j,R}(a_j)
    =\prod_{j=1}^{K}
    \operatorname{Tr}(\rho_{j,R}\, a_j).
    \label{StateofUHF}
\end{align}
For general elements of $\cC_{K,R}$, this formula extends uniquely
by linearity and normality.

Since all eigenvalues of $\rho_{j,R}$ are strictly positive for $0<\zeta<1$,  
 the corresponding state for the $j$th boson is faithful and normal.\footnote{
 For more detail, see \cite[Thm 2.4.21]{Bratteli:1979tw} for normality and \cite[Ex 2.5.5]{Bratteli:1979tw} for faithfulness of this state. In addition, the product state $\Omega_{K,R}$ is also normal and faithful. See \cite[III.3.1.2]{Blackadar:2006zz}.  } 
The modular automorphism group acts on an element of the form \eqref{a=a1x...xam} as
\begin{align}
    \sigma_t^{\Omega_{K,R}}
    (a_1\otimes\cdots\otimes a_K)
    &=
    \sigma_t^{\omega_{1,R}}(a_1)
    \otimes\cdots\otimes
    \sigma_t^{\omega_{K,R}}(a_K)
    \notag\\
    &=
    \bigotimes_{j=1}^{K}
    \left(
    \rho_{j,R}^{-it}\, a_j\, \rho_{j,R}^{it}
    \right).\label{ModularFlow}
\end{align}
In the TFD representation, with the tensor factors ordered as
$\mathcal{H}_{j,L}\otimes\mathcal{H}_{j,R}$, the modular operator
associated with the $j$th boson is, as we mentioned in footnote \ref{ModularOperator},
\begin{align}
    \Delta_j
    =\rho_{j,L}^{-1}\otimes \rho_{j,R},
\end{align}
so that the modular automorphism group is implemented by
$\Delta_j^{it}$.

So far, the discussion has assumed fixed finite values of $N$ and $K$ with $K<N$. From this point on, we  consider the $N\rightarrow\infty$ limit; more precisely, we first fix $K$ and take the large $N$ limit, and then consider the limit of $K$ approaching infinity.   After having taken the $N\to\infty$ limit, we have an increasing sequence of algebras,
\begin{align}
    \cC_{1,R}\subset \cC_{2,R}\subset\cdots,
\end{align}
with the natural unital embeddings
\begin{align}
    \cC_{K,R}&\hookrightarrow \cC_{K+1,R},
    \notag\\
    x&\mapsto x\otimes\hat{1}.
    \label{eq:A_K,R_embed}
\end{align}
Roughly,\footnote{%
More precisely, $\cC_R$ is obtained by taking the union of $\cC_{K,R}$ and its norm closure, and can be 
written as
\begin{align}
    \cC_R
    =\overline{\bigcup_{K=1}^{\infty}\cC_{K,R}}
    ^{\|\cdot\|}.\label{eq:def_C_R}
\end{align} \label{fryu10Nov26}} by taking the union of these algebras, we obtain the operator algebra for the infinite system, which we denote by $\cC_R$.  The algebra $\cC_R$ is generated by elementary tensors of the form $a_1\otimes a_2\otimes\cdots$, where $a_n\in\Xi_{n,R}$ and all but finitely many of the $a_n$ are equal to the identity operator $\hat{1}$.
Note that this construction implies that the operators in $\cC_R$ are norm limits of operators acting nontrivially on only finitely many columns. 
In addition, the product states $\Omega_{K,R}$ are compatible
with the embeddings \eqref{eq:A_K,R_embed} and therefore define a unique state~$\Omega_R$ on~$\cC_R$.

The algebra $\cC_R$ thus constructed is a C$^*$-algebra but not yet a von Neumann algebra. To obtain a von Neumann algebra, we first apply the GNS construction
to $\cC_R$ with the state $\Omega_R$, obtaining a GNS triple $(\mathcal{H},\pi,\ket{\Psi})$.
We then obtain a von Neumann algebra $\mathcal{A}_R$ by taking
the double commutant of $\pi(\cC_R)$:\footnote{
For more details, see, for example, \cite[Sec.~II.D]{Liu:2025krl}.
By the von Neumann double commutant theorem, this is the same as
taking the weak (or strong) operator closure of $\pi(\cC_R)$.
}
\begin{align}
    \mathcal{A}_R=\pi(\cC_R)''.
    \label{dblcommutant}
\end{align}
 In this construction, the resulting  $\ket{\Psi}$ describing the total system is the infinite tensor product of the individual cyclic separating vectors $\ket{\Psi_j}$: \footnote{We can say that we get a cyclic separating vector from the GNS construction of a faithful normal state. The proof of this statement is in \cite[Prop. 2.5.6, (3) and (4)]{Bratteli:1979tw}. }
\begin{align}
    \ket{\Psi}=\bigotimes_{j=1}^{\infty}\ket{\Psi_j}.
\end{align}
In our case,  we simply have $\ket{\Psi_j}=\ket{\Psi^{(s)}_j(\zeta)}$. Algebras arising in this way are called ITPFI factors (infinite tensor products of finite Type I factors).\footnote{Although each factor (the algebra of the $j$th boson) in the present bosonic construction is of Type $\mathrm{I}_\infty$ which is infinite dimensional, the resulting factor is still called an ITPFI factor; see \cite[Cor.~2.9]{Araki1968}.
}
They have been extensively studied
(see, for example,
\cite{Tomatsubook2024, Takesakibook1983, Connes1994, Araki1968,
Ueda-sanNote, Liu:2025krl}).
Recent applications to physics can be found in
\cite{Banerjee:2023eew, DiGiulio:2025xza, Miyata:2026myd}.

We have so far constructed the algebra acting on the right subsystem. The corresponding algebra acting on the left subsystem can be constructed analogously. In the GNS representation of the TFD state, the modular conjugation exchanges the left and right algebras, and they are mutual commutants:
\begin{align}
    \mathcal{A}_L=\mathcal{A}_R'.
\end{align}

To summarize, the UV ``edge algebra'', namely the algebra of $D_j,D_j^\dagger$ with $j=O(1)$, can mathematically be formulated as an ITPFI factor.
We now proceed to determine the von Neumann type of this ITPFI factor.

\subsection{Type classification}

To determine the algebraic type of the ITPFI factor $\mathcal A_R$ constructed above, we examine its Connes $S$-invariant \eqref{SpectralInv}, adapting the discussion of the fermionic case in~\cite[Chap.~XVIII, Sec.~1]{TakesakibookIII} to the bosonic setting.

Recall that the reduced density matrix at each boson index is given by
\begin{align}
    \rho_{j,R}
    &=(1-\zeta)\sum_{c_j=0}^{\infty}
    \zeta^{c_j}|c_j\rangle\langle c_j|
    \notag\\
    &=(1-\zeta)
    \begin{pmatrix}
        1 & 0 & 0 & \cdots\\
        0 & \zeta & 0 & \cdots\\
        0 & 0 & \zeta^2 & \cdots\\
        \vdots & \vdots & \vdots & \ddots
    \end{pmatrix},
    \qquad
    \zeta=\frac{\lambda}{1+\lambda}\in(0,1), 
\end{align}
and  that $\Omega_{R}$ denotes the   product state on $\mathcal A_R$ in the previous subsection, associated with the local density matrices.

As reviewed in Sec.~\ref{sec:vnprelim}, the $S$-invariant distinguishes the different subclasses of Type III factors. Its direct computation is generally nontrivial, since it involves the intersection of the spectra of modular operators associated with all cyclic separating vectors. However, for the present ITPFI factor, whose local density matrices are all identical, we can determine%
\footnote{The proof of this statement follows the argument in Lemma~1.2 and Theorem~1.1 of~\cite[Chap.~XVIII, Sec.~1]{TakesakibookIII}.
The argument in Lemma 1.2 also applies to the present case with infinite-dimensional local Hilbert spaces, since permutations of finitely many sites can be implemented by unitaries belonging to the corresponding von Neumann algebra. Since the product state is invariant under these permutations, the implementing unitaries belong to its centralizer, and the same cluster-property argument in the proof of Lemma 1.2 applies. We thank Yoshimichi Ueda for pointing this out.
} the $S$-invariant solely from  the spectrum of the modular operator $\Delta_{\Omega_R}$ associated with $\Omega_{R}$, and we do not need to take intersections over all cyclic and separating vectors. Namely,
\begin{align}
\mathcal{S}(\mathcal A_R) = \text{Spec}(\Delta_{\Omega_R}). 
\end{align}
In our case, we can read off the eigenvalues of each local density matrix as $p_{j,n}=(1-\zeta)\zeta^n$, so their ratios are
\begin{align}
    \frac{p_{j,n}}{p_{j,m}}
    =\zeta^{n-m},
    \qquad n,m=0,1,2,\ldots.
\end{align}
Since the modular operator for the product state has spectral values given by products of these local eigenvalue ratios, its spectrum is 
\begin{align}
    \operatorname{Spec}(\Delta_{\Omega_{R}})
    =\{0\}\cup\{\zeta^k\mid k\in\mathbb Z\}, 
\end{align}
where $0$ is included since $\zeta^k\to 0$ as $k\to\infty$ for $0<\zeta<1$. Therefore,
\begin{align}
    \mathcal{S}(\mathcal A_R)
    =\operatorname{Spec}(\Delta_{\Omega_{R}})
    =\{0\}\cup\{\zeta^k\mid k\in\mathbb Z\},
\end{align}
and from Table~\ref{TableofTypeIII}, we conclude that
\begin{align}
    \mathcal A_R
    \text{ is of Type III}_{\zeta},
    \qquad \text{where }\zeta=\frac{\lambda}{1+\lambda}.
    \label{III_zeta}
\end{align}
The identical construction for the left subsystem
yields the same classification for $\mathcal A_L$.

According to the algebraic ER=EPR conjecture summarized in Table \ref{tab:AlgebrasConnectedness}, von Neumann algebras that are not of Type~I are associated to quantum wormholes.  So, the fact that we found a Type III$_\zeta$ algebra is evidence that the TFD superstar represents a quantum wormhole.  We will discuss this interpretation further in Sec.~\ref{sec:discussion}.

\subsection{Special cases}

In the above we assumed that $0<\zeta<1$.  Here we discuss the values outside that range.

\paragraph{$\bullet$ $\zeta=0$ ($\lambda=0$):}
In this case, the reduced density matrix for each boson becomes a pure state,
\begin{align}
\rho_{j,R}=|0\rangle\langle 0|.
\end{align}
Consequently, the TFD state reduces to an infinite tensor product of local vacuum states, with no entanglement between the left and right subsystems. The corresponding von Neumann algebra in the GNS representation is therefore a Type I$_\infty$ factor, rather than Type~III\@.  The bulk geometry is a disconnected pair of empty AdS$_5\times S^5$ spacetimes.

\paragraph{$\bullet$ $\zeta=1$ ($\lambda=\infty$):}
This limit does not correspond to a physical spacetime.  Indeed, the partition function for the $j$th boson diverges in this limit:
\begin{align}
Z_j=\frac{1}{1-\zeta}=1+\lambda\to\infty.
\end{align}
Consequently, the reduced density matrix cannot be normalized, the TFD state is ill-defined, and the present von Neumann algebra construction is not applicable.

\section{Fine probes: fermionic realization}
\label{sec:fineFermion}

In the previous section, we constructed the fine-probe algebra of bosonic operators $D_j,D_j^\dagger$ describing excitations near the disk edge, and obtained a von Neumann algebra of Type III$_{\zeta}$ with $\zeta={\lambda\over1+\lambda}$ (see \eqref{eq:lambda_and_zeta} for the relation between the bosonic parameter $\zeta$ and the fermionic parameter $\lambda$).
This suggests that fine probes can detect quantum connectivity of the left and right universes.  In this section, we will study a different kind of edge algebra generated by the fermionic operators $C_n,C_n^\dagger$ defined in~\eqref{eq:C_algebra_def}. We construct the local algebra of these fine-grained probes and study its algebraic structure.  Similarly to the bosonic case, we will again find a Type III$_\lambda$ von Neumann algebra.

\subsection{Construction of a fermionic fine-probe algebra}

We consider the fine-grained probe algebra of near-edge excitations in
terms of fermionic operators. As in the bosonic case,
we begin with the fermionic representation of the
TFD state given in (\ref{TFD_superstar_state_2}):
\begin{align}
\Ket{\Psi^{\mathrm{(s)}}(\lambda)}
&=
\left[
\bigotimes_{n=0}^{N+N_C-1}
\frac{1}{\sqrt{1+\lambda}}
\left(
\sqrt{\lambda}\,\ket{00}_n+\ket{11}_n
\right)
\right]
\otimes
\left[\bigotimes_{n=N+N_C}^{\infty}
\ket{00}_n \right].\label{TFD_superstar_rhoPsi_3}
\end{align}
and focus on fermionic degrees of freedom near the edge of the gray disk of Fig.~\ref{fig:LLM_diag}, including levels on both sides of the edge.
Specifically, we consider operators acting nontrivially on
only finitely many levels whose separations from the edge,
measured in level number, remain finite in the large-$N$ limit.

The physical situations described by the bosonic and fermionic algebras are different.  First, the bosonic TFD state \eqref{TFD_superstar_Psi_boson_3} is associated with the canonical ensemble with fixed $N$, while the fermionic TFD state \eqref{TFD_superstar_rhoPsi_3} is associated with the grand canonical ensemble with variable $N$.  These two ensembles are expected to describe the same physics in the large~$N$ limit, but the algebras of fluctuations that they admit as $N\to\infty$ can be different depending on how the limit is defined. In particular, the bosonic operators $D_j,D_j^\dagger$ in the edge algebra studied in the previous section keep $N$, the number of fermions, fixed, while the fermionic operators in the algebra generated by $C_n,C_n^\dagger$ can increase or decrease the number of fermions.  We can define a well-defined fermionic von Neumann algebra for which these changes in $N$ are negligible by considering
products containing only $O(1)$ insertions of the operators $C_n$ and $C_n^\dagger$ as $N\to\infty$.\footnote{More precisely we consider products containing at most $K \sim O(1)$ insertions of $C_n,C_n^\dagger$, and take $N\to\infty$ first and then $K\to\infty$.}   

Even with this way of defining the fermionic algebra, if we allow the  operators $C_n,C_n^\dagger$ to act near the edge of the gray disc of the superstar in Fig.~\ref{fig:LLM_diag}, namely with energy level $n \sim O(N)$,  even a change of $N$ by $1$ will change the energy of the state by $O(N)$ because the fermions at the edge of the superstar state have energies of this order of magnitude.  In previous sections we required that the algebra of fluctuations change the energy at most by $O(1)$ so that fluctuation energy in the large $N$ limit is well-defined.  Note though that the background state as a whole has energy of $O(N^2)$ which diverges as $N\to\infty$.  So in any case, to define the energy of the fluctuation in the limit we have to work with a renormalized  Hamiltonian operator obtained by subtracting the energy of the background.  The chemical potential terms in the grand canonical ensemble effectively further renormalize the Hamiltonian so that the increase/decrease of energy due to the addition/removal of a fermion is measured relative to the maximum filled level before the fluctuation.  Thus, if consider operators $C_n, C_n^\dagger$ with $n \sim N+N_c-1 + O(1)$ the effective energy in the grand canonical ensemble is $O(1)$. This renormalization implemented by the ensemble allows us to define the fermionic edge algebra. Despite all of these differences, we will see that the fermionic algebra type is intermediate Type III---the same as in the bosonic case studied in the previous section.

\bigskip
In order to describe the fermions near the edge, for convenience, we relabel the fermion levels so that the levels immediately inside and outside the edge come first, followed by successively more distant levels on either side.
Recall that $M=N+N_C$, so that the edge of the gray disk lies between $n=M-1$ and $n=M$.
We interleave the interior levels, counted inward from $n=M-1$, with the exterior vacuum levels, counted outward from $n=M$. So after this relabeling the levels above the edge have odd labels and the levels below the edge are assigned even labels.
Thus, the original levels are placed in the order
\begin{align}
M-1,\ M,\ M-2,\ M+1,\ \ldots,\ 0,\ 2M-1,\ 2M,\ 2M+1,\ \ldots,\label{hqrw3Nov26}
\end{align}
and assigned new labels $\tilde n=0,1,2,\ldots$ in this order. Explicitly, the relabeling is given by the permutation
\begin{align}
\tilde n=
\begin{cases}
2(M-1-n), & 0\leq n<M,\\
2(n-M)+1, & M\leq n<2M,\\
n, & n\geq 2M.
\end{cases}
\end{align}
For each $l=0,\ldots,M-1$, the two sites
$\tilde n=2l$ and $\tilde n=2l+1$ therefore correspond to
\begin{align}
n_{\mathrm{in}}(l)=M-1-l,
\qquad
n_{\mathrm{out}}(l)=M+l,
\end{align}
respectively.
In particular, $\tilde n=0$ labels the outermost interior level, while $\tilde n=1$ labels the first exterior vacuum level.
The remaining vacuum levels, $n\geq 2M$, retain their original labels.
Holding $\tilde n$ fixed as $N\to\infty$ thus keeps the corresponding level a fixed number of steps from the edge. With the tensor factors ordered according to these new labels, the TFD state can be rewritten as
\begin{align}
\Ket{\Psi^{\mathrm{(s)}}(\lambda)}_{\text{(relabeled)}}
&=
\left[
\bigotimes_{l=0}^{M-1}
\left(
\Ket{\psi_\lambda}_{2l}
\otimes
\Ket{\Omega}_{2l+1}
\right)
\right]
\otimes
\bigotimes_{l=2M}^{\infty}
\Ket{\Omega}_l ,
\label{TFD_superstar_Psi_relabeled}
\end{align}
where
\begin{align}
\Ket{\psi_\lambda}_{2l}
&=
\frac{1}{\sqrt{1+\lambda}}
\left(
\sqrt{\lambda}\ket{00}_{2l}
+\ket{11}_{2l}
\right),\\
\Ket{\Omega}_{2l+1}
&=\ket{00}_{2l+1}.\label{Fermionic_TFD_vacuum}
\end{align}
In the large-$N$ limit at fixed $\lambda$, with the relabeled site indices held fixed, the state converges locally to an alternating product state,
with entangled pairs on even-numbered sites and vacuum states on odd-numbered sites.

We now construct the  operator algebra acting on
the right subsystem after the large-$N$ limit. The construction follows the same steps as in the bosonic case; the differences lie in the local algebras and their states. In what follows, we write $n$
for the relabeled site index $\tilde n$ and work with the
limiting alternating product state described above.

Each fermion level has a two-dimensional Hilbert space
$\mathcal{H}_{n,R}\cong\mathbb{C}^2$, with local matrix algebra
\begin{align}
\Xi_{n,R}
=
\mathcal{B}(\mathcal{H}_{n,R})
\cong M_2(\mathbb{C}).
\end{align}
As in the bosonic case, the algebra is constructed from the tensor products of the individual subsystem algebras. For the first $K+1$ relabeled sites, we define
the finite-site algebra \footnote{In what follows, we add the superscript (f) to the algebra and the state, in order to distinguish them from the case of bosons.}
\begin{align}
\cC_{K,R}^{\mathrm{(f)}}
=
\bigotimes_{n=0}^{K}\Xi_{n,R}.
\end{align}
The local states are defined by
\begin{align}
\omega_{n,R}(a_n)
=
\operatorname{Tr}(\rho_{n,R}\,a_n),
\qquad a_n\in\Xi_{n,R},
\end{align}
where
\begin{align}
\rho_{2l,R}
&=
\frac{1}{1+\lambda}
\left(
\lambda\ket{0}\bra{0}
+\ket{1}\bra{1}
\right),\label{fermionic_local_rho1}\\
\rho_{2l+1,R}
&=
\ket{0}\bra{0}
\label{fermionic_local_rho2}
\end{align}
with $l=0,1,2,\dots$.
Thus, the even-numbered sites carry mixed states,
while the odd-numbered sites carry pure vacuum states.

Following the bosonic construction we can define  
the state $\Omega_R^{\mathrm{(f)}}$ 
and the algebra $\cC_R^{\rm (f)}$ of the infinite system from the product of the local state $\omega_{n,R}$ and the union of $\cC_{K,R}^{\rm (f)}$, respectively.\footnote{Mathematically, as in \eqref{eq:def_C_R}, $\cC_R^{\rm (f)}$ is given by the norm closure
\begin{align}
\cC_R^{\mathrm{(f)}}
&=
\overline{
\bigcup_{K=0}^{\infty}
\cC_{K,R}^{\mathrm{(f)}}
}^{\|\cdot\|}.
\end{align}
}
The corresponding GNS representation defines
\begin{align}
\mathcal{A}_R^{\mathrm{(f)}}
&=
\pi_{\Omega_R^{\mathrm{(f)}}}
\bigl(\cC_R^{\mathrm{(f)}}\bigr)''.
\label{fermionic_vn_algebra}
\end{align}

The resulting algebra is an ITPFI factor in which each Type I factor is finite-dimensional.
Its product-state structure allows us to group the
even- and odd-numbered tensor factors separately,
yielding\footnote{Here, $\bar{\otimes}$ means the tensor product of von Neumann algebras.}$^,$\footnote{The tensor product of C$^*$-algebras gives a tensor product of von Neumann algebras by the GNS construction, if the state is also a product.}
\begin{align}
\mathcal{A}_R^{\mathrm{(f)}}
\cong
\mathcal{P}_{\lambda}
\,\bar{\otimes}\,
\mathcal{M}_{\mathrm{vac}},
\label{fermionic_factor_decomposition}
\end{align}
where $\mathcal{P}_{\lambda}$ is the infinite tensor
product factor associated with the even-numbered sites, and $\mathcal{M}_{\mathrm{vac}}$ is the factor associated with the odd-numbered vacuum sites.

We now determine the von Neumann algebra type of
$\mathcal{A}_R^{\mathrm{(f)}}$.

\subsection{Type classification}
\label{ss:type_class'n}

We first consider $0<\lambda<1$.
Recall the factorization in (\ref{fermionic_factor_decomposition}).
The factor $\mathcal{M}_{\mathrm{vac}}$ is generated by the odd-numbered sites in the product of their pure vacuum states, which is (\ref{fermionic_local_rho2}). This product state is pure, and therefore induces a Type I$_\infty$ factor.~\footnote{This statement follows from \cite[Sec.~3, p.~575]{Powers1967Bull}. }
Physically, this sector carries no entanglement between the left and right subsystems, as written in \eqref{Fermionic_TFD_vacuum}.

On the even-numbered sites, the local density matrices
in (\ref{fermionic_local_rho1}) have the common
eigenvalues
\begin{align}
    \frac{1}{1+\lambda},
    \qquad
    \frac{\lambda}{1+\lambda}.
\end{align}
Together with the product-state structure, this
identifies $\mathcal{P}_{\lambda}$ as a so-called Powers factor with parameter $\lambda$ \cite{Powers1967Bull,e6fa6f9b-294a-31d5-858f-a3dacf9aee24}.
To determine the Connes $S$-invariant of $\mathcal{P}_{\lambda}$, just as in the bosonic case, we do not need to take intersections over all cyclic and separating vectors, and it is solely determined by the modular operator associated with the even-numbered density matrices \eqref{fermionic_local_rho1}.  The spectral invariant in this case is known to be
(see, e.g., \cite[Chap.~XVIII, Sec.~1, Thm 1.1]{TakesakibookIII})
\begin{align}
    S(\mathcal{P}_{\lambda})
    =
    \{0\}\cup\{\lambda^k\mid k\in\mathbb{Z}\},
\end{align}
so $\mathcal{P}_{\lambda}$ is of Type III$_\lambda$.

Tensoring with a Type I$_\infty$ factor preserves
the Type III$_\lambda$ classification.~\footnote{For example, see \cite[III.1.5.6(ii)]{Blackadar:2006zz}.} 
Therefore,
\begin{align}
    \mathcal{A}_R^{\mathrm{(f)}}
    \cong
    \mathcal{P}_{\lambda}
    \,\bar{\otimes}\,
    \mathcal{M}_{\mathrm{vac}}
    \qquad
    \text{is of Type III}_{\lambda},
    \qquad 0<\lambda<1.
\end{align}
So, the fermionic algebra is again  intermediate Type III as for the bosonic case, although the type index is different.\footnote{In the bosonic case, we had Type III$_{\lambda\over 1+\lambda}$; see \eqref{III_zeta}.}  Again, this is consistent with the expectation that the superstar represents a quantum wormhole.

\subsection{Special cases}
\label{ss:special_lambda}

In the analysis above, we assumed that the value of $\lambda=N_C/N$ is in the range $0<\lambda<1$.  In the following, we discuss values of $\lambda$ outside this range.

\paragraph{$\bullet$ $1<\lambda<\infty\ (N<N_C)$:}
This case is exactly the same as the case with $0<\lambda<1$, if we set $\lambda\to 1/\lambda$.  So, in this case, the von Neumann algebra associated with the superstar is of Type III$_{1/\lambda}$.

\paragraph{$\bullet$ $\lambda=1$ ($N=N_{C}$):}

In this case, the expression for the state is same as \eqref{TFD_superstar_Psi_relabeled}, and we see that each fermion level inside the gray disk becomes maximally entangled between
the left and right subsystems. Using the relabeling introduced above, the TFD state in the large-$N$ limit becomes the following, in terms of the relabeled state: 
\begin{align}
\Ket{\Psi^{\mathrm{(s)}}(1)}_{\text{(relabeled)}}
=
\bigotimes_{l=0}^{\infty}
\left[
\frac{1}{\sqrt{2}}
\left(
\ket{00}+\ket{11}
\right)_{2l}
\otimes
\ket{00}_{2l+1}
\right].
\end{align}
Here, the even-numbered sites are maximally
entangled, while the odd-numbered sites
remain in the vacuum state. The reduced density matrices of the right subsystem are therefore
\begin{align}
\rho_{2l,R}
&=\frac{1}{2}I_2=\dfrac{1}{2}\begin{pmatrix}
  1  &   0           \\
 0 &  1         \\
\end{pmatrix},\\
\rho_{2l+1,R}
&=\ket{0}\bra{0}.
\end{align}
The infinite tensor product of the
maximally mixed states on the even-numbered
sites generates the hyperfinite
Type II$_1$ factor $\mathcal R$.
On the other hand, the infinite tensor
product of the pure vacuum states on the
odd-numbered sites generates a Type I$_\infty$ factor. Consequently, the corresponding von Neumann algebra is isomorphic to
\begin{align}
\mathcal{A}_R^{\mathrm{(f)}}
&\cong
\mathcal R\ 
\bar\otimes\
\mathcal{B}(\mathcal H_{\mathrm{vac}}),
\end{align}
where $\mathcal H_{\mathrm{vac}}$ is
infinite-dimensional. Since the tensor product of a
Type II$_1$ factor and a Type~I$_\infty$
factor is of Type II$_\infty$, we conclude
that $\mathcal{A}_R^{\mathrm{(f)}}
\text{ is of Type II}_{\infty}.$  This again suggests that the corresponding bulk is a quantum wormhole.

\paragraph{$\bullet$ $\lambda=0\ (N_C=0)$:}

This situation has already been discussed in the bosonic case; the reduced density matrices of the left and right subsystems become pure and there is no entanglement between the two subsystems. The state (\ref{TFD_superstar_Psi_relabeled}) in this case is
\begin{align}
\Ket{\Psi^{\mathrm{(s)}}(0)}_{\text{(relabeled)}}
=
\bigotimes_{l=0}^{\infty}
\left(
\ket{11}_{2l}
\otimes
\ket{00}_{2l+1}
\right).
\label{lambda_zero}
\end{align}
As mentioned in the vacuum part of $\mathcal{A}_R^{\mathrm{(f)}}$, 
we conclude that the corresponding von Neumann algebra is of Type I$_\infty$.  The bulk geometry is a disconnected pair of AdS$_5\times S^5$.

\paragraph{$\bullet$ $\lambda\rightarrow\infty\ (N\ll N_C)$:}

In this limit, for each fermion level inside the gray disk, the reduced density matrix approaches
\begin{align}
\rho_{2l,R}
&=
\frac{1}{1+\lambda}
\left(
\lambda\ket{0}\bra{0}
+\ket{1}\bra{1}
\right)
\xrightarrow{\lambda\rightarrow\infty}
\ket{0}\bra{0}.
\end{align}
Thus, the state restricted to any fixed finite set of relabeled levels approaches the vacuum state. The corresponding limiting TFD state is formally given by
\begin{align}
\Ket{\Psi^{\mathrm{(s)}}(\infty)}
 \overset{?}{=}
\bigotimes_{n=0}^{\infty}\ket{00}_n.
\end{align}
This is again a product state without entanglement between the left and right subsystems. Since the vacuum state is pure, by the same discussion as in the $\lambda=0$ case,
we can see that the corresponding von Neumann algebra is of Type I$_\infty$.

However, this limiting state should not be directly identified with the original superstar ensemble.
Indeed, the number of fermion levels with nontrivial occupation probability
is $M=(1+\lambda)N$, and the expected total fermion number remains
\begin{align}
\langle \widehat N_{\mathrm F}\rangle
=
\frac{M}{1+\lambda}
=
N.
\end{align}
Therefore, although the occupation probability at each individual level vanishes, the total fermion number does not vanish. In this sense, the $\lambda\rightarrow\infty$ limit loses this global information, and we conclude that $\lambda=\infty$ represents a singular endpoint of the superstar ensemble rather than a regular superstar configuration with a finite, nonzero filling fraction.

\section{Discussion}
\label{sec:discussion}

In this paper, we studied  von Neumann algebras of half-BPS probe operators of the superstar background geometry and classified their algebraic types.  According to \cite{Leutheusser:2021qhd,Engelhardt:2023xer} these types are related to the connectivity of spacetime.  Here we summarize our findings and discuss their physical interpretation and implications.

\subsubsection*{Summary of probe algebras}

The superstar ensemble of the $SU(N)$ Super-Yang-Mills theory with 16 supercharges is characterized by a parameter $\lambda=N_C/N\in [0,\infty)$, where $N_C$ is related to the number of D3-branes or  R-charge in the dual asymptotically AdS$_5$ extremal black hole geometry.  We constructed a purification of this ensemble by considering a TFD-like  configuration, described in gravity as a state entangled between two asymptotically AdS$_5$ universes. Each additive component of this entangled state is associated to a nonsingular, topologically complex geometry, but almost all of these components lie close to a typical state with a coarse-grained description as an extremal black hole. As a result, a partial trace over either universe reproduces the superstar ensemble and the associated extremal black hole geometry in the other universe.

We first considered coarse BPS probes of these states.  In the gravity description seen from either of the two entangled universes, these probes correspond to long wavelength fluctuations of the (singular) boundary conditions at the extremal black hole horizon.  The probes were designed to be insensitive to Planck-scale features of the underlying microstates in the large $N$ limit. We found that the algebra of probes is a direct integral of Type I$_1$ factors, consistent with an interpretation that the superstar geometry is a black hole with a vanishing horizon at the classical level, with no connection between the left and right universes as seen by these coarse probes \cite{Leutheusser:2021qhd,Engelhardt:2023xer}.

Then we  studied fine-grained bosonic and fermionic probes,
that can access microscopic structure of the gravitational microstates near the edge of the singular region of the extremal black hole geometry.  Although both probes represented ``edge excitations'' of this kind, they are different in the kind of perturbations they describe (see Sec.~\ref{sec:fineBoson}  and Sec.~\ref{sec:fineFermion} and below). We found various algebraic types, summarized in Fig.~\ref{fig:superstar_type}.
For generic values of $\lambda$, the algebra was of the intermediate Type III, which is consistent with the expectation that the superstar is a quantum wormhole \cite{Leutheusser:2021qhd,Engelhardt:2023xer}. The bosonic and fermionic algebras that we studied had different relationships between the parameter $\lambda = N_C/N$ and the type index: the  bosonic probes had Type III$_{\lambda/(1+\lambda)}$ and the fermionic probes had Type III$_\lambda$.  At $\lambda=1$, the fermionic algebra gave a Type II$_\infty$ algebra, again, indicative of a quantum wormhole, although this is not a special point in the bosonic algebra. At $\lambda=0$, the corresponding state is pure and the associated von Neumann algebra is of Type I$_\infty$ in both descriptions. The gravity theory then describes a pair of disconnected AdS$_5\times S^5$ spacetimes.

\begin{figure}[tbhp]
\begin{quotation}
 \begin{center}
 \begin{minipage}{0.9\linewidth}

\hspace*{1cm}
 \begin{tikzpicture}
  \path[use as bounding box] (-5,-1.5) rectangle (7,1.5);
  \draw (-3,0)--(3,0);
  \draw[fill=black] (-3,0) circle (0.1);
  \draw[fill=white] (3,0) circle (0.1);
  \draw (-3,-0.3)node[below]{I$_{\infty}$};
  \draw (-3,+0.3)node[above]{0};
  \draw (+3,+0.3)node[above]{$\infty$};
  \draw (+3,0)node[right]{$\qquad \lambda$};
  \draw (0,-0.3)node[below]{III$_{\lambda\over 1+\lambda}$};
  \draw (-3.5,-0.3)node[below left]{Type : };
\end{tikzpicture}

(a) {\sl Types of the von Neumann algebra in the bosonic description}

 \vspace{0.5cm}

\hspace*{1cm}
\begin{tikzpicture}
  \path[use as bounding box] (-5,-1.5) rectangle (7,1.5);
  \draw (-3,0)--(3,0);
  \draw[fill=black] (-3,0) circle (0.1);
  \draw[fill=white] (3,0) circle (0.1);
  \draw (0,+0.2)--(0,-0.2);
  \draw (-3,-0.3)node[below]{I$_{\infty}$};
  \draw (0,-0.3)node[below]{II$_{\infty}$};
  \draw (-1.5,-0.3)node[below]{III$_{\lambda}$};
  \draw (+1.5,-0.3)node[below]{III$_{\frac{1}{\lambda}}$};
  \draw (-3,+0.3)node[above]{$0$};
  \draw (+3,+0.3)node[above]{$\infty$};
  \draw (0,+0.3)node[above]{$1$};
  \draw (+3,0)node[right]{$\qquad\lambda$};
  \draw (-3.5,-0.3)node[below left]{Type : };
\end{tikzpicture}

(b) {\sl Types of the von Neumann algebra in the fermionic description}
\vspace{0.5cm}
\end{minipage}

\caption{ \label{fig:superstar_type}
\sl Classification of the von Neumann algebra types associated with the superstar TFD state in the bosonic (a) and fermionic (b) descriptions as a function of the parameter $\lambda$. The open circles indicate the exclusion of the endpoint $\lambda=\infty$, as discussed in the main text.
}
\end{center}
   \end{quotation}
\end{figure}

\subsubsection*{Physical interpretation of the coarse-probe algebras}

In section \ref{sec:coarse}, we studied the algebra of the coarse probes, the fluctuation fields in the right universe $A^R_{f,l}$, corresponding to the boundary condition $u(x_1,x_2)$, and found a peculiar fact in the large $N$ limit: they all commute with one another. So, the algebra of the $A^R_{f,l}$ fields is like the algebra of $\hat{x}$ without $\hat{p}$ in quantum mechanics, leading to (a direct integral of) Type~I$_1$ algebras.

Furthermore, we found that there is no right-side field $\cO^R$ canonically conjugate to $A^R_{f,l}$ satisfying $[A^R_{f,l},\cO^R]\propto \hat{1}$. This is in tension with the result of \cite{Balasubramanian:2018yjq}, which constructed a field~$\pi$ conjugate to the grayness field $u$ in the IR theory, {\it i.e.}~supergravity, with the canonical Poisson bracket $\{u(x_1,x_2),\pi(x_1',x_2') \}_{\rm PB}=\delta(x_1-x_1')\delta(x_2-x_2')$.  More precisely, 
Ref.~\cite{Balasubramanian:2018yjq} constructed $\pi$ from UV data: they assumed simple ``mesoscopic''\footnote{With structure much larger than the Planck scale but much smaller than the scale that the IR observer can resolve.} LLM diagrams consisting of alternating black-and-white bands, and identified $u$ with the density fluctuation of the bands and $\pi$ with translational modes of the bands; see Figure 7 of Ref.~\cite{Balasubramanian:2018yjq}.  The latter requires UV information, because an IR observer cannot resolve the microscopic position of the individual bands. This is much as, when constructing the $A_{f,l}$ operator \eqref{mgns28Sep26}, the IR observer does not know the value of $j$ of the $D_{j+l}^\dagger D_j$ operator with $O(1)$ precision; see the discussion below \eqref{mgns28Sep26}.
One may then wonder if one can construct a conjugate operator $\cO^R$ in terms of the fundamental operators $D_j,D_j^\dagger$ incorporating UV data not accessible to the IR observer.  However, we showed a no-go theorem showing that this is not possible on the superstar TFD state.  We do not have a resolution of this puzzle at present.

We also found that the action of the right-side operator $A^R_{f,l}$ ``induces'' change in the left-side universe just as effected by the left-side operator $A^{L\dagger}_{\fb,l}$; see \eqref{muxi14Oct26}.  This is not so surprising, because the TFD state is essentially an EPR state (when two spins---left and right---are in the EPR state, if we measure the right spin and find it to be in the up (down) state, the left spin is found to be in the down (up) state).  Because of this, we do not have independent left and right universes, as far as the IR operators $A_{f,l}$ are concerned---the left universe is like the mirror of the right universe.  Moreover, we found that we can construct an operator conjugate to $A^R_{f,l}$ by combining right and left operators.  This may be suggesting that the UV manipulation of Ref.~\cite{Balasubramanian:2018yjq} for constructing the IR operator $\pi$ is related in our TFD setting to considering some UV manipulation in the left universe.  We leave further investigation of this interesting issue for future research.

Finally, although we focused on the modes producing fluctuations in the interior of the gray disk of Fig.~\ref{fig:LLM_diag}, it is also possible to consider coarse modes living near the edge of the disk; see section \ref{ss:IR_deformations}.  
The explicit construction of those near-edge modes requires detailed knowledge about the response of IR observables to UV modes, because we need to sum over UV modes that are indistinguishable to IR observers.  We have left this also for further investigation.

\subsubsection*{Physical interpretation of the fine-probe algebras}

Having found in section \ref{sec:coarse} that the coarse-probe algebra suggests the absence of a classical wormhole, we turned in sections \ref{sec:fineBoson} and \ref{sec:fineFermion} to studying the local algebra of fine probes associated with excitations near the edge of the gray disk in Fig.~\ref{fig:LLM_diag}. In effect these excitations modify the microscopic topology of the boundary conditions for the bulk geometry in this edge region.  The idea behind studying this was that, by considering probes that have access to details of the microstates, we can incorporate quantum corrections to the classical result and explore quantum connectivity of spacetime.  Because studying the full microscopic algebra is technically difficult, we focused on the edge excitations which are under  good algebraic control.
We found that both bosonic and fermionic probe algebras are generically intermediate Type III, suggesting a quantum wormhole.

As noted above, although the bosonic and fermionic fine-probe algebras are both in intermediate Type~III, the type index as a function of the parameters of the background state ($\lambda = N_C/N$) does not generally agree, as is clear from Figure~\ref{fig:superstar_type}.\footnote{The type index agrees only for $\lambda=0$ and $\lambda={(1+\sqrt{5})/ 2}.$}  This difference can presumably be traced to the fact that the bosonic and fermionic TFD states are based on different ensembles---respectively the canonical ensemble with fixed $N$ and the grand canonical ensemble with variable $N$.\footnote{%
In the fermionic analysis in section \ref{sec:fineFermion}, we freely acted with $C_n,C_n^\dagger$, allowing operators that change $N$. It is also possible to consider the algebra of operators that do not change $N$, by imposing invariance under the $U(1)$ action $C_n\to e^{i\alpha}C_n,C_n^\dagger\to e^{-i\alpha}C_n^\dagger$.  In this case, the algebra is Type II$_\infty$ for $0<\lambda\le 1$ and Type I for $\lambda=0$ \cite{Ueda-sanNote2}.  We thank Yoshimichi Ueda for discussion on this point and sharing his result.}  These ensembles are expected to be indistinguishable in the thermodynamic (large) $N$ limit. However, fine probes are sensitive to UV data that IR probes are ignorant of, so it is probably not surprising that they are sensitive to the difference in those ensembles.
This suggests that, for crude features such as the existence or not of quantum wormholes, the finer algebraic type may not be relevant.  That said, it would be  interesting to clarify the physical meaning of the type index.

That said, considering a grand canonical ensemble with variable $N$ and studying fermion operators $C_n,C_n^\dagger$ presents some interpretational difficulties from a holographic viewpoint.   In AdS/CFT, the number of fermions $N$ is identified with the rank of the gauge group of four-dimensional $\mathcal{N}=4$ super-Yang-Mills theory.  On the gravity side, changing $N$ corresponds to changing the quantized five-form flux, which determines the AdS radius and the effective five-dimensional cosmological constant. Therefore, an ensemble with varying $N$ and operations that change $N$ cannot be straightforwardly interpreted in terms of excitations in a single gauge theory of fixed rank, or a single universe with a fixed cosmological constant. On the other hand, in the large $N$ limit, the variation of $N$ in the grand canonical ensemble is negligible compared to $N$, and operators acting nontrivially on only finitely many fermion levels can change the total fermion number by at most a finite amount; so the relative change is suppressed by $1/N$. This suggests that the background value of $N$ can be regarded as effectively fixed in the large-$N$ limit.  Processes that change $N$ to $N+n$ with $n=O(1)$ may be approximately reinterpreted, in the grand canonical ensemble, as processes of changing $N-n$ to $N$, staying in the $N$ theory after all. We have adopted this interpretation for the purposes of this paper, but more work is needed to establish definitively whether an algebra of operators that change $N$ even by a small amount can make sense in AdS/CFT.\footnote{For an example of holography in which the bulk theory corresponds to a grand canonical ensemble of boundary theories, see \cite{Aharony:2024fid}.}

We focused on the local algebra of edge excitations as a fine-probe algebra, the choice to focus on this algebra was arbitrary. There may be other algebras of operators representing microscopic manipulations that survive at large $N$. For example, as we discussed, the operators $D_j$ and $D_j^\dagger$ deep inside the gray disk, with $j=O(N)$, have $O(N)$ energies and thus are not allowed at large $N$.  However, suitable combinations of them can have $O(1)$ energies.  In section \ref{sec:coarse}, we coarse-grained such combinations to construct IR probes, but there are more general possibilities, if we do not coarse-grain.  Everything we said here also applies to the fermionic formulation.   A general question is whether one can construct a larger algebra that retains the full microscopic information of the column-height indices $j$ or the fermion level $n$, including those scaling with $N$, and how such an algebra would be related to the coarse-grained IR observables. We leave full investigation of this problem for future work.

Finally, the relation between fine-probe algebras we studied and the observables accessible in the IR supergravity description remains to  be elucidated. Even if those fine probes carry $O(1)$ excitation energies, this fact alone does not guarantee that they induce finite amplitudes in suitably normalized supergravity observables: their effects on such observables may vanish in the large-$N$ limit. Establishing whether these microscopic operations can be detected by IR observers requires a more explicit understanding of the relation between UV and IR operators.  Therefore, we do not claim that the local algebra constructed here describes the complete set of edge fluctuations visible in the supergravity description. Rather, it provides a candidate microscopic algebraic structure underlying such fluctuations.

\subsubsection*{More general local algebras}

In the case of the $\beta=0$ superstar, in section \ref{sec:fineFermion}, we zoomed in to the region near the edge of the levels excited in the background state, $n=N+N_C=(1+\lambda)N$, and constructed a local fermionic algebra in the $N\to\infty$ limit.  More precisely, setting $n=(1+\lambda)N+n'$, we focused on the levels with $n'=O(1)$ and operators acting on these levels in the $N\to\infty$ limit.  This means that the asymptotic state at levels $n'\to\infty$ is fixed: as $n'\to-\infty$, the state at level $n'$ is $(1+\lambda)^{-1/2}(\sqrt{\lambda}\ket{00}+\ket{11})_{n'}$, while as $n'\to +\infty$ the state at level $n'$ is $\ket{00}_{n'}$ (see \eqref{TFD_superstar_Psi_relabeled}).  We showed that this local algebra is of Type III$_\lambda$.\footnote{Here we are assuming that $0<\lambda<1$; for other cases separate treatments as in section \ref{ss:special_lambda} are needed.\label{fn:special_lambda}}

More generally, we can zoom in onto the region near the point $n_0=\mu\, (N+N_C)$ with arbitrary fixed $\mu=O(1)$ and study the local fermionic fine-probe algebra there.  If $0<\mu<1$, if we set $n=n_0+n'$ with $n'=O(1)$, in the $N\to\infty$ limit, the local system can be described by the state
\begin{align}
    \bigotimes_{n'=-\infty}^\infty \ket{\psi_{n'}(\lambda)},
    \qquad\text{where}\qquad
\ket{\psi_{n'}(\lambda)}   \equiv
{1\over\sqrt{1+\lambda}}(\sqrt{\lambda}\,\ket{00}+\ket{11})_{n'}.
\end{align}
Relabeling the levels just as we did in \eqref{hqrw3Nov26}, interleaving the levels with $n'\ge 0$ and the ones with $n'<0$, we can write this state alternatively as
\begin{align}
   \bigotimes_{\tilde{n}=0}^\infty
\ket{\psi_{\tilde{n}}(\lambda)}.
\label{idsd3Nov26}
\end{align}
The algebra on this state is nothing but the Powers factor $\cP_\lambda$ discussed in section \ref{ss:type_class'n}. So, the local fermionic fine-probe algebra is a Type III$_\lambda$ factor.
On the other hand, if $\mu>1$, the local state is simply the pure state,
\begin{align}
    \bigotimes_{\tilde{n}=0}^\infty\ket{00}_{\tilde{n}}=
    \bigotimes_{\tilde{n}=0}^\infty\ket{\psi_n(\infty)}
\end{align}
after relabeling the level as we did above.  Just as discussed in section \ref{ss:type_class'n}, the associated algebra is a Type I$_\infty$ factor.
Therefore, the local algebra near any gray point of the disk in Fig.~\ref{fig:LLM_diag}, including the edge, is of Type~III$_\lambda$, while the local algebra near any white point in Fig.~\ref{fig:LLM_diag} is of Type~I$_\infty$.

Instead of the $\beta=0$ superstar TFD state, we can also consider a more general state with non-uniform grayness field:
\begin{align}
 \bigotimes_{n=0}^\infty \ket{\psi_n(\lambda_n)},
\end{align}
where $\lambda_n=\lambda({n\over N})$ and the function $\lambda(x)$ is assumed to be a smooth function of $x\ge 0$.  For the expected total fermion number to be equal to $N$, $\lambda(x)$ must satisfy
\begin{align}
    \sum_{n=0}^\infty {1\over 1+\lambda_n}
 \approx   N\int_0^\infty {dx\over 1+\lambda(x)}=N.
\end{align}
For example, in the hyperstar state \eqref{TFD_hyperstar_rho}, we have
\begin{align}
    \lambda(x)={(p+1)^x\over p},
\end{align}
where we used the relation \eqref{p_ito_q,N} valid at large $N$.
The smoothness of $\lambda(x)$  means that, if we zoom in onto the region near any level $n_0$, where the $\lambda$ parameter is $\lambda_{n_0}\equiv \lambda_0$,
the local state at large $N$ is given by a TFD state of the form \eqref{idsd3Nov26} with $\lambda$ replaced by $\lambda_0$, and the local algebra is a Powers factor $\mathcal{P}_{\lambda_{0}}$ of Type III$_{\lambda_{0}}$.\footnote{See footnote \ref{fn:special_lambda}.}

We could also consider an algebra for multiple such local regions.
Let the local values of $\lambda$ in two such zoomed regions be $\lambda'$ and $\lambda''$.  Then the relevant state is
\begin{align}
 \left[\bigotimes_{n=-\infty}^\infty \ket{\psi_n(\lambda')}\right]
 \otimes
 \left[\bigotimes_{n=-\infty}^\infty \ket{\psi_n(\lambda'')}\right]
\end{align}
which, upon relabeling, can alternatively be written as
\begin{align}
 \bigotimes_{l=0}^\infty 
\Bigl[ \ket{\psi_{2l}(\lambda')}\otimes\ket{\psi_{2l+1}(\lambda'')}\Bigr].
\end{align}
For generic values of $\lambda'$ and $\lambda''$, the von Neumann algebra on this state is of Type~III$_1$~\cite{Witten:2018zxz}. So, if we take generic two points from the hyperstar state \eqref{TFD_hyperstar_rho}, the local fine-probe algebra is of Type~III$_1$.\footnote{Note that Type~III$_1$ is the one normally associated with classical wormholes.  However, because the classicality condition is not met by the superstar, the algebraic ER=EPR does not say that there is a classical wormhole.}

\subsubsection*{Connection to black hole thermodynamics}

A nontrivial feature of the fermionic type classification shown in Fig.~\ref{fig:superstar_type} is the special role of $\lambda=1$. For $0<\lambda<1$ and $\lambda>1$, the corresponding von Neumann algebras are of Type III$_{\lambda}$ and Type III$_{1/\lambda}$, respectively, whereas at $\lambda=1$, the algebra becomes Type II$_{\infty}$.

Motivated by this observation, in Appendix \ref{sec:instability}, we investigated the thermodynamic stability of single R-charged AdS$_5$ black holes, whose vanishing-horizon limit corresponds to the superstar geometry. Interestingly, we found that $\lambda=1$ also plays a distinguished role in the thermodynamic stability analysis in the grand canonical ensemble: although the superstar is thermodynamically unstable for $\lambda\neq 1$, it becomes marginally stable for $\lambda=1$.

Although the analysis in Appendix \ref{sec:instability} did not clarify a direct relation between the von Neumann algebraic type and black hole thermodynamic stability, this observation suggests a possible connection between the algebraic structure of the superstar ensemble and the thermodynamic properties of the corresponding charged black holes. Understanding whether this coincidence reflects a deeper physical mechanism would be an interesting direction for future investigation.

\section*{Acknowledgments}
We thank Yoshimichi Ueda for his lectures and comments on operator algebras, which greatly benefited the current work, and Shono Shibuya for collaboration at an early stage of the work.
VB is grateful to Evita Verheijden for many useful conversations on related topics. 
We would like to thank the organizers of The ExU-YITP Workshop on 
Holography, Gravity and Quantum Information 2023 (YITP-T-23-01) and
the YITP International workshop ``Extreme Universe 2025'' (YITP-T-25-01)
held at YITP, Kyoto University where this work was initiated and developed. 
During this work VB was supported in part by the DOE through DE-SC0013528 and the QuantISED grant DE-SC0020360, and  by the Eastman Professorship at Balliol College, University of Oxford.   
The work of RN was partly supported by JST SPRING, Grant Number JPMJSP2125. RN would like to take this opportunity to thank the ``THERS Make New Standard Program for the Next Generation Researchers.'' 
The work of MS was supported in part by MEXT
KAKENHI Grant Numbers 21H05184 and 24K00626.

\appendix

\section{Limit curves}
\label{app:limit_curves}

As we have seen in section \ref{sss:eff_thermodyn_ensembles}, the grand canonical ensemble is simple in terms of the variables $f_n$, but the limit curves are directly related to variables such as $r_i$ or $e_i$.  To go between these presentations, we can use the relation between $f_n$ and $e_i$:
\begin{align}
    i=f_0+f_1+\cdots +f_{n-1},\qquad e=n-1.
\end{align}
At large $N$, this gives
\begin{align}
    i\approx F(n), \qquad F(n)\equiv \int_0^n f_n dn,\qquad 
    \text{therefore}
    \qquad e_i=F^{-1}(i).
\end{align}
Because the $x$, $y$ coordinates are related to $e_i$ by $x\leftrightarrow i$, $y\leftrightarrow r_i=e_i-i+1$, we immediately find, at large $N$,
\begin{align}
    y=F^{-1}(x)-x.
\end{align}
Note also that
\begin{align}
    F(\infty)=N.\label{F(inf)=N}
\end{align}

For example, for the hyperstar ensemble \eqref{eq:rho(p,q)}, for which
\begin{align}
\ev{f_n}\approx {pq^{n}\over 1+pq^{n}},\qquad
F(n)\approx 
\int_0^n {pq^{n}\over 1+pq^{n}}dn
={1\over \log q}\log{1+pq^n\over p+1},\label{F(n)_hyper}
\end{align}
the limit curve is 
\begin{align}
    q^{-x}+p q^y=p+1,
\end{align}
where $p$ is related to $N$ by
\begin{align}
    q^N={1\over p+1}.
\end{align}
as can be derived from \eqref{F(inf)=N} and \eqref{F(n)_hyper}.

\section{Non-constant grayness}
\label{app:nonconst_grayness}

Consider a TFD state in bosonic language, in which the occupation number depends on $j$:
\begin{align}
  \ket{\Psi}&=\prod_{j=1}^N
  \sqrt{1-\zeta_j}
  \sum_{c_j=0}^\infty \zeta_j^{c_j/2} \ket{c_j,c_j}_j.
  \label{eq:nonconst_bos_Psi}
\end{align}
The associated reduced density matrix is, for both $\alpha=R,L$,
\begin{align}
  \rho^\alpha&
  =\prod_{j=1}^N
  (1-\zeta_j)
  \sum_{c_j=0}^\infty \zeta_j^{c_j} \ket{c_j}_j\bra{c_j}_j
  =\prod_{j=1}^N
  (1-\zeta_j)
  \zeta_j^{\hat c_j^\alpha}.
  \label{eq:nonconst_bos_rho}
\end{align}
On the state $\ket{\Psi}$, the 2-point functions of the operators $A^\alpha_{f,l},A^{\alpha\dagger}_{f,l}$, still defined by \eqref{mgns28Sep26} for $l>0$, are readily shown to be
\begin{align}
    \ev{A^{R\dagger}_{f,l}A^R_{g,m}}_\Psi
 &   \approx\ev{A^R_{g,m}A^{R\dagger}_{f,l}}_\Psi
    \approx 
        {\delta_{lm}\over N}\sum_{j=1}^N \mu_j \fb_j g_j
        \approx
        \delta_{lm}\,(f,g) ,
        \label{eq:<AR+AR>_gen}
        \\
   \ev{A^{R}_{f,l}A^L_{g,m}}_\Psi
&\approx \delta_{lm}    \,(\fb,g),
        \label{eq:<ARAL>_gen}
\end{align}
where ``$\approx$'' means an equality up to $O(1/N)$ terms, and
\begin{align}
\mu_j\equiv {\zeta_j\over (1-\zeta_j)^2}    ,\qquad
\mu(v)\equiv {\zeta(v)\over (1-\zeta(v))^2},\qquad
(f,g)\equiv \int_0^1 dv\,\mu(v)\fb(v)g(v).
\end{align}
Also, we assumed that
\begin{align}
    \zeta_j
    =
    \zeta(\tfrac{j}{N})
    ,\label{eq:zeta(x)_zeta_j}
\end{align}
where $\zeta(v)$ is a smooth function defined for $0\le v\le 1$. So, we can say that all this non-constant $\zeta(v)$ does is to change the integration measure to $dv\,\mu(v)$. The above motivates us to consider an orthonormal set of functions $\{f_k(v)\}$ satisfying
\begin{align}
    (f_k,f_{k'}) =\delta_{kk'}.
\end{align}
Just as in the constant grayness case (see \eqref{eq:large_N_fact'n}), higher correlation functions factorize into products of 2-point functions at large $N$, as long as the number of fields is finite.

The commutators in this setting are still given by \eqref{AA_comm_op}. On the state $\ket{\Psi}$, we find that
\begin{align}
    \ev{[A^R_{f,l},A^R_{g,m}]}_\Psi
    \approx\ev{[A^R_{f,l},A^{R\dagger}_{f,l}]}_\Psi
    \approx 0.
\end{align}
The commutators vanish at large $N$ in the IR Hilbert space $\cH^R_{\rm IR}$, where $\cH^R_{\rm IR}$ is the right-side Hilbert space generated by finite products of the IR operators $A^R_{f,l},A^{R\dagger}_{g,m}$, generalizing the one in the main text to non-constant $\zeta_j$.  
Therefore, as in the case with constant grayness, the IR algebra of right-side operators is that of commuting operators $A^R_{f,l},A^{R\dagger}_{g,m}$.  Namely, it is a direct integral of Type I$_1$ factors.

The ``no-go theorem'' \eqref{hotu2Oct26} can also be generalized.  We find that
\begin{align}
    [A^R_{f,l}\,,\rho_R]
    &\approx
    -{l\over N}A^R_{f\zeta'/\zeta,l}\,\rho_R.
\end{align}
Therefore, for any right-side operator $\cO^R$,
\begin{align}
 \ev{[A^R_{f,l}\,,\cO^R]}_{\rho_R}
 &= \Tr_R( \rho_R [A^R_{f,l}\,,\cO^R])
 = \Tr_R (\cO^R[\rho_R\,,A^R_{f,l}])\notag\\
&\approx {l\over N}\ev{\cO^R A^R_{f\zeta'/\zeta,l}}_{\rho_R},
\end{align}
which means that the commutator expectation value of $A^R_{f,l}$ with any right-side operator $\cO^R$ vanishes on $\ket{\Psi}$ at large $N$.

Behind the correlators \eqref{eq:<AR+AR>_gen} and \eqref{eq:<ARAL>_gen} is the fact that $A^R_{f,l}$ and $A^{L\dagger}_{\fb,l}$ behave the same on the TFD state $\ket{\Psi}$ at large $N$:
\begin{align}
    A^R_{f,l}\ket{\Psi}\approx
    A^{L\dagger}_{\fb,l}\ket{\Psi},\qquad
    A^{R\dagger}_{f,l}\ket{\Psi}\approx
    A^{L}_{\fb,l}\ket{\Psi}.
    \label{mdxh10Nov26}
\end{align}
Here, unlike the case with constant grayness (cf.~\eqref{muxi14Oct26}), these relations are not exact and hold only up to $O(1/N)$ corrections. For example, we find
\begin{align}
    \|(A^R_{f,l}-
    A^{L\dagger}_{\fb,l})\ket{\Psi}\|^2
    ={l^2\over 4N^2}\int dv |f(v)|^2
    {\zeta'(v)^2\over \zeta(v)^2}
    \mu(v)
    +O\Bigl({1\over N^3}\Bigr),
\end{align}
which vanishes at large $N$. The relation \eqref{mdxh10Nov26} means  that, as in the constant grayness case, the replacements $A_{f,l}^{R}\to A_{\fb,l}^{L\dagger}$ and $A_{f,l}^{R\dagger}\to A_{\fb,l}^{L}$ leave any correlation function in the state $\ket{\Psi}$ unchanged in the large-$N$ limit if the number of operators is $O(1)$.

The construction of the operators that do have non-vanishing commutators with $A^R,A^{R\dagger}$ goes exactly the same way as in the case with constant grayness.
The operator $\Delta$ defined in \eqref{eq:Delta_def} still satisfies the key relations \eqref{itu7Oct26}, which in the present case read
\begin{align}
 \Delta\ket{\Psi}=0,\qquad
 \Delta A_\alpha \ket{\Psi}=A_\alpha\ket{\Psi},\qquad
 \Delta A_\alpha^\dagger\ket{\Psi}=A_\alpha^\dagger\ket{\Psi},
\end{align}
where $\alpha=R,L$ and $A_\alpha=A^\alpha_{f,l},A_\alpha^\dagger=A^{\alpha\dagger}_{f,l}$.  If we define $B_\alpha,B_\alpha^\dagger$ by \eqref{eq:B_alpha_def}, we get the commutators
\begin{align}
\ev{[A_R,B_R^\dagger]}_{\Psi}
\approx\ev{[A_L,B_L^\dagger]}_{\Psi}
\approx\ev{[A_R,B_L]}_{\Psi}
\approx\ev{[A_L,B_R]}_{\Psi}
\approx 2i\kappa,
\qquad    \kappa=(f,f).
\end{align}
We again have the equivalence
\begin{align}
    A_R\cong A_L^\dagger,\qquad
    A_R^\dagger\cong A_L,\qquad
    B_R\cong B_L^\dagger,\qquad 
    B_R^\dagger\cong B_L.
\end{align}
So, we have only one field theory common to left and right, and the IR algebra generated by  $A,B$  operators is Type I$_\infty$.

\subsection{$l=0$ modes}
\label{app:l=0_modes}

The $l=0$ modes require a separate treatment.  Here we discuss the case with general non-constant grayness, but the result is applicable also to the case with constant grayness, by setting $\zeta_j=\zeta(v)=\zeta$.

Let us define
\begin{align}
    A_{f,0}^\alpha
    ={1\over\sqrt{N}}\sum_{j=1}^N f_j D_j^{\alpha\dagger}D_j^\alpha
    ={1\over\sqrt{N}}\sum_{j=1}^N f_j \hat{c}^\alpha_j,
\end{align}
which is an Hermitian operator for real $f_j$.
For the 1-point function
\begin{align}
    \ev{A_{f,0}^\alpha}_\Psi
    ={1\over\sqrt{N}}\sum_{j=1}^N f_j \ev{\hat{c}^\alpha_j}_\Psi
    ={1\over\sqrt{N}}\sum_{j=1}^N f_j {\zeta_j\over 1-\zeta_j}
    \approx \sqrt{N}\int_0^1 dv\, {\zeta(v)\over 1-\zeta(v)}f(v)
\end{align}
to vanish at large $N$, we require that $f$ satisfy\footnote{Alternatively, we can define 
\begin{align}    
A_{f,0}'^\alpha
    ={1\over\sqrt{N}}\sum_{j=1}^N f_j (\hat{c}^\alpha_j -\ev{\hat{c}^\alpha_j }),
\end{align}
whose 1-point function vanishes.
The sets $\{A_{f,0}^\alpha\}$ and $\{A_{f,0}'^\alpha\}$ are not exactly the same, because
$A_{f=1,0}'^\alpha$ is not included in the former.
}
\begin{align}
   \int_0^1 dv\, {\zeta(v)\over 1-\zeta(v)}f(v)=0,
   \label{jtit7Nov26}
\end{align}
which we assume henceforth.
Assuming \eqref{jtit7Nov26}, the 2-point function is
\begin{align}
    \ev{A_{f,0}^\alpha A_{g,0}^\alpha}_\Psi
   \approx{1\over N}\sum_{j=1}^N {\zeta_j\over (1-\zeta_j)^2}f_j g_j
    \approx (f,g).
\end{align}

The field canonically conjugate to $A_{f,0}^R$ is given by the following expression that contains both the left and right:
\begin{align}
    B^R_{g,0}=-{i\over \sqrt{N}}\sum_{j=1}^{N-1}{G_j}
    {(1-\zeta_j)(1-\zeta_{j+1})\over \sqrt{\zeta_j \zeta_{j+1}}}
    (T_j^L T_j^R-T_j^{L\dagger}T_j^{R\dagger}),\qquad
    T_j^\alpha\equiv D_{j+1}^{\alpha\dagger}D_j^\alpha,
\end{align}
where $G_j$ is defined by
\begin{align}
    G_j=\sum_{k=1}^j \mu_k g_k.
\end{align}
Using relations such as
\begin{align}
\begin{gathered}
    [\hat{c}_j^\alpha,T_k^{\alpha}]
    =-\delta_{jk}T_j^\alpha +\delta_{j,k+1}T_{j-1}^\alpha,\qquad
    [\hat{c}_j^\alpha,T_k^{\alpha\dagger}]
    =
\delta_{jk}T_j^{\alpha\dagger}-\delta_{j,k+1}T_{j-1}^{\alpha\dagger},
\\
\ev{T_j^L T_j^R}_\Psi=\ev{T_j^{L\dagger} T_j^{R\dagger}}_\Psi
={\sqrt{\zeta_j \zeta_{j+1}}\over (1-\zeta_j)(1-\zeta_{j+1})},
\end{gathered}
\end{align}
after some tedious manipulations, it is straightforward to show that
\begin{align}
    \ev{[A^R_{f,0},B^R_{g,0}]}_\Psi
    &=-{2i\over N}\sum_{j=1}^{N-1}(f_{j+1}-f_j)G_j
    \notag\\
    &={2i\over N}
    \left[-G_{N-1}f_N+f_{N-1}(G_{N-1}-G_{N-2})+\cdots+f_2(G_2-G_1)+f_1G_1\right]
    \label{jjgr7Nov26}
\end{align}
If we define $G_0=0$ and also require that $\{g_j\}$ satisfy
\begin{align}
    0=G_N=\sum_{j=1}^N \mu_j g_j
    \approx N\int_0^1 dv\, \mu(v) g(v),
\end{align}
then \eqref{jjgr7Nov26} can be written as
\begin{align}
    \ev{[A^R_{f,0},B^R_{g,0}]}_\Psi
    &={2i\over N}\sum_{j=1}^N f_j (G_{j}-G_{j-1})\notag\\
    &={2i\over N}\sum_{j=1}^N \mu_j f_j g_j 
    \approx {2i}(f,g).
\end{align}
This is the $l=0$ analogue of the $l>0$ relation \eqref{wji7Oct26}, showing that $B^R_{g,0}$ is the field conjugate to~$A_{g,0}^R$.

\section{$A_{f,l}$ are grayness fluctuation fields}
\label{app:A_are_fluct}

We would like to study the fluctuation that the operator
\begin{align}
 A^R_{f,l}={1\over\sqrt{N}}\sum_j f(\tfrac{j}{N}) D_{j+l}^{R\dagger} D^R_j
\end{align}
induces on the LLM plane, when acting on the TFD state $\ket{\Psi}$ \eqref{eq:nonconst_bos_Psi}
with non-constant grayness.  The functions $f(v)$ and $\zeta(v)$ (see Eq.~\eqref{eq:zeta(x)_zeta_j}) are assumed to be smooth functions having no structure over  distances $\Delta v=O(1/N)$.

As discussed in section \ref{sss:bulk_side}, the classical grayness field $u(x_1,x_2)$ can be regarded as the fermion density on the single-fermion phase space \cite{Balasubramanian:2005mg, Balasubramanian:2018yjq} (see also \cite{Takayama:2005yq}) with coordinates
$(x_1,x_2)=(q,p)$.  To define the quantum mechanical version of $u(q,p)$, let us first introduce the second-quantized fermion field by
\begin{align}
 \hat\psi(q)=\sum_n f_n(q)\hat C_n,\qquad
 \{\hat\psi(q),\hat\psi(q')^\dagger \}=\delta(q-q'),
\end{align}
where $C_n$ is the fermion annihilation operator defined in \eqref{eq:C_algebra_def},
while $f_n(q)$ are the one-particle harmonic oscillator wavefunctions in
position space:
\begin{align}
 f_n(q)=\frac{1}{\sqrt{\pi^{1/2}2^n n!}} H_n(q)e^{-q^2/2}
\end{align}
with $H_n(q)$ the Hermite polynomial. Then the fermion density operator (Wigner distribution) is defined as
\begin{align}
 \hat u(q,p)=2\int_{-\infty}^\infty dx\, e^{2ipx/\hbar }
 \hat\psi(q-x)^\dagger  \hat\psi(q+x).
\end{align}
Actually $\hat{u}(q,p)$ can also be written in terms of the wavefunction of the
2d harmonic oscillator in polar coordinates.  The relevant Schr\"odinger
equation is
\begin{align}
 \left[-{\hbar^2\over 2\mu}\left(
 {1\over r}\partial_r (r\partial_r)+{1\over r^2}\partial_\phi^2
 \right)+{\mu\omega^2\over 2}r^2\right]\psi = E\psi,
\end{align}
and its orthonormal set of solutions is
\begin{align}
 \psi_{nm}(r,\phi)={1\over a}\sqrt{n!\over \pi (|m|+n)!}
 \left({r\over a}\right)^{|m|}
 e^{-{r^2\over 2a^2}}(-1)^n L_n^{(|m|)}(\tfrac{r^2}{a^2})\,e^{im\phi},
\qquad   a=\sqrt{\hbar\over \mu\omega},
\label{llgt1Sep26}
\end{align}
where $L_n^{(\alpha)}(z)$ is the associated Laguerre polynomial,
with energy eigenvalues
\begin{align}
  E_{nm}=\hbar\omega (2n+|m|+1),\qquad
n\in \bbZ_{\ge 0}, \quad m\in\bbZ.
\end{align}
$\psi_{nm}$ satisfy the relations
\begin{gather}
 \int r\,dr\, d\phi\, \psi_{nm}^* \psi_{n'm'}=\delta_{nn'}\delta_{mm'},
 \qquad
 \sum_{n=0}^\infty \sum_{m\in\bbZ}
 \psi_{nm}(r,\phi)\, \psi_{nm}(r',\phi')^*
 =\delta(x_1-x_1')\delta(x_2-x_2').
\end{gather}
In terms of $\psi_{nm}$, the fermion density $\hat{u}$ can be written as
\begin{align}
 u(r,\phi)=\sqrt{2\pi \hbar}\sum_{m,n}C_m^{\dagger} C_n
 \psi_{{\rm min}(n,m),n-m}(r,\phi),
 \qquad a=\sqrt{\hbar\over 2},
 \label{hbpw10Nov26}
\end{align}
where $(r,\phi)$ are the polar coordinates on the $(q,p)$ plane.

The fluctuation on the LLM plane with polar coordinates $(r,\phi)$ can be computed from the two-point amplitude
\begin{align}
 \cF_{f,l}(r,\phi)\equiv \ev{u^R(r,\phi) A^R_{f,l}}_\Psi.
\label{iqez22Oct26}
\end{align}
More precisely, if we define the coherent state
\begin{align}
 \ket{\chi}={1\over\sqrt{Z}}\ket{\chi_0},\qquad
 \ket{\chi_0}=e^{z A_{f,l}}\ket{\Psi},
\end{align}
then the change in $u^R$ is given at leading order by
\begin{align}
 \delta u^R = \ev{u^R}_\chi- \ev{u^R}_\Psi
 = 2\Re(z\cF_{f,l}).
\end{align}
Using \eqref{hbpw10Nov26}, the amplitude is
\begin{align}
 \cF^R_{f,l}=\sqrt{2\pi \hbar\over N}\sum_{m,n}\sum_j 
 f(\tfrac{j}{N})\,
 \psi_{{\rm min}(n,m),n-m}(r,\phi)
 \,\ev{\Psi|C_m^{R\dagger} C^R_n  D_{j+l}^{R\dagger} D^R_j |\Psi}.
\label{jfsy22Oct26}
\end{align}

\begin{figure}[tbh]
\begin{quotation}
\begin{center}
\scalebox{0.8}{
 \begin{tikzpicture}[scale=0.2]
 \draw (0,25) -- (0,0) -- (1,0) -- (1,25);
 \draw[latex-latex] (-0.5,0) -- +(0,2) node [left,midway] {$c_N$};
 \draw[fill=black] (0,2) rectangle +(1,1);
 \node at (1,2.5) [right] {$n_N=c_N$};
 \draw[latex-latex] (-0.5,3) -- +(0,3) node [left,midway] {$c_{N-1}$};
 \draw[fill=black] (0,6) rectangle +(1,1);
 \node at (1,6.5) [right] {$n_{N-1}=1+c_{N-1}+c_N$};
 \node at (0,10) [left] {$\vdots$};
 \node at (5,10) {$\vdots$};
 \draw[fill=black] (0,12) rectangle +(1,1);
 \draw[latex-latex] (-0.5,13) -- +(0,2) node [left,midway] {$c_{2}$};
 \draw[fill=black] (0,15) rectangle +(1,1);
 \node at (1,15.5) [right] {$n_2=N-2+c_2+\cdots+c_N$};
 \draw[latex-latex] (-0.5,16) -- +(0,3) node [left,midway] {$c_{1}$};
 \draw[fill=black] (0,19) rectangle +(1,1);
 \node at (1,19.5) [right] {$n_1=N-1+c_1+\cdots+c_N$};
 \draw[-latex] (0.5,26) -- +(0,2) node [right] {$n$};
\end{tikzpicture}}
\caption{\sl The relation between the  boson index (column-height index $j$) and the fermion level $n_j$ via the boson occupation number $c_j$.  The
vertical position~$(n)$ corresponds to the fermion level.  The black boxes represent
levels occupied by fermions.\label{fig:n_and_j_via_cj}}
\end{center}
\end{quotation}
\end{figure}
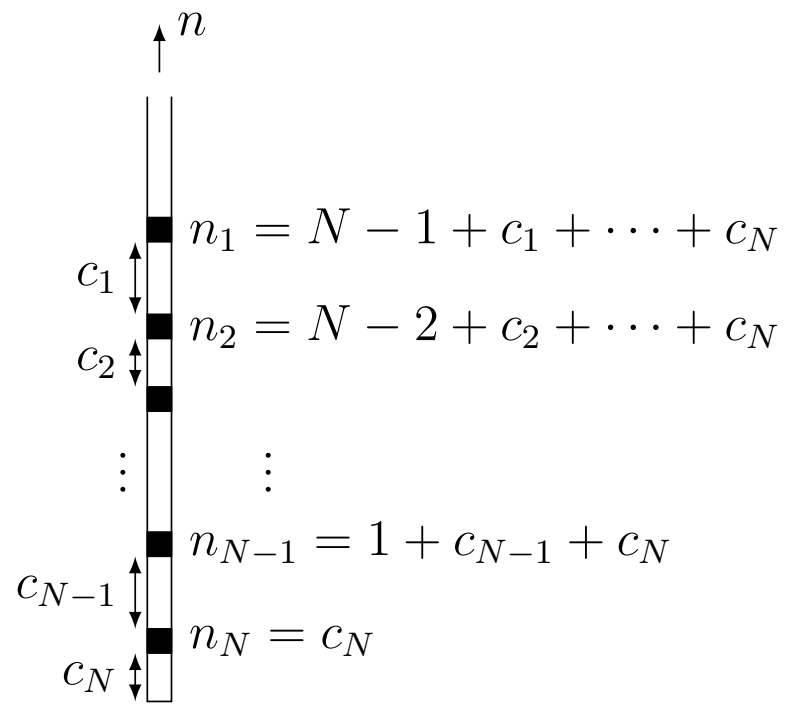

In order to go between the bosonic and fermionic descriptions, define $n_j$
to be the occupied level of the $j$th fermion counted {\it from the
top}.  This way, $j$ serves simultaneously as the fermion index and the boson index (the column-height index).
Using \eqref{isko28Sep26}--\eqref{mgra14Nov23}, we find
\begin{align}
 n_j=N-j+\sum_{j'=j}^N c_{j'}.\label{mcfj22Oct26}
\end{align}
See also Figure \ref{fig:n_and_j_via_cj}.

\begin{figure}[tb]
\begin{center}
  \begin{tikzpicture}[scale=0.3]
 \draw (0,0) 
 -- ++(3,0) -- ++(0,1)
 -- ++(3,0) -- ++(0,4)
 -- ++(4,0) -- ++(0,1)
 -- ++(3,0);
 \draw (0,8) -- +(13,0);
 \draw[densely dotted] (6,1) rectangle +(1,4);
 \draw[latex-latex] (8,5) -- +(0,3) node [midway,left] {$j$};
 \draw[latex-latex] (4.5,1) -- +(0,7) node [midway,left] {$j+l$};
 \end{tikzpicture}
 \caption{\sl The action of the operator $D_{j+l}^\dagger D_{j}$ for $c_{j+1}=\cdots=c_{j+l-1}=0$. \label{fig:DdaggerD2}}
\end{center}
\end{figure}
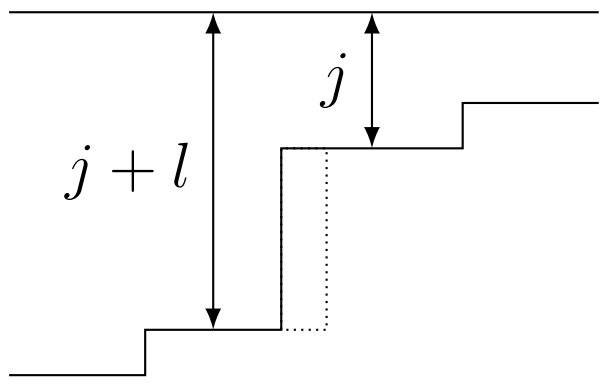

Now let us examine $\ev{\Psi|C_m^{R\dagger} C^R_n  D_{j+l}^{R\dagger} D^R_j |\Psi}$ in \eqref{jfsy22Oct26}.  Acting on the state
$\ket{\Psi}$, which involves various states represented by various
Young diagrams characterized by $\{c_j\}$, the action of the operator
$D^{R\dagger}_{j+l}D^R_{j}$ is as shown in Figure~\ref{fig:DdaggerD}.
As represented by the $l$ dotted boxes in the figure, this generically involves annihilating $l$ fermions and creating $l$ fermions, and acting on the resulting state further with $C^{R\dagger}_m C^R_n$ does not give a state with nonzero overlap with the original
state $\ket{\Psi}$, leading to zero contribution to $\cF_{f,l}^R$.
However, for the state with
\begin{align}
 c_{j+1}=\cdots=c_{j+l-1}=0,\label{lmce22Oct26}
\end{align}
the Young diagram is as in Figure \ref{fig:DdaggerD2}, and the action of $D^{R\dagger}_{j+l}D^R_{j}$ corresponds to annihilating one fermion
at level $n_{j+l}$ and creating one fermion at level $n_{j+1}+1=n_{j+l}+l$.
Acting on the resulting state further with $C^{R\dagger}_m C^R_n$ 
gives a state with a nonzero overlap with the original state
$\ket{\Psi}$ for $m=n_{j+l},n=n_{j+l}+l$.
As a result, the amplitude is
\begin{align}
 \cF^R_{f,l}=\sqrt{2\pi\hbar\over N}\sum_j f(\tfrac{j}{N})\,
 \psi_{n_{j+l}^{},l}^{}(r,\phi)\,(-1)^{l-1}
 (1-\zeta_{j+1})\cdots (1-\zeta_{j+l-1})
 \Ev{\sqrt{\hat{c}_j(\hat{c}_{j+l}+1)}}_\Psi.
\end{align}
The sign $(-1)^{l-1}$ comes from acting with
$C^{R\dagger}_{n_{j+l}^{}}C^{R}_{n_{j+l}^{}+l}$ on
$C^{R\dagger}_{n_{j+l}^{}+l}\cdots C^{R\dagger}_{n_{j+l}^{}+1}$
representing the fermionic state after the action of
$D^{R\dagger}_{j+l}D^R_j$, while the factor
$(1-\zeta_{j+1})\cdots(1-\zeta_{j+l-1})$ comes from the factor
$\prod^{j'=j+l-1}_{j+1}(1-\zeta_{j'})$ in \eqref{eq:nonconst_bos_Psi} uncanceled
due to the projection onto \eqref{lmce22Oct26} for those values of $j'$.
Also using
\begin{align}
\begin{split}
 \ev{\sqrt{\hat{c}_j}}&=(1-\zeta_j)\sum_{c_j=0}^\infty \zeta^{c_j}\sqrt{c_j}
 =(1-\zeta_j){\rm Li}_{-\half}(\zeta_j),
\\
 \ev{\sqrt{\hat{c}_{j+l}^{}+1}}&=(1-\zeta_{j+l})\sum_{c_{j+l}^{}=0}^\infty \zeta_{j+l}^{c_{j+l}^{}}\sqrt{c_{j+l}^{}+1}
 =(1-\zeta_{j+l}){{\rm Li}_{-\half}(\zeta_{j+l})\over \zeta_{j+l}},
 \end{split}
\end{align}
we find 
\begin{align}
 \cF^R_{f,l}(r,\phi)\approx (-1)^{l-1}\sqrt{2\pi\hbar\over N}\sum_j f(\tfrac{j}{N})\,
 \psi_{n_j^{},l}(r,\phi)\,
 (1-\zeta_{j})^{l+1}{{\rm Li}_{-\half}(\zeta_j)^2\over \zeta_j},\label{mrrj22Oct26}
\end{align}
where we dropped subleading terms at large $N$ with $l=O(1)$.

We would like to extract the $r$ dependence of \eqref{mrrj22Oct26}.  For
that, let us first relate the radial coordinate $r$ and the index $j$,
or $v\equiv {j\over N}$. On the one hand, $r$ is related to the fermion
level $n$ by
\begin{align}
 r^2=2n\hbar.\label{mcwc22Oct26}
\end{align}
On the other hand, at large $N$, the relation \eqref{mcfj22Oct26} can be
written as
\begin{align}
 n\approx N-j+N\int_{j\over N}^1 dv\,\ev{c(v)}
 =N-j+N\int_{j\over N}^1 dv\,{\zeta(v)\over 1-\zeta(v)}\label{mcwg22Oct26}
\end{align}
where $\ev{c({j\over N})}=\ev{\hat{c}_j}={\zeta_j\over 1-\zeta_j}$ is
the expectation value of the occupation number of the $j$th boson, and recall that $\zeta({j\over N})=\zeta_j$. Combining
\eqref{mcwc22Oct26} and \eqref{mcwg22Oct26}, we find that $r^2$ and $v$
are related by
\begin{align}
 r^2=2N\hbar\int_v^1 {dv\over 1-\zeta(v)}.\label{mthi22Oct26}
\end{align}

Next, note that the function $\psi_{n,l}(r,\phi)$ is sharply peaked
around
\begin{align}
 r^2=2n\hbar
\end{align}
in the large $n$, fixed $l$ limit (this limit is justified because we
are interested in $n=O(N)$). More precisely, away from this point, the
function oscillates rapidly with wavelength $\Delta (r^2)\sim \hbar$ on
the smaller $r$ side, and exponentially decays on the larger $r$
side. So, if we consider integrals against smooth test functions, such as $f(v)$ and $\zeta(v)$,
$\psi_{n,l}(r,\phi)$ behaves as a delta function:
\begin{align}
 \psi_{n,l}(r,\phi)\sim \sqrt{2\hbar\over \pi}\delta(r^2-2n\hbar)\,e^{i\phi l}.\label{nebe22Oct26}
\end{align}
The overall coefficient was determined by evaluating $\int_0^r
dr\,\psi_{n,l}(r,\phi)$ for large $n$. Therefore, \eqref{mrrj22Oct26}
can be written as
\begin{align}
 \cF^R_{f,l}(r,\phi)\approx (-1)^{l-1} {2\hbar\over \sqrt{N}} e^{il\phi}\sum_j f(\tfrac{j}{N})\,
 (1-\zeta_{j})^{l+1}{{\rm Li}_{-\half}(\zeta_j)^2\over \zeta_j}
\delta(r^2-2n_j\hbar).
\end{align}
Now we replace $\sum_j\to N\int_0^1 dv$ with $v={j\over N}$.  Then
$\delta(r^2-2n_j\hbar)$ can be regarded as a delta function for $j$,
through the $j$-dependence of $n_j$.   If we carry out the $v$ integral, taking into
account the Jacobian ${dn\over dv}={2N\hbar\over 1-\zeta(v)}$ found from
\eqref{mcwg22Oct26}, we obtain
\begin{align}
 \cF^R_{f,l}(r,\phi)\approx {\cC(v)\over\sqrt{N}} f(v)\, e^{il\phi},
 \qquad 
\cC(v)= (-1)^{l-1}(1-\zeta(v))^{l+2}
{{\rm Li}_{-\half}(\zeta(v))^2\over \zeta(v)},
\end{align}
where $v$ is written in terms of $r$ using \eqref{mthi22Oct26}.  This
shows that the operator $A^R_{f,l}$ indeed corresponds to
grayness fluctuation, with $f(v)$ related to the profile along $r$ (up to modulation by $\cC(v)$)
and with~$l$ giving the angular frequency.

In particular, for the case with constant grayness, $\zeta(v)=\zeta$, namely for the $\beta=0$ superstar, we find
\begin{align}
 v=1-{r^2\over R^2},\qquad  R^2={2N\hbar \over 1-\zeta},
\end{align}
and
\begin{align}
 \cF_{f,l}(r,\phi)\approx {\cC\over\sqrt{N}} f(1-\tfrac{r^2}{R})\, e^{il\phi},
 \qquad 
\cC= (-1)^{l-1}(1-\zeta)^{l+2}
{{\rm Li}_{-\half}(\zeta)^2\over \zeta}.
\end{align}
Therefore, $f(v)$ precisely corresponds to the radial fluctuation profile.

\section{Instabilities and algebraic type}
\label{sec:instability}

In the main text, we found that the type of the fine-probe algebra depends on the parameter~$\lambda$.  In this appendix, we point out a possible connection between the algebra type and the (in)stability of the corresponding bulk black hole solutions. 

\subsection{Single R-charge black holes and the superstar solution}

We begin by studying the stability of the superstar solution, seen as an extremal limit of a family of black holes with finite horizon area.  Compactifying 10D Type IIB supergravity on $S^5$ gives $d=5$, $\cN=8$ gauged supergravity \cite{Gunaydin:1985cu}, which can be truncated to gauged $SO(6)$ supergravity.  This theory can be further truncated to $U(1)^3$ supergravity which contains  scalar fields charged under the three $U(1)$s. Charged black hole solutions in the $U(1)^3$ theory without charged scalar hair were constructed by Behrndt, Cveti\v{c}, and Sabra (BCS) in \cite{Behrndt:1998jd} for the non-rotating case and in \cite{Cvetic:2004ny, Wu:2011gq} for the rotating case.  The non-rotating solution with only one charge is (see also \cite{Myers:2001aq}) 
\begin{align}
\begin{split}
 ds^2&=h^{1/3}\left(-{f\over h}dt^2+{dr^2\over f}+r^2d\Omega_3^2\right),\\
 h&=1+{q\over r^2},\qquad
 f=1-{r_0^2\over r^2}+ {r^2\over \RAdS^2}h
\qquad A=(h^{-1}-1)dt,
\end{split}\label{mcyp14Jul24}
\end{align}
If we define a dimensionless radial coordinate $y={r/R_{\rm AdS}}$ and a dimensionless charge parameter $\lambda$ by
\begin{align}
    \lambda\equiv {q\over \RAdS^2}\ge 0,
\end{align}
the parameter $r_0$ can be related to the horizon radius $r_+$ as $y_0=y_+\sqrt{y_+^2 + 1+\lambda}$, where $y_0=r_0/R_{\rm AdS}$ and $y_+=r_+/R_{\rm AdS}$. As we will see, in the extremal, supersymmetric limit, the charge parameter $\lambda$ here is identified  with the $\lambda$ parameter in the superstar ensemble at $\beta=0$, \eqref{TFD_superstar_rho_2}.
The energy and charge of the solution are \cite{Dias:2022eyq}
\begin{align}
 E&={N^2\over \RAdS}{3y_+^2(1+y_+^2)+\lambda(2+3y_+^2)\over 4},\qquad
 Q={N^2\over \RAdS}\sqrt{\lambda(y_+^2+1)(y_+^2+\lambda)\over 4},
\end{align}
where $N$ is the rank of the dual super-Yang-Mills theory, and
the chemical potential, temperature, and entropy are
\begin{align}
 \mu&=\sqrt{\lambda(y_+^2+1)\over y_+^2+\lambda},\qquad
 \RAdS T={2y_+^2+\lambda+1\over 2\pi \sqrt{y_+^2+\lambda}},\qquad
 {S\over N^2}=\pi y_+^2\sqrt{y_+^2+\lambda}.
 \label{mrtp19May25}
\end{align}
This solution is supersymmetric $(E=Q)$ only for $y_+=0$, since
\begin{align}
 E^2-Q^2={N^4\over 16\RAdS ^2}y_+^2(1+y_+^2+\lambda)(8\lambda+9y_+^2+9\lambda y_+^2+9y_+^4)\ge 0.
\end{align}
Also, because $y_+^2\ge 0$, the first equation in \eqref{mrtp19May25} implies that
\begin{align}
    \lambda<\mu^2<1,\qquad\text{or}\qquad
    \lambda>\mu^2>1.
    \label{mu_and_range_of_lambda}
\end{align}
If $\mu=1$, $\lambda$ can take any positive value.

\begin{figure}[tb]
\begin{center}
\vspace{0.5cm}
\begin{tabular}{c@{\hspace{1.5ex}}c@{\hspace{1.5ex}}c@{\hspace{1.5ex}}c@{\hspace{1.5ex}}c}
\includegraphics[height=3cm]{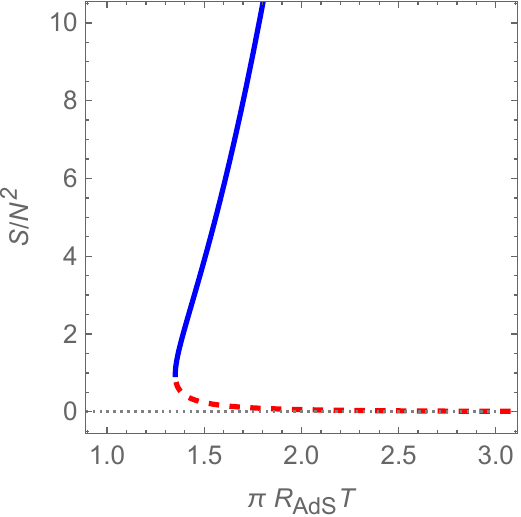}
&\includegraphics[height=3cm]{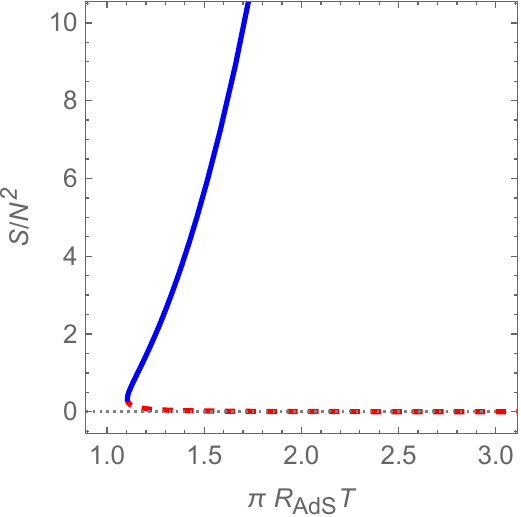}
&\includegraphics[height=3cm]{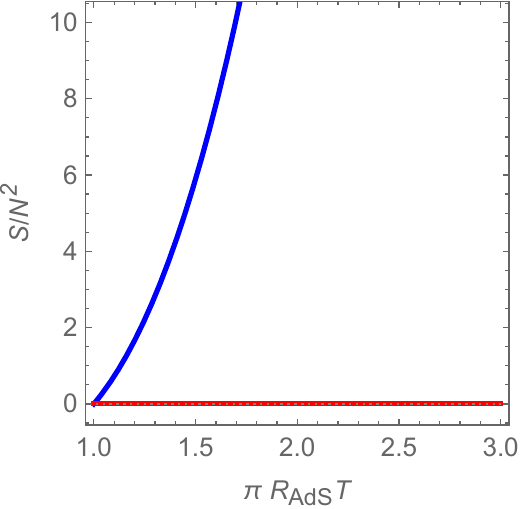}
&\includegraphics[height=3cm]{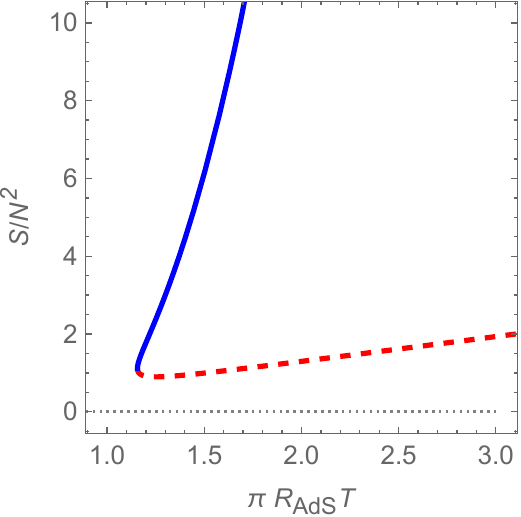}
&\includegraphics[height=3cm]{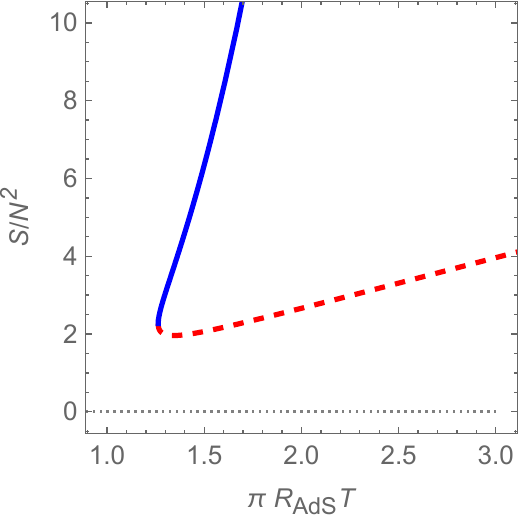}
\\
(a) $\mu=0.5$
&(b) $\mu=0.95$
&(c) $\mu=1$
&(d) $\mu=1.05$
&(e) $\mu=1.1$
\end{tabular}
\caption{\sl $S$ versus $T$ for the single R-charge black holes for various values of the chemical potential $\mu$. The blue lines are for the large black holes (stable) while the red dashed ones are for the small black holes (unstable).
The temperature is bounded from below by $T_{\rm min}={1/(\pi \RAdS)}$. For $\mu<1$ ((a) and (b)) the allowed values of $\lambda$ satisfy $\lambda<1$, while for $\mu>1$ ((d) and (e)) the allowed values satisfy $\lambda>1$ (see \eqref{mu_and_range_of_lambda}).  Panel (c) corresponds to the special case $\mu=1$, for which all $\lambda>0$ are allowed.  In this case, the large black hole branch (blue) corresponds to $\lambda=1$.  In the small black hole branch (red), for a given value of $T$, there are two  branches for  $\lambda$, one with $\lambda>1$ and another with $\lambda<1$; see \eqref{mu=0:E,Q,T}.}
\label{fig:SvsT}
\end{center}
\end{figure}

In the grand canonical ensemble in which $\mu$ and $T$ are fixed, there are two branches of solutions \cite{Dias:2022eyq,Henriksson:2019zph}.
The branch with larger entropy is dubbed the ``large'' BCS black hole and the one with smaller entropy the ``small'' BCS black hole \cite{Dias:2022eyq}.
It was shown in \cite{Dias:2022eyq} that the large black hole is thermodynamically stable while the small one is thermodynamically unstable, based on the Hessian of the entropy with respect to $E,Q$.
In Figure~\ref{fig:SvsT}, we show the values of $S$ and $T$ for large and small black hole branches for various values of $\mu$.

Let us consider the special case where the chemical potential is $\mu=1$.  This includes the supersymmetric solution, i.e., the superstar solution.  From 
\eqref{mrtp19May25}, we see that when $\mu=1$, there are two branches:
\begin{itemize}

  \item[{\it A)}] 
{\it Large black hole branch: $\lambda=1$}\\
This corresponds to the blue curve in Figure \ref{fig:SvsT}(c). The various quantities are
\begin{align}
\begin{split}
 h&=1+{1\over y^2},\qquad f=2+y^2-{y_+^2(y_+^2+2)\over y^2},\\
 E-Q&={N^2\over 4\RAdS }y_+^2(3y_+^2+4)>0,\qquad
 \pi \RAdS T=\sqrt{y_+^2+1}\ge 1,\\
 {S\over N^2}&=\pi y_+^2\sqrt{y_+^2+1}=\pi((\pi \RAdS T)^2-1)(\pi \RAdS  T),\label{jvvh15May24}
\end{split}
\end{align}
where $y_+\ge 0$.
This branch is thermodynamically stable, except at $y_+=0$ where it meets the small black hole branch and one eigenvalue of the Hessian becomes zero \cite[Appendix C]{Simon:2018laf}.

 \item[{\it B)}] 
{\it Small black hole branch: $y_+=0$}\\
This corresponds to the red line in Figure \ref{fig:SvsT}(c).  The various quantities are
\begin{align}
 h&=1+{\lambda\over y^2},\qquad f=y^2+\lambda+1,\label{mqxs15May24}\\
 E&=Q={N^2\over 2\RAdS}\lambda,\qquad
\pi \RAdS T={1\over 2}\biggl(\sqrt{\lambda}+{1\over \sqrt{\lambda}}\biggr)\ge 1,\qquad
 S=0.\label{mu=0:E,Q,T}
\end{align}
This solution is supersymmetric because $E=Q$ although, curiously, it has a finite temperature $T>0$.\footnote{This is presumably because the superstar ensemble wants to decay into the more natural hyperstar ensemble but is prevented from doing so by the constraint \eqref{iwnc20Nov23}.}  The minimum temperature is reached at
$\lambda=1$, at which $T=T_{\rm min}={1/(\pi \RAdS)}$ and $E=Q={N^2/(2\RAdS) }$.  
In  this branch, for a given value of $T$, there are two solutions with different values of the charge parameter $\lambda$, one with $\lambda<1$ and another with $\lambda>1$. The $\lambda<1$ branch is the $\mu\nearrow 1$ limit of the small black hole branch in the $\mu<1$ case shown in Figure~\ref{fig:SvsT}(a,b), while the $\lambda>1$ branch is the  $\mu\searrow 1$ limit of the small black hole branch in the $\mu>1$ case shown in Figure~\ref{fig:SvsT}(d,e).
This branch is thermodynamically unstable, even though supersymmetric, except at $\lambda=1$ where it meets the large black hole branch and the negative eigenvalue of the Hessian becomes zero.

This branch is nothing but the superstar solution discussed in section~\ref{sss:bulk_side}.
Indeed, if we substitute the data \eqref{z,V_superstara} into the LLM solution \eqref{lnej18Jul24} and redefine coordinates as \cite{Simon:2018laf}
\begin{align}
\begin{split}
    \eta&\to R_{\rm AdS} \,r \sin\theta,\qquad
    r\to \sqrt{r^2+(\lambda+1)R_{\rm AdS}^2}\cos\theta,
    \\
    \phi&\to \phi -{t\over R_{\rm AdS}},\qquad
    t\to {t\over R_{\rm AdS}},
\end{split}
\end{align}
we find that its 5d part precisely reduces to the metric \eqref{mcyp14Jul24} with \eqref{mqxs15May24}.  The full 10d solution can be also shown to agree with the uplift of the 5d solution \eqref{mcyp14Jul24}.
\end{itemize}

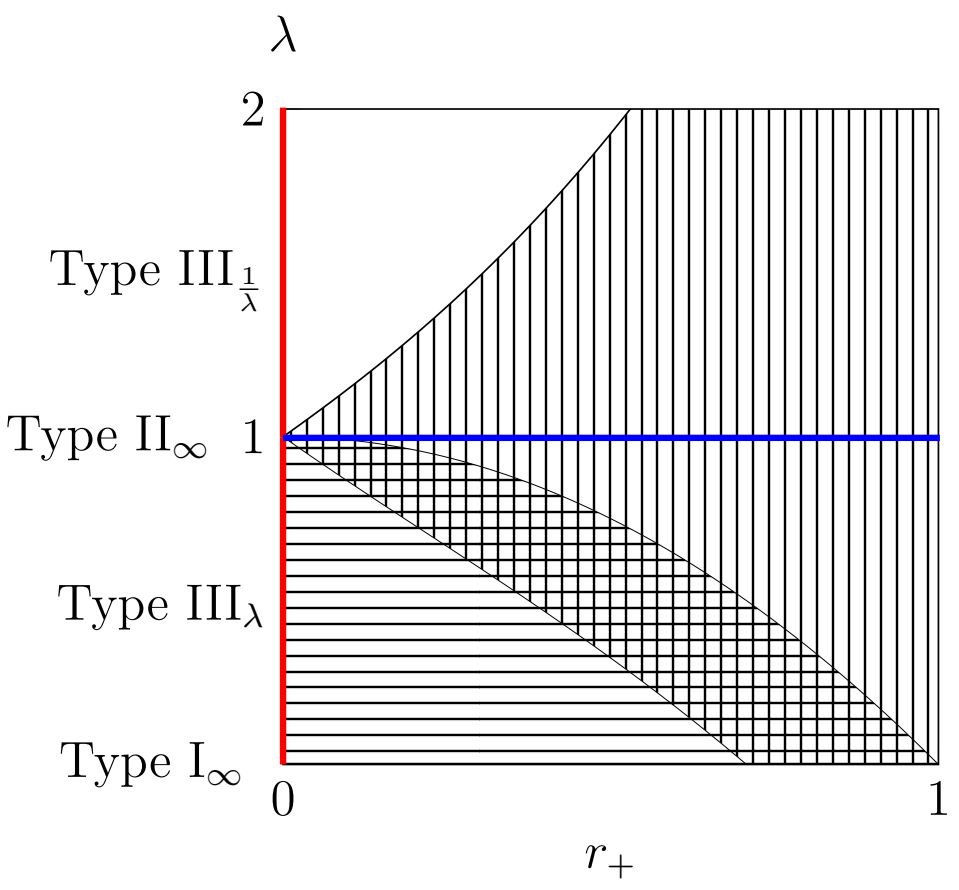
\begin{figure}[htbp]
\centering
\scalebox{0.8}{
   \begin{tikzpicture}
    \begin{axis}[
        width=7cm,
        height=7cm,
        xmin=0, xmax=1,
        ymin=0, ymax=2,
        xlabel={$r_+$},
        xtick={0,1},
        xticklabels={$0$,$1$},
        ytick={1,2},
        legend pos=north east,
        domain=0:1,
        samples=100
    ]
        
        \addplot[name path=upper, black, thin] {1 + x^2/2 + sqrt(8*x^2 + 9*x^4)/2};
        
        \addplot[name path=middle, black, very thin] {1 - x^2};
        
        \addplot[name path=lower, black, very thin] {1 + x^2/2 - sqrt(8*x^2 + 9*x^4)/2};

        \addplot[name path=zero, black, thick]{0};

        \addplot [pattern=vertical lines,pattern color=black] fill between[of=lower and upper];
        \addplot [pattern=horizontal lines,pattern color=black] fill between[of=zero and middle];

        \draw[thick] (axis cs:0,0) -- (axis cs:0,2);

        \node at (axis cs: -0.1, 1) [left] {($\lambda = 1$)};
        \node at (axis cs: -0.1, 0) [left] {($\lambda = 0$)};
    \end{axis}
    
    \draw[line width = 1.5,color=red]  (0,2.7)--(0,5.43);
    \draw[line width = 1.5,color=red]  (0,0)--(0,2.7);
    \draw[line width = 1.5,color=blue]  (0,2.7)--(5.43,2.7);

    \node at (0,0)  [left,xshift=-5] {Type I$_{\infty}$} ;
    \node at (0,1.3)[left] {Type III$_{\lambda}$};
    \node at (0,2.7)[left,xshift=-13] {Type II$_{\infty}$};
    \node at (0,4.0)[left] {Type III$_{1\over \lambda}$};
    \node at (0, 6) {\makecell{$\lambda$}};
\end{tikzpicture}}
\caption{\sl The phase diagram of the single R-charged black hole in 5d.  The correspondence with the algebra type of the fermionic fine-probe algebra  is discussed in section \ref{sec:conn_thermo_instability}.}
\label{CveticFigure1a}
\end{figure}

\bigskip\bigskip
We can also analyze stability in terms of  the Gibbs Euclidean action\footnote{If the usual Euclidean action is $I_E$, then the Gibbs Euclidean action is $I_G=I_E-\beta q\phi$, where $q$ is the R-charge and $\phi$ is the conjugate electric potential.} \cite{Cvetic:1999ne}. 
Fig.~\ref{CveticFigure1a} (which is mostly the same as Figure 1a in \cite{Cvetic:1999ne}) shows the phase diagram on the $r_+$-$\lambda$ plane.  The black hole solution exists everywhere on the plane, but it is a local minimum of the action only in the vertically hatched region; in other regions it is thermodynamically unstable.  In the horizontally hatched region, empty AdS$_5$ is the global minimum.  In the region hatched both horizontally and vertically, the black hole solution is metastable. Branch A above (the $\lambda=1$ large black hole) corresponds to the blue line in the figure,
while branch B, i.e., the superstar, corresponds to the red line in the figure.  So, again, we see that the superstar is unstable for $\lambda\neq 1$ and marginally stable at $\lambda=1$.

\subsection{Thermodynamic instability and algebraic Type}
\label{sec:conn_thermo_instability}

\begin{figure}[htbp]
\centering
\scalebox{0.9}{
   \begin{tikzpicture}[scale=1.0]
  \draw[->,>=stealth,semithick] (0,0)--(3.5,0) node[right]{$ \pi R_{\text{AdS}}T
  $
  }; 
\draw[->,>=stealth,semithick] (0,0)--(0,4) node[left]{$S/ N^{2}$}; 
\draw (0,0) node[below left]{$0$}; 
\draw (.9,0) node[below]{$1$}; 
\draw [very thick,domain=1:2,blue]  plot(\x,{pow(\x,3)-pow(\x,2)});
\draw [very thick, red] (1,0)--(3.4,0);

\node at (2.8,2.5)  {\makecell{\color{blue} A (stable)}};
\node at (2.2,-0.5)  {\makecell{\color{red} B \\[-.5ex] \color{red}(unstable)}};

\node at (-1.3,2.5)  {\makecell{Type II$_{\infty}$\\($\lambda=1$)}};
\node at (5.5,1.2)  {\makecell{%
Type III$_{\lambda}$ $(0<\lambda<1)$\\
or Type III$_{1/\lambda}$ $(1<\lambda)$\\
}};
\draw [thick,-latex] (-1.3,1.85)--(1,0);
\draw [thick,-latex] (3.5,1.2)--(2.7,0);
\end{tikzpicture}}
          \caption{\sl $S$ versus $T$ for the $\mu=1$ black holes (same as Figure \ref{fig:SvsT}(c)) and the algebra type--stability correspondence.  The blue branch (branch A) is a large black hole, which is thermodynamically stable, while the red branch (branch B) is a small black hole, which is thermodynamically unstable and is in this case a superstar solution.}
    \label{DiasFigure6}
\end{figure}
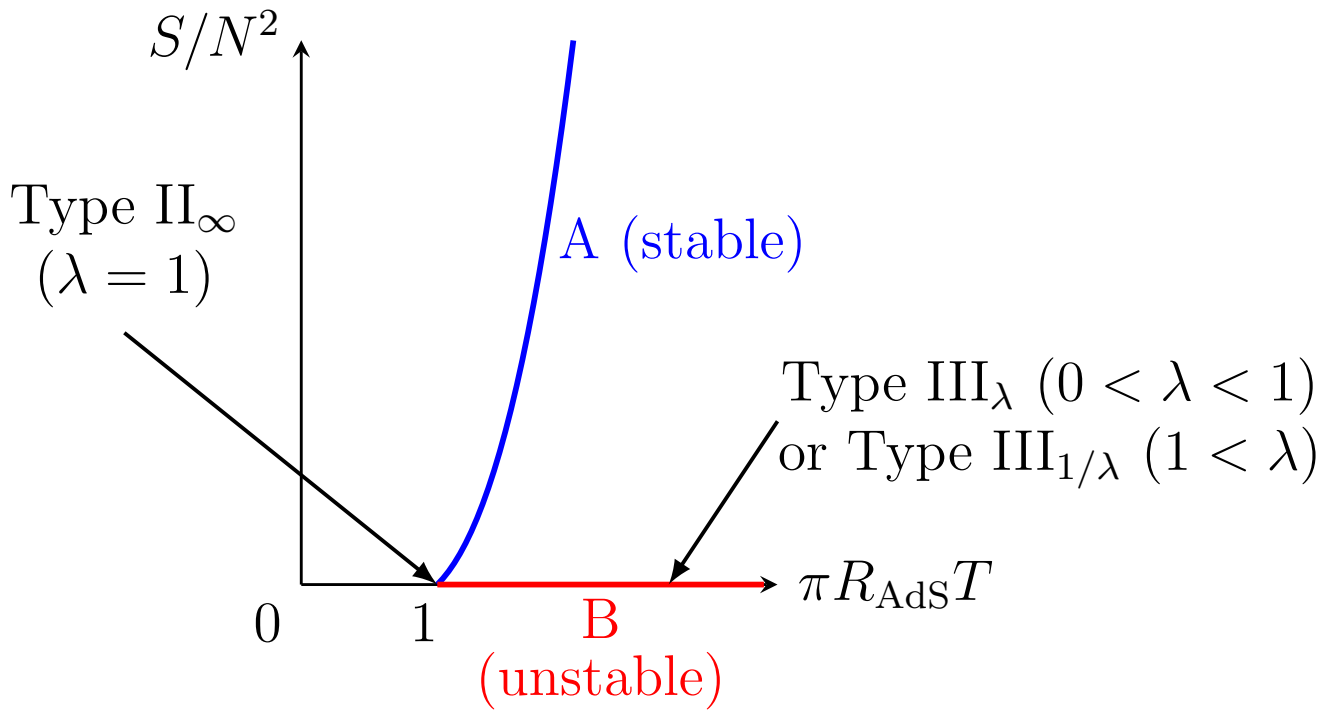

Above we saw that the superstar solution is thermodynamically unstable for $\lambda\neq 1$ and marginally stable for $\lambda = 1$.
Combining this fact with the algebraic classification of fermionic fine probes in section \ref{sec:fineFermion}, we find a correspondence between the algebra type and stability:  Type III$_\lambda$ corresponds to unstable solutions (unstable superstars),
Type II$_\infty$ corresponds to marginally stable solutions (the marginally stable superstar) in gravity, while Type I$_\infty$ corresponds to a stable solution (empty AdS$_5\times S^5$).
In Fig.~\ref{CveticFigure1a}, we displayed the algebra types that correspond to the superstar solution (red line).
In Fig.~\ref{DiasFigure6}, we present the $S$ versus $T$ plot of the $\mu=1$ solution that explains this algebra type--stability correspondence.   In Table \ref{tab:alg-stab}, we summarize the algebra type--stability correspondence for the superstar solutions.   Again, we should keep in mind that we are only looking at the algebra of operators whose action keeps us within the space of half-BPS states.

\begin{table}[htbp]
  \begin{center}
\begin{tabular}{c|c|c|c} \hline 
   $\lambda$   &  types & ER bridges & stability of spacetime \\ \hline\hline
   $\lambda=1$ &  II$_{\infty}$ & quantum & marginally stable  \\ 
   $0<\lambda<1$ & III$_{\lambda}$ &quantum &unstable \\ 
   $\lambda>1$ & III$_{1/\lambda}$ &quantum &unstable \\ 
   $\lambda=0$ & I$_{\infty}$ & (no ER bridge) & (stable) \\ \hline
 \end{tabular}
 \end{center} 
 \caption{\sl The algebraic Type--stability correspondence for the superstar solutions.\label{tab:alg-stab}}
\end{table}

The thermodynamic stability of the hyperstar solution is not known, because an explicit solution has not been written down and analyzed. But we expect it to be stable, because its boundary dual is just the Fermi-Dirac distribution that maximizes entropy at effective inverse temperature $\beta$.  We showed that the hyperstar ensemble corresponds to a Type III$_1$ algebra, so this would add an entry to Table \ref{tab:alg-stab} as a stable quantum ER bridge corresponding to Type~III$_1$.

\bibliography{refs}
\bibliographystyle{JHEP}


\end{document}